\documentclass[paper,toc,nohyper]{JHEP3}
\usepackage{epsfig}
\usepackage{amsbsy,ams} 
\usepackage{subfigure,epsfig}
\usepackage{graphicx}
\usepackage{pifont}
\usepackage{amsmath}
\usepackage{booktabs}
\usepackage{bm}
                 
\def\dsigma{{\rm d} \hat\sigma}
\def\ph#1{\phantom{.}}
\def\JET{J}

\def\Finite{{\cal F}inite}
\def\bom#1{{\mbox{\boldmath $#1$}}}

\def\doubletilde#1{\widetilde{\vphantom{\raise 1.5pt \hbox{#1}}\smash{\kern -2pt\widetilde{#1}}}}
\def\e{\epsilon}

\def\eps{\epsilon}

\def\h{\mathrm{H}}

\def\spa#1.#2{\left\langle\mskip-1mu#1\,#2\mskip-1mu\right\rangle}
\def\spb#1.#2{\left[\mskip-1mu#1\,#2\mskip-1mu\right]}

\def\NLO{\rm NLO}

\def\sinner{s^{\prime}}
\def\souter{s}

\def\l{\left}
\def\r{\right}

\def\b#1{\bar{#1}}

\def\h{\mathrm{H}}

\newcommand{\beq}{\begin{equation}}
\newcommand{\eeq}{\end{equation}}
\newcommand{\bea}{\begin{eqnarray}}
\newcommand{\eea}{\end{eqnarray}}
\def\z#1{{\zeta_{#1}}}

\def\n2f{{n^{\,2}_{\! f}}}
\def\pgg(#1){p_{\rm{gg}}(#1)}
\def\H(#1){{\rm{H}}_{#1}}
\def\Hh(#1,#2){{\rm{H}}_{#1,#2}}

\def\NLO{{\cal N}_{LO}}

\def\NNNLO{{\cal N}_{NNLO}}

\def\NNNLORR{{\cal N}_{NNLO}^{RR}}
\def\NNNLORV{{\cal N}_{NNLO}^{RV}}
\def\NNNLOVV{{\cal N}_{NNLO}^{VV}}
\def\NNLOV{{\cal N}_{NLO}^{V}}

\def\XX{\mathbb{X}}
\def\FF{\mathbb{X}}

\def\GG#1{\Gamma^{(#1)}}
\def\barGG#1{\overline{\Gamma}^{(#1)}}

\def\dPS#1#2{{\rm d}\Phi_{#1}(p_3,\ldots,p_{#2};p_1,p_2)\,\frac{1}{S_{#1}}}

\def\dPSxxyy#1#2{\int{\rm d}\Phi_{#1}(p_3,\ldots,p_{#2};x_1 y_1 p_1,x_2 y_2 p_2)\frac{1}{S_{#1}}\,\frac{{\rm d}x_1}{x_1}\frac{{\rm d}x_2}{x_2}\frac{{\rm d}y_1}{y_1}\frac{{\rm d}y_2}{y_2}}

\def\dPSzz#1#2{\int{\rm d}\Phi_{#1}(p_3,\ldots,p_{#2};z_1 p_1,z_2 p_2)\frac{1}{S_{#1}}\,\frac{{\rm d}z_1}{z_1}\frac{{\rm d}z_2}{z_2}}

\def\dPStwozz{\int{\rm d}\Phi_{2}(p_3,\ldots,p_{4};z_1 p_1,z_2 p_2)\frac{1}{2!}\,\frac{{\rm d}z_1}{z_1}\frac{{\rm d}z_2}{z_2}}
\def\dPStwo{\int{\rm d}\Phi_{2}(p_3,\ldots,p_{4};p_1,p_2)\frac{1}{2!}\,}

\allowdisplaybreaks

\preprint{
  IPPP/12/81\\
  \\
  \today}

\title{Double Virtual corrections for gluon scattering at NNLO}

\author{Aude Gehrmann-De Ridder$^a$, Thomas Gehrmann$^b$, E.W.N. Glover$^c$, Joao Pires$^a$
	\\
$^a$Institute for Theoretical Physics, ETH, CH-8093 Z\"urich, Switzerland\\
$^b$Institut f\"ur Theoretische Physik, Universit\"at Z\"urich, Wintherturerstrasse 190,\\ CH-8057 Z\"urich, Switzerland\\
	$^c$Institute for Particle Physics Phenomenology, University of Durham,
South Road,\\ Durham DH1 3LE, England
}	

\abstract{We use the antenna subtraction method to isolate the double virtual infrared singularities present in gluonic scattering amplitudes at next-to-next-to-leading order. In previous papers, we derived the subtraction terms that rendered (a) the double real radiation tree-level process finite in the single and double unresolved regions of phase space and (b) the mixed  single real radiation one-loop process both finite and well behaved in the unresolved regions of phase space.  
Here, we show how to construct the double virtual subtraction term using antenna functions with both initial- and final-state partons which remove the explicit infrared poles present in the two-loop amplitude.  As an explicit example, we write down the subtraction term for the four-gluon two-loop process. The infrared poles are explicitly and locally cancelled in all regions of phase space leaving a finite remainder that can be safely evaluated numerically in four-dimensions.
\\
\today
}

\begin{document}
\tableofcontents
\newpage
\section{Introduction}
\label{sec:intro}

In hadronic collisions the most basic form of the strong interaction at short
distances is the scattering of a coloured parton off another coloured parton.
Experimentally, such scattering can be observed via the production of one or
more jets of hadrons with large transverse energy. In QCD, the  inclusive
cross section has the form,
\begin{equation}
\label{eq:totsig}
{\rm d}\sigma(H_1,H_2) =\sum_{i,j} \int   
\frac{d\xi_1}{\xi_1} \frac{d\xi_2}{\xi_2} f_i(\xi_1,\mu^2) f_j(\xi_2,\mu^2) \dsigma_{ij}(\xi_1H_1,\xi_2H_2,\alpha_s(\mu)) \nonumber
\end{equation}
where the probability of finding a parton of type $i$ carrying a momentum fraction $\xi$ of the parent proton momentum $H_i$ is described by the parton distribution function $f_i(\xi,\mu^2)d\xi$ and the partonic scattering cross section ${\rm d}\hat\sigma_{ij}$  for parton $i$ to scatter off a parton $j$ normalised to the hadron-hadron flux\footnote{The partonic cross section normalised to the parton-parton flux is obtained by absorbing the inverse factors of $\xi_1$ and $\xi_2$ into $\dsigma_{ij}$.} is summed over the possible parton types $i$ and $j$. To simplify the discussion, we have set the renormalisation and factorisation scales to be equal, $\mu_R = \mu_F = \mu$.

For example, let us consider the $m$-jet cross section.  
The (renormalised and mass factorised) partonic cross section for a parton of type $i$ scattering off parton of type $j$, $\dsigma_{ij}$ to form $m$ jets has the perturbative expansion 
\begin{eqnarray}
\label{eq:sigpert}
\dsigma_{ij}(\xi_1H_1,\xi_2H_2,\alpha_s(\mu)) &=& \dsigma_{ij,LO}(\xi_1H_1,\xi_2H_2)
+\left(\frac{\alpha_s(\mu)}{2\pi}\right)\dsigma_{ij,NLO}(\xi_1H_1,\xi_2H_2)\nonumber \\
&+&\left(\frac{\alpha_s(\mu)}{2\pi}\right)^2\dsigma_{ij,NNLO}(\xi_1H_1,\xi_2H_2)
+{\cal O}(\alpha_s^3)
\end{eqnarray}
where the next-to-leading order (NLO) and next-to-next-to-leading order (NNLO) strong corrections are identified.

The simplest jet cross sections at hadron colliders are the single jet inclusive
and  the exclusive or inclusive dijet cross sections. In the single jet
inclusive cross section, each identified jet in an event contributes
individually. The exclusive  dijet cross section consists of all events with
exactly two identified jets, while events  with two or more identified jets
contribute to the inclusive dijet cross section. These  cross sections have been
studied as functions of different kinematical variables:  the transverse
momentum and rapidity of the jets (of any jet for the single jet  inclusive
distribution, or of the two largest transverse momentum jets for the  dijet
distributions). Precision measurements of single jet and dijet cross sections 
have been performed by CDF~\cite{cdfjet} and D0~\cite{d0jet} at the Tevatron and
by ATLAS~\cite{atlasjet,atlasjet2} and CMS~\cite{cmsjet,cmsjet2,cmsjet3} at the LHC.

The single jet inclusive jet cross section has been analysed in view  of a
determination of the parton distributions in the proton~\cite{asjet}.  Tevatron
data  on this observable are included in nearly all global fits of parton
distributions,  where they provide important constraints on the large-$x$
behaviour of the  gluon distribution. Likewise, measurements of the jet and
dijet cross section  can be used to extract the strong coupling constant
$\alpha_s$ up to scales  that are otherwise unattainable with other collider
measurements~\cite{Giele:1995kb}.

All these precision studies rely at present on NLO theory predictions for jet
cross sections~\cite{eks,jetrad,nlojet1,nlojet2,powheg2j,meks,Dittmaier:2012kx}.  The uncertainty inherent to these
predictions is the dominant  source of error in the extraction of $\alpha_s$. A
consistent inclusion of jet data  in global fits of parton distributions at a
given order also requires the perturbative  description of the jet cross section
to this order. Consequently, the inclusion of jet data in the NNLO parton
distribution fits introduces an unquantified systematic  error. The NNLO
corrections to jet production  are therefore of crucial importance for 
precision physics studies of QCD and of the structure of the  proton at hadron
colliders. 

At LO, the $m$-jet cross section is obtained by evaluating the tree-level cross section for processes with $m$-partons in the final state (i.e., the processes involving $(m+2)$-partons with two partons in the initial state), and requiring that each final state parton is identified as a jet by some jet algorithm.   In other words,
\begin{equation} 
\dsigma_{ij,LO}(\xi_1H_1,\xi_2H_2) = \int_{{\rm{d}}\Phi_{N}}
\dsigma_{ij,LO}^B(\xi_1H_1,\xi_2H_2)
\end{equation}
where the Born-level partonic cross section is integrated over the 
$N$-parton final state ${\rm{d}}\Phi_{N}$ subject to the constraint that precisely $m$ jets are observed by the jet algorithm $J_m^{(N)}$,
\begin{equation}
\int_{{\rm{d}}\Phi_{N}} \equiv \int {\rm{d}}\Phi_{N} J_m^{(N)}.
\end{equation}
In this case of course, $N=m$.

Suppose we now want to compute the $m$-jet cross section to NLO. For this, we have to consider the real radiation cross section $\dsigma_{NLO}^R$ with ($m+1$) partons in the final state, the one-loop correction $\dsigma_{NLO}^V$ with $m$ partons in the final state, and a mass factorisation counter-term $\dsigma_{NLO}^{MF}$ that absorbs the divergences arising from initial state collinear radiation into the parton densities:
\begin{eqnarray}
\dsigma_{ij,NLO}=\int_{{\rm d}\Phi_{m+1}}\dsigma^R_{ij,NLO}+\int_{{\rm d}\Phi_{m}}\dsigma^V_{ij,NLO}
+\int_{{\rm d}\Phi_{m}}\dsigma^{MF}_{ij,NLO}.\label{eq:NLOxsec}
\end{eqnarray}
The mass factorisation counter term is related to the LO cross section 
through
\begin{eqnarray}
\label{eq:MFNLO}
\dsigma_{ij,NLO}^{MF}(\xi_1 H_1,\xi_2 H_2) = 
-\int \frac{{\rm d}z_1}{z_1} \frac{{\rm d}z_2}{z_2}
\,\GG1_{ij;kl}(z_1,z_2)\,
\dsigma_{kl,LO}(z_1 \xi_1 H_1,z_2\xi_2 H_2),
\end{eqnarray}
where 
\begin{equation}
\GG1_{ij;kl}(z_1,z_2)=
 \delta(1-z_2) \,\delta_{lj}\,\Gamma^{(1)}_{ki}(z_1)   
+\delta(1-z_1)\,\delta_{ki}\,\Gamma^{(1)}_{lj}(z_2), 
\end{equation}
and $\GG1_{ij}(z)$ is the one-loop Altarelli-Parisi kernel.

The terms on the right hand side of (\ref{eq:NLOxsec}) are separately divergent although their sum is finite. To write a Monte Carlo program to compute the NLO contribution to the cross section,  we must first isolate and cancel the singularities of the different pieces and then numerically evaluate the finite remainders.

Subtraction schemes are a well established solution to this problem. They work by finding a suitable counter-term $\dsigma^S_{NLO}$ for $\dsigma^R_{NLO}$. This subtraction term has to satisfy two properties, namely it must have the same singular behaviour in all appropriate unresolved limits as ${\rm d}\hat\sigma^R_{NLO}$ and yet be simple enough to be integrated analytically over all singular regions of the $(m+1)$-parton phase space in $d$ dimensions. We proceed by rewriting (\ref{eq:NLOxsec}) in the following form:
\begin{eqnarray}
{\rm d}\hat\sigma_{ij,NLO}&=&\int_{{\rm d}\Phi_{m+1}}\left({\rm d}\hat\sigma_{ij,NLO}^R-{\rm d}\hat\sigma_{ij,NLO}^S\right)\nonumber \\
&
+&\int_{{\rm d}\Phi_m}\left(\int_1{\rm d}\hat\sigma_{ij,NLO}^S+{\rm d}\hat\sigma_{ij,NLO}^V+{\rm d}\hat\sigma_{ij,NLO}^{MF}\right).
\label{eq:NLOxsec2} 
\end{eqnarray}
In its unintegrated form the subtraction term, $\dsigma_{ij,NLO}^S$, has the same singular behaviour as $\dsigma_{ij,NLO}^R$ such that the first integral is finite by definition and can be integrated numerically in four dimensions over the $(m+1)$-parton phase space. The integrated form of the counter-term $\int_1 \dsigma_{ij,NLO}^S$  analytically cancels the explicit singularities of the virtual contribution $\dsigma_{ij,NLO}^V$ and the mass factorisation counter-term $\dsigma_{ij,NLO}^{MF}$. After checking the cancellation of the pole pieces, we can take the finite remainders of these contributions and perform the last integral on the right hand side of (\ref{eq:NLOxsec2}) numerically over the $m$-parton phase space.

At NNLO, there are three distinct contributions due to double real radiation radiation $\dsigma_{ij,NNLO}^{RR}$, mixed real-virtual radiation $\dsigma_{ij,NNLO}^{RV}$ and double virtual radiation $\dsigma_{ij,NNLO}^{VV}$,
plus two NNLO mass factorisation counter terms, $\dsigma_{ij,NNLO}^{MF,1}$ and 
$\dsigma_{ij,NNLO}^{MF,2}$, such that
\begin{eqnarray}
\dsigma_{ij,NNLO}&=&\int_{{\rm{d}}\Phi_{m+2}} \dsigma_{ij,NNLO}^{RR}
+\int_{{\rm{d}}\Phi_{m+1}} \dsigma_{ij,NNLO}^{RV} 
+\int_{{\rm{d}}\Phi_m}\dsigma_{ij,NNLO}^{VV}\nonumber \\
&+&\int_{{\rm{d}}\Phi_{m+1}}\dsigma_{ij,NNLO}^{MF,1}
+\int_{{\rm{d}}\Phi_m}\dsigma_{ij,NNLO}^{MF,2}
\end{eqnarray}
where the integration is again over the appropriate $N$-particle final state subject to the constraint that precisely $m$-jets are observed. As usual the individual contributions in the $m$-, $(m+1)$- and $(m+2)$-parton final states are all separately infrared divergent although, after renormalisation and factorisation, their sum is finite.

The NNLO 
mass factorisation counter-terms contributing to the $(m+1)$- and $m$-particle final states are given by,
\begin{eqnarray}
\label{eq:MFNNLOone}
\lefteqn{\dsigma_{ij,NNLO}^{MF,1}(\xi_1 H_1,\xi_2 H_2)=
-\int\frac{{\rm d}z_{1}}{z_{1}} \frac{{\rm d}z_{2}}{z_{2}}}\nonumber \\
&& \times\, \GG1_{ij;kl}(z_1,z_2)\,
\bigg[\dsigma_{kl,NLO}^{R}
-\dsigma_{kl,NLO}^S\bigg] (z_1\xi_1 H_1,z_2\xi_2 H_2), 
\end{eqnarray}
and
\begin{eqnarray}
\label{eq:MFNNLOtwo}
\lefteqn{\dsigma_{ij,NNLO}^{MF,2}(\xi_1 H_1,\xi_2 H_2)=
-\int\frac{{\rm d}z_{1}}{z_{1}} \frac{{\rm d}z_{2}}{z_{2}}}\nonumber \\
&\times&\Bigg[\GG2_{ij;kl}(z_1,z_2)\,
\dsigma_{kl,LO}(z_1\xi_1 H_1,z_2\xi_2 H_2)
\nonumber\\
&&+\GG1_{ij;kl}(z_1,z_2)\,
\left[\int_1\dsigma_{kl,NLO}^{S}
+\dsigma_{kl,NLO}^V+\dsigma_{kl,NLO}^{MF}\right](z_1 \xi_1 H_1,z_2 \xi_2 H_2) \Bigg],
\end{eqnarray}
respectively, where 
\begin{equation}
\GG2_{ij;kl}(z_1,z_2)= \delta(1-z_{2})\,  \delta_{lj}\Gamma_{ki}^{(2)}(z_{1})
+\delta(1-z_{1})\,  \delta_{ki}\Gamma_{lj}^{(2)}(z_{2}) +\Gamma_{ki}^{(1)}(z_{1})\Gamma_{lj}^{1}(z_2)
\end{equation}
and
$\Gamma^{(2)}_{ij}(z)$ is the two-loop Altarelli-Parisi kernel.

Following the same logic as at NLO, we introduce subtraction terms.  The first, $\dsigma^S_{NNLO}$, must correctly describe the single and double unresolved regions of the double real radiation contribution for $\dsigma^{RR}_{NNLO}$.  The second, $\dsigma^{VS}_{NNLO}$, must cancel the explicit poles in $\dsigma^{RV}_{NNLO}$ as well as reproducing the single unresolved limits.
The general form for the subtraction terms for an $m$-particle final state at NNLO is therefore:
\begin{eqnarray}
\label{eq:signnlogeneral}
\dsigma_{ij,NNLO}&=&\int_{{\rm{d}}\Phi_{m+2}}\left(\dsigma_{ij,NNLO}^{RR}-\dsigma_{ij,NNLO}^S\right)
+\int_{{\rm{d}}\Phi_{m+2}}\dsigma_{ij,NNLO}^S\nonumber\\
&+&\int_{{\rm{d}}\Phi_{m+1}}\left(\dsigma_{ij,NNLO}^{RV}-\dsigma_{ij,NNLO}^{VS}\right)
+\int_{{\rm{d}}\Phi_{m+1}}\dsigma_{ij,NNLO}^{VS}
+\int_{{\rm{d}}\Phi_{m+1}}\dsigma_{ij,NNLO}^{MF,1}\nonumber\\
&+&\int_{{\rm{d}}\Phi_m}\dsigma_{ij,NNLO}^{VV}
+\int_{{\rm{d}}\Phi_m}\dsigma_{ij,NNLO}^{MF,2}.
\end{eqnarray}
Note that because the analytic integration of the subtraction term $\dsigma_{ij,NNLO}^S$ over the single and double unresolved regions of phase space gives contributions to both the $(m+1)$- and $m$-parton final states, we can explicitly decompose the integrated double real subtraction term into two pieces that are integrated over the phase space of either one or two unresolved particles respectively,
\begin{equation}
\int_{{\rm{d}}\Phi_{m+2}}\dsigma_{ij,NNLO}^S
= \int_{{\rm{d}}\Phi_{m+1}} \int_1 \dsigma_{ij,NNLO}^{S,1}
+\int_{{\rm{d}}\Phi_{m}} \int_2 \dsigma_{ij,NNLO}^{S,2}.
\end{equation}
We can therefore rewrite Eq.~\eqref{eq:signnlogeneral} such that each line corresponds to a different number of final state particles,
\begin{eqnarray}
\dsigma_{ij,NNLO}&=&\int_{{\rm{d}}\Phi_{m+2}}\left[\dsigma_{ij,NNLO}^{RR}-\dsigma_{ij,NNLO}^S\right]
\nonumber \\
&+& \int_{{\rm{d}}\Phi_{m+1}}
\left[
\dsigma_{ij,NNLO}^{RV}-\dsigma_{ij,NNLO}^{T}
\right] \nonumber \\
&+&\int_{{\rm{d}}\Phi_{m\phantom{+1}}}\left[
\dsigma_{ij,NNLO}^{VV}-\dsigma_{ij,NNLO}^{U}\right],
\end{eqnarray}
where each of the square brackets is finite and well behaved in the infrared singular regions.  More precisely,
\begin{eqnarray}
\label{eq:Tdef}
\dsigma_{ij,NNLO}^{T} &=& \phantom{ -\int_1 }\dsigma_{ij,NNLO}^{VS}
- \int_1 \dsigma_{ij,NNLO}^{S,1} - \dsigma_{ij,NNLO}^{MF,1},  \\
\label{eq:Udef}
\dsigma_{ij,NNLO}^{U} &=& -\int_1 \dsigma_{ij,NNLO}^{VS}
-\int_2 \dsigma_{ij,NNLO}^{S,2}
-\dsigma_{ij,NNLO}^{MF,2}.
\end{eqnarray}

So far, the discussion has been quite general, and the form of the subtraction terms not specified.   At NLO, there are several very well defined approaches  
for systematically constructing NLO subtraction terms, notably those due to
Catani and Seymour \cite{Catani:1996vz} and Frixione, Kunzst and Signer \cite{Frixione:1995ms}.

Several subtraction schemes for NNLO calculations have been proposed in the 
literature, where they are worked out to a varying level of detail. Up to now,
successful applications of subtraction at NNLO to specific observables were
accomplished with sector
decomposition~\cite{Binoth:2000ps,Heinrich:2002rc,Anastasiou:2003gr,
Binoth:2004jv,Anastasiou:2010pw},  $q_T$-subtraction~\cite{qtsub}, antenna
subtraction~\cite{GehrmannDeRidder:2005cm} and  most recently with an approach
based on sector-improved  residue subtraction~\cite{stripper1,stripper2}. The
sector decomposition approach relies on an iterated decomposition of the final
state phase space and matrix  element, allowing an expansion in distributions,
followed by a numerical  evaluation of the sector integrals. It has been
successfully applied to  Higgs
production~\cite{babishiggs1,babishiggs2,babishiggs3} and vector boson 
production~\cite{kirilldy} at NNLO. The $q_T$-subtraction method is  restricted
to processes with colourless final states at leading order, it is  based on the
universal infrared structure of the real emissions, which can be  inferred from
transverse momentum resummation. This method has been applied  at NNLO to the
Higgs production~\cite{grazzinihiggs}, vector boson 
production~\cite{grazzinidy1,grazzinidy2}, associated
$VH$-production~\cite{grazziniwh}, photon pair production~\cite{grazzinigg} and
in modified form to  top quark decay~\cite{scettop}. Antenna subtraction is
described in detail below and has been applied to three-jet production in
electron-positron
annihilation~\cite{our3j1,our3j2,weinzierl3j1,weinzierl3j2,weinzierl3j3} and
related  event shape
distributions~\cite{ourevent1,ourevent2,weinzierlevent1,ourevent3,weinzierlevent2}. 
The sector-improved residue subtraction  extends NLO residue
subtraction~\cite{Frixione:1995ms}, combined with a numerical  evaluation of the
integrated subtraction terms. It has been applied to  compute the NNLO
corrections to top quark pair production~\cite{czakontop1,czakontop2,Czakon:2012pz}.

Here, we are interested in the application of the antenna subtraction
scheme to hadron-hadron collisions, and specifically to processes such as $pp \to {\rm jet}+X$, $pp \to V+{\rm jet}+X$ (with $V=W^\pm$ or $Z$) and $pp \to H+{\rm jet}+X$.  The formalism has been fully worked out
for processes with massless partons and initial state hadrons at NLO~\cite{Daleo:2006xa} and
NNLO~\cite{Daleo:2009yj,Boughezal:2010mc,Gehrmann:2011wi,GehrmannDeRidder:2012ja}. 
The extension of antenna subtraction for the production of heavy particles at hadron colliders has been studied in Refs.~\cite{Abelof:2011jv,Abelof:2011ap,Abelof:2012rv,Abelof:2012he}.
Within this approach, 
the subtraction terms are constructed from so-called antenna functions which
describe all unresolved partonic radiation (soft and collinear) between a hard pair of radiator partons. The hard radiators may be in the initial or in the final state, and
in the most general case, final-final (FF), initial-final  (IF) and initial-initial  (II) antennae need to be considered. The subtraction terms and therefore the antennae
also need to be integrated analytically over the unresolved phase space, which is different in the three configurations.
All of the integrals relevant for processes at NNLO with massless quarks have now been completed \cite{Daleo:2009yj,Boughezal:2010mc,Gehrmann:2011wi,GehrmannDeRidder:2012ja}.
		  
In previous papers~\cite{Glover:2010im,GehrmannDeRidder:2011aa}, the subtraction terms $\dsigma_{NNLO}^S$ and $\dsigma_{NNLO}^T$
corresponding to the leading colour pure gluon contribution to dijet production at hadron colliders were derived.   $\dsigma_{NNLO}^S$ was shown to reproduce the singular behaviour present in $\dsigma_{NNLO}^{RR}$  in all of the single and double unresolved limits, while    
$\dsigma_{NNLO}^T$ analytically cancelled the explicit infrared poles present in $\dsigma_{NNLO}^{RV}$ and also reproduced the singular behaviour in the single unresolved limits.

It is the purpose of this paper to construct the appropriate subtraction term
$\dsigma_{NNLO}^{U}$ to render the leading colour four-gluon double virtual contribution
$\dsigma_{NNLO}^{VV}$ explicitly finite and numerically well behaved in all
of phase space.  

Our paper is organised in the following way.    In
Sect.~\ref{sec:VVsub}, the general structure of $\int_2\dsigma^{S,2}_{NNLO}$
and $\int_1 \dsigma^{VS}_{NNLO}$ is discussed and analysed, together with the detailed structure for
gluon scattering at leading-order in the number of colours.  

We review gluon scattering at LO and NLO in Sect.~\ref{sec:gluonscattering}. There are
two separate configurations relevant for $gg \to gg$ scattering depending on whether the
two initial state gluons are colour-connected or not. We denote the configuration where
the two initial state gluons are colour-connected (i.e. adjacent) by IIFF, while the
configuration where the two initial state gluons are not colour-connected is denoted by
IFIF.  Our notation for gluonic amplitudes is summarised in
Sect.~\ref{subsec:gluonamps} while the one-loop, unintegrated and integrated
subtraction contributions and mass factorisation terms at NLO are reviewed in
Sect.~\ref{subsec:gluonNLO}

In Sect.~\ref{sec:VVgluon} we turn our attention to the specific process of gluon
scattering at NNLO.  We consider the infrared pole structure of the two-loop four-gluon
amplitudes in Sect.~\ref{subsec:four-VV}.  The integrated subtraction terms
$\int_2\dsigma^{S,2}_{NNLO}$ and $\int_1 \dsigma^{VS}_{NNLO}$ are presented in 
Sect.~\ref{subsec:four-intsub} and the mass factorisation term discussed in
Sect.~\ref{subsec:four-MF2}.  Together, these combine to give the  the double virtual
subtraction term $\dsigma_{NNLO}^{U}$ and explicit forms for the IIFF and IFIF
configurations are given in Sect.~\ref{sec:dsigmaU} where the cancellation of explicit
poles between  $\dsigma_{NNLO}^{VV}$ and  $\dsigma_{NNLO}^{U}$ is made manifest.   Finally, our findings are summarised in Sect.~\ref{sec:conc}.

Two appendices are also enclosed.  Appendix~A summarises the phase space
mappings for the final-final, initial-final and initial-initial configurations.
Appendix~B collects the splitting functions relevant to gluon scattering. 
Appendix~C gives details of the convolutions of integrated tree-level antennae appearing in $\dsigma_{NNLO}^{U}$.

\section{Double virtual antenna subtraction at NNLO}
\label{sec:VVsub}
In this paper, we focus on the scattering 
of two massless coloured partons to produce massless coloured partons, 
and particularly the production of jets from gluon scattering in hadronic collisions.
We follow closely the notation of Refs.~\cite{Glover:2010im,GehrmannDeRidder:2011aa}.
The leading-order parton-level contribution from the $(m+2)$-parton processes to the $m$-jet cross section at LO in $pp$ collisions,
\begin{equation}
\label{eq:process}
pp \to m~{\rm jets}
\end{equation}
is given by  
\begin{eqnarray}
{\rm d}\hat\sigma_{LO}&=&\NLO \sum_{\textrm{perms}}{\rm d}\Phi_{m}(p_3,\hdots,p_{m+2};p_1,p_2)\frac{1}{S_{m}}\nonumber\\
&&\times|{\cal M}_{m+2}(\hat{1},\hat{2}\ldots,m+2)|^2 J_{m}^{(m)}(p_3,\hdots,p_{m+2})
\label{eq:LOcross}.
\end{eqnarray}
We denote a
generic tree-level $(m+2)$-parton colour ordered amplitude by the symbol 
${\cal M}_{m+2}(\hat{1},\hat{2}\ldots,m+2)$,
where $\hat{1}$ and
$\hat{2}$ denote the initial state partons of momenta $p_1$ and  $p_2$
while the $m$-momenta in the final state are labeled 
$p_3,\hdots,p_{m+2}$. For convenience, and where the order of momenta does not matter, we will often denote the set of $(m+2)$-momenta $\{p_1,\hdots,p_{m+2}\}$ by $\{p\}_{m+2}$. The symmetry factor $S_{m}$ accounts for the production of identical particles in the final-state. 

This squared matrix element is decomposed in a colour-ordered manner into leading and sub-leading colour contributions (as for example
derived explicitly in \cite{our3j1}). At leading colour, $|{\cal M}_{m+2}(\ldots)|^2$  consists of the squares of the
colour-ordered amplitudes, while at sub-leading colour, it is made from the appropriate sum of interference terms of colour-ordered amplitudes. Both at leading and sub-leading colour, $|{\cal M}_{m+2}(\ldots)|^2$ can be decomposed such that each potentially unresolved parton is colour connected to only two other partons.

For gluonic amplitudes the permutation sum runs over the group of non-cyclic permutations of $n$ symbols.   
The normalisation factor, ${\cal N}_{LO}$, includes the hadron-hadron flux-factor, spin and colour summed and averaging factors as well as the dependence on the renormalised QCD coupling
constant $\alpha_s$. 

The $2\to m$ particle phase space ${\rm d}\Phi_{m}$ is given by
\begin{eqnarray}
&&{\rm d}\Phi_{m}(p_3,\ldots,p_{m+2}; p_1, p_2)=\nonumber\\
&&\frac{{\rm d}^{d-1}p_3}{2E_3(2\pi)^{d-1}}\ldots\frac{{\rm d}^{d-1}p_{m+2}}{2E_{m+2}(2\pi)^{d-1}}
(2\pi)^d\delta^d(p_1+p_2-p_3-\ldots-p_{m+2}).
\end{eqnarray}
The jet function $J_{m}^{(n)}(\{p\}_{n+2})$ defines the
procedure for building $m$ jets from $n$ final state partons.  
The key property of
$J_{m}^{(n)}$ is that the jet observable is collinear 
and infrared safe.

In a previous paper~\cite{Glover:2010im}, we discussed the NNLO
contribution coming from processes where two additional partons are
radiated, the double real contribution  $\dsigma^{RR}_{NNLO}$ and its
subtraction term $\dsigma^{S}_{NNLO}$. $\dsigma^{RR}_{NNLO}$ involves
the $(m+4)$-parton process at tree level and is given by, 
\begin{eqnarray}
{\rm d}\hat\sigma^{RR}_{NNLO}&=&\NNNLO^{RR}
\sum_{\textrm{perms}}\dPS{m+2}{m+4} \nonumber \\
&&\times|{\cal M}_{m+4}(\hat{1},\hat{2}\ldots,m+4)|^2 J_{m}^{(m+2)}(\{p\}_{m+4})
\label{eq:RRcross}.
\end{eqnarray}

Subsequently in~\cite{GehrmannDeRidder:2011aa} we discussed the NNLO contribution from one-loop processes 
where one additional parton is radiated, the real-virtual contribution $\dsigma^{RV}_{NNLO}$
and its subtraction term $\dsigma^{T}_{NNLO}$. $\dsigma^{RV}_{NNLO}$ involves the
$(m+3)$-parton process at one-loop and is given by,
\begin{eqnarray}
\dsigma^{RV}_{NNLO}
&=& \NNNLO^{RV} \sum_{\textrm{perms}}\dPS{m+1}{m+3} \nonumber \\ &&\times
|{\cal M}^1_{m+3}(\hat{1},\ldots,m+3)|^2\;
J_{m}^{(m+1)}(\{p\}_{m+3})\;,
\label{eq:nnloonel}
\end{eqnarray}
where we introduced a shorthand notation for the interference of one-loop and tree-amplitudes, 
\begin{equation}
\label{eq:m1def}
|{\cal M}^1_{m+3}(\hat{1},\ldots,m+3)|^2 
= 2 \,{\rm Re}\, \left({\cal M}^{1}_{m+3}(\hat{1},\ldots,m+3)\,
{\cal M}^{{0},*}_{m+3}(\hat{1},\ldots,m+3)\right)\;,
\end{equation}
which explicitly 
captures the colour-ordering of the leading colour contributions.

In this paper, we are concerned with the NNLO contribution coming from two-loop processes, i.e., the  $(m+2)$-parton process, $\dsigma^{VV}_{NNLO}$
and the remaining subtraction and mass factorisation contributions that are collectively denoted $\dsigma_{NNLO}^{U}$.

In our notation, the two loop $(m+2)$-parton contribution to $m$-jet final states at NNLO in hadron-hadron collisions is given by
\begin{eqnarray}
\dsigma^{VV}_{NNLO}
&=& \NNNLO^{VV} \sum_{\textrm{perms}}\dPS{m}{m+2} \nonumber \\ &&\times
|{\cal M}^2_{m+2}(\hat{1},\ldots,m+2)|^2\;
J_{m}^{(m)}(\{p\}_{m+2})\;,
\label{eq:nnlotwol}
\end{eqnarray}
where again we use a shorthand notation for the interference of two-loop and tree-amplitudes plus the one-loop squared contribution, 
\begin{eqnarray}
\label{eq:m2def}
|{\cal M}^2_{m+2}(\hat{1},\ldots,m+2)|^2 
&=&2 \,{\rm Re}\, \left({\cal M}^{2}_{m+2}(\hat{1},\ldots,m+2)\,
{\cal M}^{{0},*}_{m+2}(\hat{1},\ldots,m+2)\right)\nonumber\\ 
&\phantom{2 \,{\rm Re}}&+\,\left({\cal M}^{1}_{m+2}(\hat{1},\ldots,m+2)\,
{\cal M}^{{1},*}_{m+2}(\hat{1},\ldots,m+2)\right).
\end{eqnarray}
The subleading contributions in colour are implicitly included in \eqref{eq:RRcross}, \eqref{eq:m1def} and
\eqref{eq:m2def} but will not be considered in 
detail in this paper. 

The normalisation factor $\NLO$ depends on the specific process and parton channel under consideration. Nevertheless, at leading colour $\NNNLOVV$,
$\NNNLORV$ and $\NNNLORR$ are simply related to $\NLO$ for any number of jets and for any partonic process by 
\begin{eqnarray}
\NNNLOVV &=& \NLO \left(\frac{\alpha_s N}{2\pi}\right)^2 \bar{C}(\e)^2,\nonumber\\
\NNNLORV&=&{\cal N}_{LO}  \left(\frac{\alpha_s N}{2\pi}\right)^{2}
\frac{\bar{C}(\epsilon)^2}{C(\epsilon)},\\
\NNNLORR&=&{\cal N}_{LO}  \left(\frac{\alpha_s N}{2\pi}\right)^{2}
\frac{\bar{C}(\epsilon)^2}{C(\epsilon)^2},
\end{eqnarray}
where
\begin{eqnarray}
\label{eq:Cdef}
C(\epsilon)=(4\pi)^{\epsilon}\frac{e^{-\epsilon\gamma}}{8\pi^2},\\
\label{eq:Cbar}
\bar{C}(\e)=(4\pi)^{\e}e^{-\e\gamma}.
\end{eqnarray}
Note that each power of the (bare) coupling is accompanied by a factor of $\bar{C}(\e)$.
In this paper, we are mainly concerned with the NNLO corrections to \eqref{eq:process} when
$m=2$ and for the pure gluon channel at leading order in the number of colours.
For this special case, the leading order normalisation factor ${\cal N}_{LO}$ is given by
\begin{eqnarray}
\label{eq:NLOdef}
{\cal N}_{LO} &=& \frac{1}{2s} \times \frac{1}{4(N^2-1)^2}\times \left(g^2 N\right)^{2} (N^2-1)
\end{eqnarray}
where $s$ is the invariant mass squared of the colliding hadrons.

The renormalised two-loop virtual correction  ${\cal M}^{{2}}_{m+2}$  to the
$(m+2)$-parton matrix element in Eq.~\eqref{eq:nnlotwol} contains explicit global infrared  poles, which can be expressed
using the infrared singularity operators defined  in
\cite{Catani:1998bh,Sterman:2002qn}. As discussed in Sect.~\ref{sec:intro}, in order
to carry out the numerical integration over the $m$-parton phase,
weighted by the appropriate jet function, we have to construct an 
infrared subtraction term\footnote{Strictly speaking, $\dsigma^{U}_{NNLO}$ is not a subtraction term since its made up of integrated subtractions terms from the double real radiation
and real-virtual radiation. Nevertheless, since it contains all the terms needed to render the $m$-particle final state finite, it is convenient to call it the double virtual subtraction term.} $\dsigma^{U}_{NNLO}$ which 
removes these explicit infrared poles of the double virtual 
$(m+2)$-parton matrix element.

The subtraction term has three components given by Eq.~\eqref{eq:Udef}. 
$\int_2 \dsigma_{NNLO}^{S,2}$ is derived from the double real radiation subtraction term $\dsigma_{NNLO}^{S}$ integrated over the phase space of two unresolved particles
and $\int_1 \dsigma_{NNLO}^{VS}$ is derived from the real-virtual subtraction term $\dsigma_{NNLO}^{VS}$ integrated over the single unresolved
phase space.
These two contributions partially cancel the explicit poles in the  double virtual matrix element. The remaining poles are associated with the initial state collinear
singularities and are absorbed by the mass factorisation counter-term
$\dsigma_{NNLO}^{MF,2}$.  

One of the key features of any infrared subtraction scheme is the factorisation of the matrix elements and phase space in the singular limits where one or more particles are unresolved. In the antenna subtraction scheme, this factorisation is achieved through an appropriate phase space mapping described in
Appendix~\ref{sec:appendixA}, such that the singularities are isolated in an antenna that multiplies matrix elements which involve only hard partons with redefined momenta.  In determining the various contributions to $\dsigma_{NNLO}^{U}$, we shall therefore specify the integrated antennae and the reduced colour ordered matrix-element squared involved.     For conciseness, only the redefined hard radiator momenta will be specified in the functional dependence
of the matrix element squared. The other momenta will simply be denoted by ellipsis. 

In order to combine the integrated subtraction terms and the  double virtual matrix elements, it is convenient to slightly modify the phase space, such that
\begin{eqnarray}
\dsigma^{VV}_{NNLO}
&=& \NNNLO^{VV}\sum_{\textrm{perms}}\dPSzz{m}{m+2} \nonumber \\ &&\times
|{\cal M}^2_{m+2}(\hat{1},\ldots,m+2)|^2\;\delta(1-z_1)\delta(1-z_2)\,
J_{m}^{(m)}(\{p\}_{m+2})\;.\nonumber\\
\label{eq:nnloonelalt}
\end{eqnarray}
The integration over $z_1$ and $z_2$ reflects the fact that the subtraction terms contain contributions due to radiation from the initial state such that the
parton momenta involved in the hard scattering carry only a fraction $z_i$ of the incoming momenta. 
In general, there are three regions:   
the soft ($z_1 = z_2 = 1$), collinear ($z_1 = 1$, $z_2 \neq 1$ and $z_1 \neq 1$, $z_2 = 1$) 
and hard ($z_1 \neq 1$, $z_2 \neq 1$).  The double virtual matrix elements only contribute in the soft region, as indicated by the two delta functions.

In Sections~\ref{subsec:intS2} and \ref{subsec:intVS}, we discuss the first two terms that contribute to $\dsigma_{NNLO}^{U}$ given in Eq.~\eqref{eq:Udef},
namely $\int_2 \dsigma_{NNLO}^{S,2}$ and $  \int_1 \dsigma_{NNLO}^{VS}$. The final contribution $\dsigma_{NNLO}^{MF,2}$ is discussed in Section~\ref{sec:dsigmaU}.

\subsection{Contributions from integration of double real subtraction term: $\int_{2}\dsigma_{NNLO}^{S,2}$}
\label{subsec:intS2}
There are five different types of contributions to ${\rm{d}}\hat\sigma_{NNLO}^S$
according to the colour connection of the unresolved partons \cite{GehrmannDeRidder:2005cm,our3j1}:
\begin{itemize}
\item[(a)] One unresolved parton but the experimental observable selects only
$m$ jets.
\item[(b)] Two colour-connected unresolved partons (colour-connected).
\item[(c)] Two unresolved partons that are not colour-connected but share a common
radiator (almost colour-connected).
\item[(d)] Two unresolved partons that are well separated from each other 
in the colour 
chain (colour-unconnected).
\item[(e)] Compensation terms for the over subtraction of large angle soft emission.
\end{itemize}
In the antenna subtraction approach, each type of contribution takes the form of antenna functions multiplied by colour ordered matrix elements.   The various types of contributions are summarised in Table~1.  
\begin{table}[t!]
\begin{center}
{\small
\begin{tabular}{|c|ccccc|}
\hline
 & $a$ & $b$ & $b,c$ & $d$ & $e$   \\\hline
$\dsigma_{NNLO}^{S}$ 
& $X_3^0 |{\cal M}^0_{m+3}|^2$ 
& $X_4^0 |{\cal M}^0_{m+2}|^2$ 
& $X_3^0 X_3^0 |{\cal M}^0_{m+2}|^2$
& $X_3^0 X_3^0 |{\cal M}^0_{m+2}|^2$ 
& $S X_3^0 |{\cal M}^0_{m+2}|^2$   \\
$\int_1 \dsigma_{NNLO}^{S,1}$ 
& ${\cal X}_3^0 |{\cal M}^0_{m+3}|^2$  
& -- 
&  ${\cal X}_3^0 X_3^0 |{\cal M}^0_{m+2}|^2$ 
& -- 
& ${\cal S} X_3^0 |{\cal M}^0_{m+2}|^2$ \\
$\int_2 \dsigma_{NNLO}^{S,2}$ 
& -- 
&  ${\cal X}_4^0 |{\cal M}^0_{m+2}|^2$ 
& -- 
&${\cal X}_3^0 {\cal X}_3^0 |{\cal M}^0_{m+2}|^2$ 
&  --   \\ \hline
\end{tabular}
}
\end{center}
\label{tab:S1breakdown}
\caption{Type of contribution to the double real subtraction term ${\rm{d}}\hat\sigma_{NNLO}^{S}$, together with the integrated form of each term.   The unintegrated antenna and soft functions are denoted as $X_3^0$, $X_4^0$ and $S$ while their integrated forms are ${\cal X}_3^0$, ${\cal X}_4^0$ and ${\cal S}$ respectively.  ${\cal M}^0_{n}$ denotes an $n$-particle tree-level colour ordered amplitude.  }
\end{table}
We see that on the one hand the $a$, $c$ and $e$ types of subtraction terms, as well as the $b$-type terms that are products of three-particle antennae, 
can be integrated over a single unresolved particle phase space and therefore contribute to the $(m + 1)$-particle final state of the real-virtual contribution. 
This was described in detail in~\cite{GehrmannDeRidder:2011aa}. On the other hand the double unresolved antenna functions $X_4^0$ contribution to $\dsigma_{NNLO}^{S,b}$ (which we denote by $\dsigma_{NNLO}^{S,b,4}$)
and the colour-unconnected $X_3^0 X_3^0$ terms of $\dsigma_{NNLO}^{S,d}$ can immediately be integrated over the phase space of both unresolved particles and appear directly in the $m$-particle final state of the double virtual contribution, 
\begin{equation}
\label{eq:S2}
\int_2 \dsigma_{NNLO}^{S,2} = 
\int_2 \dsigma_{NNLO}^{S,b,4}
+\int_2 \dsigma_{NNLO}^{S,d}.
\end{equation}
We now turn to a detailed discussion of each of the terms in Eq.~\eqref{eq:S2}.

\subsubsection
[Integration of colour connected double unresolved antennae: $\\\int_2\dsigma_{NNLO}^{S,b,4}$]
{Integration of colour connected double unresolved antennae: $\int_2\dsigma_{NNLO}^{S,b,4}$}
In the antenna subtraction approach, the colour-connected double unresolved configuration coming from the tree-level process with two additional particles, 
i.e., the double real process involving ($m+4$) partons, is subtracted using a four-particle antenna function - two hard radiator partons emitting two
unresolved partons. Once it is integrated over the unresolved phase space of both partons, one recovers an ($m+2$)-parton contribution that cancels explicit
pole contributions in the virtual two-loop ($m+2$)-parton matrix element.

The integrated subtraction term formally written as $\int_{2}\dsigma_{NNLO}^{S,b,4}$ is split into three different contributions, depending on whether the hard radiators are in
the initial or final state.

When both hard radiators $i$ and $l$ are in the final state, then $X^0_{ijkl}$ is a final-final (FF) antenna function that describes all colour connected double unresolved
singular configurations (for this colour-ordered amplitude) where partons $j,k$ are unresolved. The subtraction term, summing over all possible positions of the unresolved
partons, reads,
\begin{eqnarray}
&&\dsigma_{NNLO}^{S,b,4(FF)}=\NNNLO^{RR}\sum_{\textrm{perms}}\dPS{m+2}{m+4}\nonumber\\
&&\times\sum_{j,k} X_{ijkl}^0 |{\cal M}_{m+2}(\hdots,I,L,\hdots)|^2
J_{m}^{(m)}(\{p\}_{m+2}).
\label{eq:sub1}
\end{eqnarray}
Besides the four parton antenna function $X^0_{ijkl}$ which depends only on
$p_i$, $p_j$, $p_k$ and $p_l$,
the subtraction term involves an $(m+2)$-parton amplitude depending on 
the redefined on-shell momenta $p_I$ and $p_L$, whose
definition in terms of the original momenta is given in Appendix~\ref{sec:appendixAFF}. 
The $(m+2)$-parton amplitude also depends
on the other final state momenta which, in the final-final map, are not redefined and
on the two initial state momenta $p_1$ and $p_2$. This dependence is
manifest as the ellipsis in \eqref{eq:sub1}.  The jet function is applied to the $m$ final state momenta that remain after the mapping,
i.e., $\{p_3, \ldots, p_I, p_L, \ldots, p_{m+4}\}\equiv\{p\}_{m+2}$.

To perform the integration of the subtraction term in Eq.~\eqref{eq:sub1} and make its infrared poles explicit, we exploit the following factorisation of the phase space,
\begin{eqnarray}
\label{eq:psx1}
\lefteqn{{\rm d} \Phi_{m+2}(p_{3},\ldots,p_{m+4};p_1,p_2)  }
\nonumber \\ &=&
{\rm d} \Phi_{m}(p_{3},\ldots,p_{m+2};z_1p_1,z_2p_2) \,\frac{{\rm d}z_1}{z_1}\frac{{\rm d}z_2}{z_2} \nonumber \\
&&\times \delta(1-z_1)\delta(1-z_2)
{\rm d} \Phi_{X_{ijkl}} (p_i,p_j,p_k,p_{l};p_i+p_j+p_k+p_{l}), 
\label{eq:FFPSfact}
\end{eqnarray}
where we have simply relabelled the final-state momenta.
In \eqref{eq:FFPSfact} the antenna phase space ${\rm d} \Phi_{X_{ijkl}}$ is proportional 
to the four-particle phase space relevant to a $1\to 4$ decay, and one
can define the integrated final-final antenna by
\begin{equation}
\label{eq:x4intff}
{\cal X}^0_{ijkl}(s_{IL},z_1,z_2) = \frac{1}{\l[C(\epsilon)\r]^2} 
\int {\rm d} \Phi_{X_{ijkl}}\;X^0_{ijkl}\,\delta(1-z_1)\,\delta(1-z_2),
\end{equation}
where $C(\e)$ is given by~\eqref{eq:Cdef}.
The integrated double unresolved contribution in this final-final
configuration then reads,  
\begin{eqnarray}
\int_2 \dsigma_{NNLO}^{S,b,4(FF)}
&=&   \NNNLOVV\,
\sum_{\textrm{perms}}\dPSzz{m}{m+2}
\nonumber \\
&\times& \,
\sum_{il}\;  
{\cal X}^0_{ijkl}(s_{il},z_1,z_2)\,
|{\cal M}_{m+2}(\ldots,i,l,\ldots)|^2\,
\JET_{m}^{(m)}(\{p\}_{m+2})\;
 \label{eq:subv1b4ff}\nonumber \\
\end{eqnarray}
where the sum runs over all colour-connected pairs of final state
momenta $(p_i,p_l)$ and the final state momenta $I,L$ have been
relabelled as $i,l$.
Expressions for the integrated final-final four-parton antennae are available in Ref.~\cite{GehrmannDeRidder:2005cm}.

When only one of the hard radiator partons is in the initial state,
$X^0_{i,jkl}$ is a initial-final (IF) antenna function that describes 
all colour connected double unresolved configurations (for this colour-ordered amplitude) 
where partons $j,k$ are unresolved between the initial state parton
denoted by $\hat{i}$ (where $\hat{i}=\hat{1}$ or $\hat{i}=\hat{2}$) 
and the final state parton $l$. The antenna only depends on these
four parton momenta $p_{i},p_{j},p_{k}$ and $p_{l}$.  
The subtraction term, summing
over all possible positions of the unresolved partons, reads,
\begin{eqnarray}
\dsigma_{NNLO}^{S,b,4(IF)}&=&\NNNLORR\sum_{\textrm{perms}}\dPS{m+2}{m+4}\nonumber\\
&&\times \sum_{ i=1,2} \sum_{j,k} X_{i,jkl}^0 |{\cal M}_{m+2}(\ldots,\hat{I},L,\ldots)|^2
J_{m}^{(m)}(\{p\}_{m+2}).\nonumber\\
\label{eq:sub2}
\end{eqnarray}
As in the final-final case, the reduced $(m+2)$-parton matrix element
squared involves the mapped momenta $\hat{I}$ and $L$ which are defined in Appendix~\ref{sec:appendixAIF}. Likewise, the jet algorithm acts on the $m$-final
state momenta that remain after the mapping has been applied.

In this case, the phase space in \eqref{eq:sub2} can be 
factorised into the convolution 
of an $m$-particle phase space, involving only the redefined momenta, 
with a $2\to 3$ particle phase space \cite{Daleo:2006xa}. For
the special case $i=1$, it reads,
\begin{eqnarray}
\label{eq:psx2}
\lefteqn{
{\rm d}\Phi_{m+2}(p_3,\dots,p_{m+4};p_1,p_2)}\nonumber \\
&=&
{\rm d} \Phi_{m}(p_3,\ldots,p_{m+2};z_1p_1,z_2p_2) \,\frac{{\rm d} z_1}{z_1}\,\frac{{\rm d}z_2}{z_2}\, \delta(z_1-\hat{z}_1)\,\delta(1-z_2)\frac{Q^2}{2\pi}{\rm d}\Phi_{3}(p_j,p_k,p_l;p_1,q)\nonumber \\
\label{eq:IFPSfact}
\end{eqnarray}
with $Q^2=-q^2$ and $q=p_j+p_k+p_l-p_1$ and where we have also relabelled the final-state momenta.
The quantity $\hat{z}_1$ is defined in Eq.~\eqref{eq:xiIF}.

Using this factorisation property, one can carry out the 
integration over the unresolved phase space of the antenna function in $\eqref{eq:sub2}$ analytically. We define 
the integrated initial-final antenna function by,
\begin{equation}
\label{eq:x4intif}
{\cal X}_{1,jkl}^0(s_{\bar{1}L},z_1,z_2)=\frac{1}{\l[C(\epsilon)\r]^2}\int {\rm d}\Phi_3 \frac{Q^2}{2\pi}\,\delta(z_1-\hat{z}_1)\,\delta(1-z_2) X_{1,jkl}^0\,,
\end{equation}
where $C(\epsilon)$ is given in \eqref{eq:Cdef}. Similar expressions are obtained when $i=2$ via exchange of $z_1$ and $z_2$,  
\begin{equation}
{\cal X}_{2,jkl}^0(s_{\bar{2}L},z_1,z_2)=\frac{1}{\l[C(\epsilon)\r]^2}\int {\rm d}\Phi_3 \frac{Q^2}{2\pi}\,\delta(z_2-\hat{z}_2)\,\delta(1-z_1) X_{2,jkl}^0\,.
\end{equation}

The integrated double unresolved contribution in this initial-final
configuration then reads, 
\begin{eqnarray}
\lefteqn{\int_2 \dsigma_{NNLO}^{S,b,4(IF)}=
  \NNNLOVV\,
\sum_{\textrm{perms}}\dPSzz{m}{m+2}}\nonumber\\
&&\phantom{+} 
\sum_{i=1,2}\,\sum_{l}\;  {\cal X}^0_{i,jkl}(s_{\bar{i}l},z_1,z_2)
|{\cal M}_{m+2}(\ldots,\hat{\bar{i}},l,\ldots)|^2\,
\JET_{m}^{(m)}(\{p\}_{m+2}).\;\nonumber
\label{eq:subv1b4if} 
\end{eqnarray}
In this expression, the redefined final-state momentum $L$ (relabelled
as $l$)
and the rescaled initial state radiator $\hat{I} (=\hat{\bar{i}} = z_i p_i)$ 
appears in the functional dependence of the
integrated antenna and in the matrix-element squared.
The other rescaled initial state momentum $\hat{\bar{r}} = z_r p_r$ with 
$r=1,2$ with $r \neq i$,
does not appear explicitly but forms part of the ellipsis. 
Explicit expressions for the integrated initial-final four-parton antennae 
are available in Ref.~\cite{Daleo:2009yj}.

If we consider the case where the two hard radiator partons $i$ and
$l$ are both in the initial state, 
then $X^0_{il,jk}$ is a initial-initial (II) antenna function that describes all colour connected double unresolved singular configurations (for this colour-ordered amplitude) 
where partons $j,k$ are unresolved.  
The subtraction term, summing over all possible positions of the unresolved partons, reads,
\begin{eqnarray}
&&\dsigma_{NNLO}^{S,b,4,(II)}=\NNNLORR\sum_{\textrm{perms}}\dPS{m+2}{m+4}\nonumber\\
&&\times\sum_{i,l=1,2} \sum_j X_{il,jk}^0 |{\cal M}_{m+2}(\ldots,\hat{I},\hat{L},\ldots)|^2
J_{m}^{(m)}(\tilde{p}_3,\ldots,\tilde{p}_{m+4}).
\label{eq:sub3}
\end{eqnarray}
where as usual we denote momenta in the initial state with a hat. The radiators
$\hat{i}$ and $\hat{l}$ are replaced by new rescaled initial state partons $\hat{I}$ and $\hat{L}$ and all other spectator momenta
are Lorentz boosted to preserve momentum conservation as described in Appendix~\ref{sec:appendixAII}.

For the initial-initial configuration the phase space in $\eqref{eq:sub3}$ factorises into the convolution of an $m$-particle phase space, involving only redefined momenta, 
with the phase space of partons $j,k$ \cite{Daleo:2006xa} so that when 
$i=1$ and $l=2$,
\begin{eqnarray}
\label{eq:psx3}
\lefteqn{{\rm d}\Phi_{m+2}(p_3,\dots,p_{m+4};p_1,p_2)}\nonumber \\
&=&
{\rm d}\Phi_{m}(\tilde{p}_3,\ldots,\tilde{p}_{m+4};z_1 p_1, z_2 p_2)
\times\,z_1z_2\, \delta(z_1-\hat{z}_1)\,\delta(z_2-\hat{z}_2)\,[{\rm d} p_j]\,[{\rm d} p_k] \,\frac{{\rm d} z_1}{z_1}\,\frac{{\rm d}z_2}{z_2},\nonumber \\
\end{eqnarray}
where the single particle phase space measure is $[{\rm d} p_j]={{\rm d}^{d-1}p_j}/{2E_j/(2\pi)^{d-1}}$ and $\hat{z}_i$ is defined in  Eq.~\eqref{eq:xiII}.

The only dependence on the original momenta lies in the antenna function $X^0_{il,jk}$ and the antenna phase space.  One can therefore carry out the integration over the unresolved phase space analytically, to find the integrated antenna function,
\begin{equation}
\label{eq:x4intii}
{\cal X}^0_{12,jk}(s_{\bar{1}\bar{2}},z_1,z_2) = \frac{1}{\l[C(\epsilon)\r]^2}\int
[{\rm d} p_j]\,[{\rm d} p_k]\,  z_1\,  z_2\,\delta(z_1-\hat{z}_1)\,\delta(z_2-\hat{z}_2)\,X_{12,jk}^0\;
\end{equation}
where $C(\epsilon)$ is given in 
Eq.~\eqref{eq:Cdef}.
Explicit forms for the integrated initial-initial four-parton
antennae are available in Refs.~\cite{Boughezal:2010mc,GehrmannDeRidder:2012ja}.  
The integrated double unresolved contribution in this initial-initial
configuration then reads, 
\begin{eqnarray}
\lefteqn{\int_2 \dsigma_{NNLO}^{S,b,4(II)}
= \NNNLOVV\, 
\sum_{\textrm{perms}}\dPSzz{m}{m+2}}\nonumber \\
&\times& \,  \sum_{i,l=1,2}\;   {\cal X}^0_{il,jk}(s_{\bar{i}\bar{l}},z_i,z_l) 
|{\cal M}_{m+2}(\ldots,\hat{\bar{i}},\hat{\bar{l}},\ldots)|^2\,
\JET_{m}^{(m)}(\{p\}_{m+2})\;. 
\label{eq:subv1b4ii}
\end{eqnarray}
As in the initial-final case, the redefined initial state momenta are
$\hat{I}=\hat{\bar{i}} = z_i p_i$ and $\hat{L}=\hat{\bar{l}}=z_l p_l$. 
All of the final-state momenta affected by the initial-initial mapping are simply relabelled, $\tilde{p}_k \to p_k$.

Note that in general, there may be several different four-parton antennae depending on which particles have been crossed into the initial state, and on whether these particles are directly colour connected (adjacent) or nearly colour connected (non-adjacent).  In the 
gluonic case, there are two possibilities, the adjacent case (bearing in mind the cyclic properties of the gluonic matrix elements),  $F_4^0(1,j,k,2)$, and the nearly colour connected (or non-adjacent) case, $F_4^0(1,j,2,k)$. Upon integration, both of the gluonic antennae will depend on $s_{\bar{1}\bar{2}}$ and are labelled ${\cal F}^0_{4,adj}$ and ${\cal F}^0_{4,nonadj}$ respectively.

So far, we have discussed the individual terms cascading down from the double real subtraction term $\int_2\dsigma_{NNLO}^{S,b,4}$.   Collecting up terms proportional to a single colour ordering denoted by $(\ldots)$,  we find a contribution of the form,
\begin{eqnarray}
&&\int_2 \dsigma_{NNLO}^{S,b,4}= \NNNLOVV\, 
\dPSzz{m}{m+2}\,
\JET_{m}^{(m)}(\{p\}_{m+2})
\nonumber \\
&& \hspace{2cm}\times \XX_4^0(\ldots;z_1,z_2)\,|{\cal M}_{m+2}(\ldots)|^2\;\nonumber \\
\label{eq:subv1b4sum}
\end{eqnarray}
where the ellipsis refers to the colour ordering of the partons and for gluon scattering 
\begin{equation}
\XX_4^0(\ldots;z_1,z_2)=
\sum_{ab}\;   \frac{1}{S^{X_4}_{ab}}
{\cal X}^0_{4,ab}(s_{ab},z_1,z_2) 
\end{equation}
where the sum runs over the colour connected pairs $(a,b)$ in the cyclic list $(\ldots)$.   The integrated double unresolved antenna ${\cal X}^0_{4,ab}(s_{ab},z_1,z_2)$ represents the sum over the antennae obtained by inserting two unresolved particles between and around the hard radiator pair $(a,b)$. 
$S^{X_4}_{ab}$ is the symmetry factor associated with the integrated antenna.  For gluons, 
$S^{F_4}_{ab} = 4$ for the final-final case, $S^{F_4}_{ab} = 2$ for the initial-final case and 
$S^{F_4}_{ab} = 1$ and $S^{F_4}_{ab} = 2$ for the adjacent and non-adjacent initial-initial cases respectively.  For example, in the four-gluon case, 
we find that when the initial state gluons are colour-adjacent (IIFF), 
\begin{eqnarray}
\label{eq:f4sumIIFF}
\XX_4^0(\bar{1}_g,\bar{2}_g,i_g,j_g;z_1,z_2)
&=&\phantom{+}
{\cal F}^0_{4,adj}(s_{\bar{1}\bar{2}},z_1,z_2) 
+    
\frac{1}{2}
{\cal F}^0_{4,nonadj}(s_{\bar{1}\bar{2}},z_1,z_2) \nonumber \\
&& +
\frac{1}{2}
{\cal F}^0_{4}(s_{\bar{2}i},z_1,z_2) 
 +
\frac{1}{4}
{\cal F}_{4}(s_{ij},z_1,z_2)
+
\frac{1}{2}
{\cal F}^0_{4}(s_{j\bar{1}},z_1,z_2), 
\end{eqnarray}
and when the initial state gluons are not colour-adjacent (IFIF),
\begin{eqnarray}
\label{eq:f4sumIFIF}
\XX_4^0(\bar{1}_g,i_g,\bar{2}_g,j_g;z_1,z_2)
&=&\phantom{+}
\frac{1}{2}
{\cal F}^0_{4}(s_{\bar{1}i},z_1,z_2) 
 +
\frac{1}{2}
{\cal F}^0_{4}(s_{i\bar{2}},z_1,z_2) \nonumber \\
&&
 +
\frac{1}{2}
{\cal F}^0_{4}(s_{\bar{2}j},z_1,z_2) 
 +
\frac{1}{2}
{\cal F}^0_{4}(s_{j\bar{1}},z_1,z_2).
\phantom{ \frac{1}{2}
{\cal F}^0_{4}(s_{j\bar{1}},z_1,z_2)~~}
\end{eqnarray}

\subsubsection
[Integration of colour unconnected double unresolved antennae: $\\\int_2\dsigma_{NNLO}^{S,d}$]
{Integration of colour unconnected double unresolved antennae: $\int_2\dsigma_{NNLO}^{S,d}$}
\label{sec:intSd}
As discussed earlier in Sect.~\ref{subsec:intS2}, contributions to the double real subtraction term due to colour-unconnected
hard radiators that have the generic form $X_3 \times X_3$ must also be added back in integrated form.  Each antenna is fully independent and the $(m+2)$-particle phase space
of the double real contribution factorizes into the $m$-particle phase space of the double virtual contribution times independent antenna phase spaces for both the inner and 
outer antennae. Note that each antenna phase space introduces integrals over the momentum fractions carried by the incoming partons. We label these momentum fractions
as $x_1$ and $x_2$ for the inner antenna and $y_1$ and $y_2$ for the outer antenna. This contribution is then integrated over both of the antenna phase spaces to make the
infrared pole structure explicit.

This leads to the following structure for the integrated double unresolved colour unconnected contribution,
\begin{eqnarray}
&&\int_2 \dsigma_{NNLO}^{S,d}
\label{eq:dint}
=   \NNNLOVV\,\dPSxxyy{m}{m+2}\nonumber \\
&&\hspace{1.0cm}\times\sum_{\souter,\sinner}{\cal X}^0_{3}(\souter,x_1,x_2)\,{\cal X}^0_{3}(\sinner,y_1,y_2)\,|{\cal M}_{m+2}(\{p\}_{m+2})|^2\,\JET_{m}^{(m)}(\{p\}_{m+2})\nonumber
\end{eqnarray}
where all momenta lie in the set $\{p\}_{m+2}$. Note that the invariant masses of the integrated antenna $\souter$ and $\sinner$, always involve different momenta, 
and the type of integrated antennae are fixed by the corresponding term in $\dsigma_{NNLO}^{S,d}$.

The integration over $x_{1}, x_{2}, y_{1}, y_{2}$ reflects the fact that the integrated subtraction terms contain contributions due to radiation from the initial
state such that the parton momenta involved in the hard scattering carry only a fraction $x_i y_i$ of the incoming momenta. In order to explicitly show
the cancellation of $\e$-poles we need to combine this contribution with the other integrated subtraction terms present in $\int_2 \dsigma_{NNLO}^{S,b,4}$,
$\int_1 \dsigma_{NNLO}^{VS,a}$ as well as the two-loop matrix elements themselves. We therefore identify the momentum fraction carried by the incoming partons $x_iy_i p_i$ with the momentum fraction carried by the double virtual contribution $z_ip_i$ by imposing the following constraint,
\begin{equation}
\label{eq:constraint}
1=\int {\rm d}z_{1} {\rm d}z_{2}\,\delta(z_{1}-x_{1}y_{1})\delta(z_{2}-x_{2}y_{2})\,.
\end{equation}
This leads to two-dimensional convolutions of integrated antennae that we perform analytically of the type,
\begin{eqnarray}
\left[{\cal X}_{3}^{0}\otimes{\cal X}_{3}^{0}\right](\souter,\sinner;z_1,z_2)&=&\int {\rm d}x_{1}{\rm d}x_{2}{\rm d}y_{1}{\rm d}y_{2}\,{\cal X}_{3}^0(\souter,x_1,x_2)
{\cal X}_{3}^0(\sinner,y_1,y_2)
\nonumber\\
&\times&\delta(z_{1}-x_{1}y_{1})\delta(z_{2}-x_{2}y_{2})\,.
\label{eq:X3cX3}
\end{eqnarray}
Explicit expressions for the convolutions of the gluonic antennae relevant to this paper are given in Appendix~\ref{sec:appendixC}.

We note that the ${\cal X}_{3}^{0}\otimes{\cal X}_{3}^{0}$ terms coming from $\int_2 \dsigma_{NNLO}^{S,d}$ combine with similar terms coming from the integration of the $\int_1 \dsigma_{NNLO}^{VS,(b,c)}$ real-virtual subtraction term, and play a particular role in the infrared  structure of the double virtual subtraction term $\dsigma_{NNLO}^{U}$.  This will be discussed more fully in section~\ref{subsec:intVSbc}.

\subsection{Contributions from integrating the real-virtual subtraction term: $\int_{1}\dsigma_{NNLO}^{VS}$}
\label{subsec:intVS}
There are three different contributions to the real-virtual subtraction term $\dsigma_{NNLO}^{VS}$ which have different physical origin:
\begin{itemize}
\item[(a)] One unresolved parton in the single radiation real-virtual contribution. Each subtraction term takes
the form of a tree-level or one-loop antenna function multiplied by the
one-loop or tree-level colour ordered matrix elements respectively. It is  
 denoted by $\dsigma_{NNLO}^{VS,a}$.   
\item[(b)] Terms of the type ${\cal X}_3^0 X_3^0$ that cancel the explicit poles introduced by one-loop matrix elements and one-loop antenna functions present in $\dsigma_{NNLO}^{VS,a}$. This term is named $\dsigma_{NNLO}^{VS,b}$. 
\item[(c)] Terms of the type ${\cal X}_3^0 X_3^0$ that compensate for
any remaining poles in the real-virtual channel labelled $\dsigma_{NNLO}^{VS,c}$,
\end{itemize}
so that \begin{equation}
 \dsigma_{NNLO}^{VS}=
 \dsigma_{NNLO}^{VS,a}
+ \dsigma_{NNLO}^{VS,b} 
+ \dsigma_{NNLO}^{VS,c}.
\end{equation}
The types of contributions present in each of these terms are summarised in Table~2 and were described in a previous paper~\cite{GehrmannDeRidder:2011aa}. We see
that $a,b$ and $c$ types of subtraction terms can be integrated over a single unresolved particle phase space and therefore contribute to the
$m$-particle final state, so that
\begin{equation}
\int_{1}\dsigma_{NNLO}^{VS}=\int_{1}\dsigma_{NNLO}^{VS,a}+\int_{1}\dsigma_{NNLO}^{VS,(b,c)}.
\label{eq:intVS}
\end{equation}
We now turn to a detailed discussion of each of the terms in Eq.~\eqref{eq:intVS}.
\begin{table}[t!]
\begin{center}
\begin{tabular}{|c|ccc|}
\hline
& \multicolumn{3}{c|}{$\dsigma_{NNLO}^{VS}$}\\\hline
Final State Particles&   $a$ & $a$  & $(b,c)$ \\\hline
$m+1$   & $ X_3^1 |{\cal M}^0_{m+2}|^2$    & $ X_3^0 |{\cal M}^1_{m+2}|^2$ & ${\cal X}_3^0 X_3^0 |{\cal M}^0_{m+2}|^2$   \\
$m$ &  ${\cal X}_3^1 |{\cal M}^0_{m+2}|^2$& ${\cal X}_3^0 |{\cal M}^1_{m+2}|^2$ & ${\cal X}_3^0 {\cal X}_3^0 |{\cal M}^0_{m+2}|^2$ \\ \hline
\end{tabular}
\end{center}
\label{tab:VStypes2}
\caption{Type of contribution to the real-virtual subtraction term ${\rm{d}}\hat\sigma_{NNLO}^{VS}$, together with the integrated form of each term.   The unintegrated antenna functions are denoted as $X_3^0$ and $X_3^1$ while their integrated forms are ${\cal X}_3^0$ and  ${\cal X}_3^1$ respectively.  $|{\cal M}^1_{n}|^2$ denotes the interference of the tree-level and one-loop $n$-particle colour ordered amplitude while $|{\cal M}^0_{n}|^2$ denotes the square of an $n$-particle tree-level colour ordered amplitude.  }
\end{table}

\subsubsection[Integration of single unresolved real-virtual subtraction: $\int_1\dsigma_{NNLO}^{VS,a}$]
{Integration of single unresolved real-virtual subtraction: $\int_1\dsigma_{NNLO}^{VS,a}$}
\label{subsec:intVSa}
In the antenna subtraction approach, the single unresolved configuration coming from the one-loop process
with one additional particle, i.e., the real-virtual process involving $(m+3)$ partons, is subtracted using a single unresolved tree level three
parton antenna multiplied by a $(m+2)$-parton one-loop amplitude and a single unresolved one-loop three parton antenna multiplied by a 
$(m+2)$-parton tree-level amplitude. Once the unresolved phase space is integrated over, one recovers an $(m+2)$-parton contribution
that cancels explicit poles in the virtual two-loop $(m+2)$-parton matrix element.

The integrated subtraction term, formally written as $\int_1 \dsigma_{NNLO}^{VS,a}$, is split into three different contributions,
depending on whether the hard radiators are in the initial or final state.

In the final-final configuration, the integrated subtraction term is given by,
\begin{eqnarray}
\int_{1}\dsigma_{NNLO}^{VS,a,(FF)}&=&\NNNLOVV \dPSzz{m}{m+2}\Bigg\{\nonumber\\
&\phantom{+}&\sum_{ik}{\cal X}_{ijk}^{0}(s_{ik},z_1,z_2)|{\cal M}_{m+2}^{1}(\hdots,i,k,\hdots)|^2\,\JET_{m}^{(m)}(\{p\}_{m+2})\nonumber\\
&+&\sum_{ik}{\cal X}_{ijk}^{1}(s_{ik},z_1,z_2)|{\cal M}_{m+2}^{0}(\hdots,i,k,\hdots)|^2\,\JET_{m}^{(m)}(\{p\}_{m+2})\Bigg\}
\label{eq:intVSaff}
\end{eqnarray}
where the notation for the integrated one-loop three parton antenna function follows Eq.~\eqref{eq:x4intff},
\begin{equation}
\label{eq:x31intff}
{\cal X}^1_{ijk}(s_{IK},z_1,z_2) = \frac{1}{C(\epsilon)} 
\int {\rm d} \Phi_{X_{ijk}}\;X^1_{ijk}\,\delta(1-z_1)\,\delta(1-z_2),
\end{equation}
and in \eqref{eq:intVSaff} the final state momenta $I,K$ have been relabelled as $i,k$. Explicit integrated forms for ${\cal X}^1_{ijk}$  
are available in~\cite{GehrmannDeRidder:2005cm}.

It should be noted that  $X^1_{ijk}$ is renormalised at a scale corresponding 
to the invariant mass of the antenna partons, $s_{ijk}$, while the one-loop
parton matrix element are renormalised at a scale $\mu^2$. To 
ensure correct subtraction of terms arising from renormalisation, 
we have to substitute
\begin{equation}
X^1_{ijk} \to X^1_{ijk} + \frac{\beta_0}{\e} \, X^0_{ijk} \left(
\left(\frac{s_{IK}}{\mu^2}\right)^{-\e} - 1 \right)  
\end{equation}
which in integrated form reads,
\begin{equation}
{\cal X}^1_{ijk}(s_{ik},z_1,z_2) \to {\cal X}^1_{ijk}(s_{ik},z_1,z_2) + \frac{\beta_0}{\e}\,  {\cal X}^0_{ijk}(s_{ik},z_1,z_2)\left(
\left(\frac{s_{ik}}{\mu^2}\right)^{-\e} - 1 \right).
\end{equation}
The terms arising from this substitution will in general be kept apart in the construction of the 
colour ordered subtraction terms, since they all share a  common colour structure $\beta_0$. 

Similar integrated subtraction terms are appropriate in the initial-final and initial-initial configurations. In the initial-final case we have,
\begin{eqnarray}
\label{eq:IntVSaif}
\int_{1}\dsigma_{NNLO}^{VS,a,(IF)}&=&\NNNLOVV \dPSzz{m}{m+2}\Bigg\{\nonumber\\
&\phantom{+}&\sum_{i=1,2}\sum_{k}{\cal X}_{i,jk}^{0}(s_{\bar{i}k},z_{1},z_{2})|{\cal M}_{m+2}^{1}(\hdots,\hat{\bar{i}},k,\hdots)|^2\,\JET_{m}^{(m)}(\{p\}_{m+2})\nonumber\\
&+&\sum_{i=1,2}\sum_{k}{\cal X}_{i,jk}^{1}(s_{\bar{i}k},z_{1},z_{2})|{\cal M}_{m+2}^{0}(\hdots,\hat{\bar{i}},k,\hdots)|^2\,\JET_{m}^{(m)}(\{p\}_{m+2})\Bigg\}\nonumber\\
\end{eqnarray}
where the integrated one-loop three parton initial-final antenna function for the special case $i=1$ is defined as,
\begin{equation}
\label{eq:x31intif}
{\cal X}^1_{1,jk}(s_{\bar{1}K},z_{1},z_{2}) = \frac{1}{C(\epsilon)} 
\int {\rm d} \Phi_{2}\frac{Q^2}{2\pi}\;\delta(z_1-\hat{z}_1)\delta(1-z_2)\,X^1_{1,jk}.
\end{equation}
In \eqref{eq:IntVSaif} the final state momentum $K$ is relabelled as $k$ while the rescaled initial state radiator $\hat{I}=z_i p_i$.  Explicit integrated forms for ${\cal X}^1_{1,jk}$
are available in~\cite{Daleo:2009yj}.

Finally, the initial-initial integrated subtraction term is
\begin{eqnarray}
\int_{1}\dsigma_{NNLO}^{VS,a,(II)}&=&\NNNLOVV \dPSzz{m}{m+2}\Bigg\{\nonumber\\
&\phantom{+}&\sum_{i,k=1,2}{\cal X}_{ik,j}^{0}(s_{\bar{i}\bar{k}},z_{1},z_{2})|{\cal M}_{m+2}^{1}(\hdots,\hat{\bar{i}},\hat{\bar{k}},\hdots)|^2\,\JET_{m}^{(m)}(\{p\}_{m+2})\nonumber\\
&+&\sum_{i,k=1,2}{\cal X}_{ik,j}^{1}(s_{\bar{i}\bar{k}},z_{1},z_{2})|{\cal M}_{m+2}^{0}(\hdots,\hat{\bar{i}},\hat{\bar{k}},\hdots)|^2\,\JET_{m}^{(m)}(\{p\}_{m+2})\Bigg\}\nonumber\\
\label{eq:intVSaif}
\end{eqnarray}
where the integrated one-loop three parton initial-initial antenna function is defined as,
\begin{equation}
\label{eq:x31intii}
{\cal X}^1_{12,j}(s_{\bar{1}\bar{2}},z_{1},z_{2}) = \frac{1}{C(\epsilon)}
\int
[{\rm d} p_j]\,  z_1\,  z_2\,\delta(z_1-\hat{z}_1)\,\delta(z_2-\hat{z}_2)\,X_{12,j}^1\;.
\end{equation}

As in the initial-final case, the redefined initial state momenta
are $\hat{I}=\bar{i} = z_i p_i$ and $\hat{K}=\bar{k}=z_kp_k$. Explicit integrated forms for ${\cal X}^1_{12,j}$
are available in~\cite{Gehrmann:2011wi}.

So far, we have discussed the individual terms cascading down from the real-virtual subtraction term $\int_1\dsigma_{NNLO}^{VS,a}$.   Collecting up the terms associated with a single colour ordering,  we find a contribution of the form,
\begin{eqnarray}
&&\int_1 \dsigma_{NNLO}^{VS,a}
= \NNNLOVV\, 
 \dPSzz{m}{m+2}\, \JET_{m}^{(m)}(\{p\}_{m+2})\nonumber \\
&& \times \Bigg[   
\XX_3^0(\ldots,z_1,z_2) \,
\left(
|{\cal M}^1_{m+2}(\ldots)|^2\,
-\frac{b_0}{\epsilon}|{\cal M}_{m+2}(\ldots)|^2\right)
 \;\nonumber \\
&& \phantom{\times }+ \frac{b_0}{\epsilon}\XX_{s,3}^0(\ldots;z_1,z_2)
\, |{\cal M}_{m+2}(\ldots)|^2
+ \XX_3^1(\ldots,z_1,z_2)\,
|{\cal M}_{m+2}(\ldots)|^2\,
\Bigg]\; 
\label{eq:subvsasum}
\end{eqnarray}
where $b_0$ is the leading colour coefficient of $\beta_0$ and, for gluon scattering 
\begin{eqnarray}
\label{eq:x3nsumdef}
\XX_3^n(\ldots;z_1,z_2) &=&\sum_{ab}\;   \frac{1}{S^{X_3}_{ab}}
{\cal X}^n_{3,ab}(s_{ab},z_1,z_2),\\
\label{eq:x3ssumdef}
\XX_{s,3}^0(\ldots;z_1,z_2) &=&\sum_{ab}\;   
\left(\frac{|s_{ab}|}{\mu^2}\right)^{-\e}\, \frac{1}{S^{X_3}_{ab}}
{\cal X}^0_{3,ab}(s_{ab},z_1,z_2), 
\end{eqnarray}
and as in Eq.~\eqref{eq:subv1b4sum}, the sum runs over the colour connected pairs $(a,b)$ in the cyclic list $(\ldots)$.  Each integrated antenna represents
the contribution  obtained by inserting one unresolved particle between the hard radiator pair $(a,b)$.   
$S^{X_3}_{ab}$ is the symmetry factor associated with the integrated antenna.  For gluons, 
$S^{F_3}_{ab} = 3$ for the final-final case, $S^{F_3}_{ab} = 2$ for the initial-final case and 
$S^{F_3}_{ab} = 1$ for the initial-initial case.  Assembling terms for the four gluon case, we find that when the initial state gluons are colour connected, i.e. in 
the IIFF topology, we have 
\begin{eqnarray}
\label{eq:f3sumIIFF}
\XX_{3}^n(\bar{1}_g,\bar{2}_g,i_g,j_g;z_1,z_2) 
&=&\phantom{+\frac{1}{3}}
{\cal F}^n_{3}(s_{\bar{1}\bar{2}},z_1,z_2) +    
\frac{1}{2}
{\cal F}^n_{3}(s_{\bar{2}i},z_1,z_2) \nonumber \\
&&+
\frac{1}{3}
{\cal F}^n_{3}(s_{ij},z_1,z_2)
+
\frac{1}{2}
{\cal F}^n_{3}(s_{j\bar{1}},z_1,z_2),
\end{eqnarray}
for $n=0,1$, 
while in the IFIF case we have,
\begin{eqnarray}
\label{eq:f3sumIFIF}
\XX_{3}^n(\bar{1}_g,i_g,\bar{2}_g,j_g;z_1,z_2) 
&=&\phantom{+}
\frac{1}{2}
{\cal F}^n_{3}(s_{\bar{1}i},z_1,z_2) +    
\frac{1}{2}
{\cal F}^n_{3}(s_{i\bar{2}},z_1,z_2) \nonumber \\
&&+
\frac{1}{2}
{\cal F}^n_{3}(s_{\bar{2}j},z_1,z_2)
+
\frac{1}{2}
{\cal F}^n_{3}(s_{j\bar{1}},z_1,z_2). 
\end{eqnarray}
$\XX_{s,3}^0$ for these two processes is obtained in an obvious manner.

\subsubsection
[Integration of explicit poles subtraction in the real-virtual channel: $\\\int_1\dsigma_{NNLO}^{VS,(b,c)}$]
{Integration of explicit poles subtraction in the real-virtual channel: $\int_1\dsigma_{NNLO}^{VS,(b,c)}$}
\label{subsec:intVSbc}
As discussed earlier, contributions to the real-virtual subtraction term that have the generic form ${\cal X}_3\times X_3$ must also be integrated over the phase space
of the unintegrated antenna. As for $\int_2\dsigma_{NNLO}^{S,d}$, each term present in $\dsigma_{NNLO}^{VS,(b,c)}$ produces a contribution of the form,
\begin{eqnarray}
&&\int_1 \dsigma_{NNLO}^{VS,(b,c)}
\label{eq:bcintFFv1}
=   \NNNLOVV\,\dPSxxyy{m}{m+2}\nonumber \\
&&\hspace{1.0cm}\times\sum_{\souter,\sinner}{\cal X}^0_{3}(\souter,x_1,x_2)\,{\cal X}^0_{3}(\sinner,y_1,y_2)\,|{\cal M}_{m+2}(\{p\}_{m+2})|^2\,\JET_{m}^{(m)}(\{p\}_{m+2})\;.
\end{eqnarray}

The infrared structure of these terms for gluon scattering takes a very particular form.  For example, if we consider a particular colour ordering of gluons $(\ldots)$, 
the structure of the $\int_1 \dsigma_{NNLO}^{VS,b}$ contribution is given by,
\begin{eqnarray}
\lefteqn{\int_1 \dsigma_{NNLO}^{VS,b}}\nonumber \\
&
=& \NNNLOVV\, 
 \dPSzz{m}{m+2}\, \JET_{m}^{(m)}(\{p\}_{m+2})\nonumber \\
&\times& \Bigg[ 
\left[ \XX_3^0 \otimes \XX_3^0 \right](\ldots; z_1,z_2)  
-2\,\overline{\left[ \XX_3^0 \otimes \XX_3^0 \right]}(\ldots; z_1,z_2)\Bigg]
|{\cal M}_{m+2}(\ldots)|^2\,
\end{eqnarray}
where, 
\begin{eqnarray}
\left[ \XX_3^0 \otimes \XX_3^0 \right](\ldots; z_1,z_2)
&=&
\int  {\rm d}x_{1}{\rm d}x_{2}{\rm d}y_{1}{\rm d}y_{2}\,
\delta(z_{1}-x_{1}y_{1})\delta(z_{2}-x_{2}y_{2})\nonumber \\
&&\times
\left[\sum_{ab}\; \frac{1}{\left(S^{X_3}_{ab}\right)}{\cal X}^0_{3,ab}(s_{ab},x_1,x_2)\right]
\left[\sum_{cd}\; \frac{1}{\left(S^{X_3}_{cd}\right)}{\cal X}^0_{3,cd}(s_{cd},y_1,y_2)\right]\nonumber\\
&=&
\int  {\rm d}x_{1}{\rm d}x_{2}{\rm d}y_{1}{\rm d}y_{2}\,
\delta(z_{1}-x_{1}y_{1})\delta(z_{2}-x_{2}y_{2})\nonumber \\
&&\times \XX_3^0(\ldots;x_1,x_2)
         \XX_3^0(\ldots;y_1,y_2)\;,\\
\overline{\left[ \XX_3^0 \otimes \XX_3^0 \right]}(\ldots; z_1,z_2)
&=&
\int  {\rm d}x_{1}{\rm d}x_{2}{\rm d}y_{1}{\rm d}y_{2}\,
\delta(z_{1}-x_{1}y_{1})\delta(z_{2}-x_{2}y_{2})\nonumber \\
&& \times \sum_{ab}\; \frac{1}{\left(S^{X_3}_{ab}\right)^2}
{\cal X}^0_{3,ab}(s_{ab},x_1,x_2)
{\cal X}^0_{3,ab}(s_{ab},y_1,y_2)\nonumber \\
&=& 	\sum_{ab}\; \frac{1}{\left(S^{X_3}_{ab}\right)^2}
\left[{\cal X}_3^0 \otimes {\cal X}_3^0\right](s_{ab},s_{ab};z_1,z_2) 
\end{eqnarray}
where, as in Eqs.~\eqref{eq:subv1b4sum} and \eqref{eq:subvsasum}, the sums runs over the adjacent pairs $(a,b)$ and $(c,d)$ in the cyclic list $(\ldots)$. 
The ${\left[ \XX_3^0 \otimes \XX_3^0 \right]}$ term is produced by the part of $\dsigma_{NNLO}^{VS,(b)}$ that absorbs the poles in the one-loop matrix element, $A^1_{m+2}$, while the $\overline{\left[ \XX_3^0 \otimes \XX_3^0 \right]}$ contribution comes from the antenna that corrects for the pole structure of $X_3^1$. 

Similarly, the contribution from the sum of $\int_2 \dsigma_{NNLO}^{S,d}$ and
$\int_1 \dsigma_{NNLO}^{VS,c}$ is given by,
\begin{eqnarray}
\lefteqn{\int_2 \dsigma_{NNLO}^{S,d}+ \int_1 \dsigma_{NNLO}^{VS,c}}\nonumber \\
&
=& \NNNLOVV\, 
 \dPSzz{m}{m+2}\, \JET_{m}^{(m)}(\{p\}_{m+2})\nonumber \\
&\times& \Bigg[ 
-\frac{1}{2} \left[ \XX_3^0 \otimes \XX_3^0 \right](\ldots; z_1,z_2)  
+\overline{\left[ \XX_3^0 \otimes \XX_3^0 \right]}(\ldots; z_1,z_2)\Bigg]
|{\cal M}_{m+2}(\ldots)|^2\, .
\label{eq:subXXsum}
\end{eqnarray}
Here, all of $\overline{\left[ \XX_3^0 \otimes \XX_3^0 \right]}$ and the three terms in ${\left[ \XX_3^0 \otimes \XX_3^0 \right]}$ where the invariants have at least one common hard radiator momentum are produced by $\int_1 \dsigma_{NNLO}^{VS,(c)}$. The remaining terms in ${\left[ \XX_3^0 \otimes \XX_3^0 \right]}$, where there is no overlap between the hard radiator momenta, come from $\int_2 \dsigma_{NNLO}^{S,d}$.

In our example of four gluon scattering, we see that in the IIFF case,
\begin{eqnarray}
\lefteqn{\overline{\left[ \XX_3^0 \otimes \XX_3^0 \right]}(\bar{1}_g,\bar{2}_g,i_g,j_g;z_1,z_2)}\nonumber \\
&=&\phantom{+\frac{1}{9}}
\left[{\cal X}_{3}^{0}\otimes{\cal X}_{3}^{0}\right](s_{\bar{1}\bar{2}},s_{\bar{1}\bar{2}};z_1,z_2)
+\frac{1}{4}
\left[{\cal X}_{3}^{0}\otimes{\cal X}_{3}^{0}\right](s_{\bar{2}i},s_{\bar{2}i};z_1,z_2)\nonumber \\
&&
+\frac{1}{9}
\left[{\cal X}_{3}^{0}\otimes{\cal X}_{3}^{0}\right](s_{ij},s_{ij};z_1,z_2)
+\frac{1}{4}
\left[{\cal X}_{3}^{0}\otimes{\cal X}_{3}^{0}\right](s_{j\bar{1}},s_{j\bar{1}};z_1,z_2),\nonumber \\
\end{eqnarray}
while in the IFIF topology,
\begin{eqnarray}
\lefteqn{\overline{\left[ \XX_3^0 \otimes \XX_3^0 \right]}(\bar{1}_g,i_g,\bar{2}_g,j_g;z_1,z_2)}\nonumber \\
&=&\phantom{+}\frac{1}{4}
\left[{\cal X}_{3}^{0}\otimes{\cal X}_{3}^{0}\right](s_{\bar{1}i},s_{\bar{1}i};z_1,z_2)
+\frac{1}{4}
\left[{\cal X}_{3}^{0}\otimes{\cal X}_{3}^{0}\right](s_{i\bar{2}},s_{i\bar{2}};z_1,z_2)\nonumber \\
&&
+\frac{1}{4}
\left[{\cal X}_{3}^{0}\otimes{\cal X}_{3}^{0}\right](s_{\bar{2}j},s_{\bar{2}j};z_1,z_2)
+\frac{1}{4}
\left[{\cal X}_{3}^{0}\otimes{\cal X}_{3}^{0}\right](s_{j\bar{1}},s_{j\bar{1}};z_1,z_2).\nonumber \\
\end{eqnarray}
Explicit expressions for the convolutions of the gluonic antennae relevant to this paper are given in Appendix~\ref{sec:appendixC}.

\section{Gluon scattering at LO and NLO}
\label{sec:gluonscattering}
In this section we discuss gluon amplitudes and gluon scattering up to NLO.

\subsection{Gluonic amplitudes}
\label{subsec:gluonamps}

The leading colour contribution to the 
$m$-gluon $n$-loop amplitude can be written as~\cite{Berends:1987cv,Kosower:1987ic,Mangano:1987xk,Mangano:1990by,Bern:1994zx},
\begin{eqnarray}
\lefteqn{{\bf A}^n_m(\{p_i,\lambda_i,a_i\})}\nonumber \\
&=&2^{m/2} g^{m-2} 
\left(\frac{g^2 N\, C(\epsilon)}{2}\right)^n \sum_{\sigma\in S_m/Z_m}\textrm{Tr}(T^{a_{\sigma(1)}}\cdots T^{a_{\sigma(m)}})
{\cal A}^n_m(\sigma(1),\cdots,\sigma(m)).\nonumber \\
\label{eq:cdecomploop}
\end{eqnarray}
where the  permutation sum, $S_m/Z_m$ is the group of non-cyclic permutations of $m$ symbols.  
We systematically extract a loop factor of $C(\epsilon)/2$ per loop with
$C(\epsilon)$ defined in \eqref{eq:Cdef}.
The helicity information is not relevant to the discussion of the subtraction terms and from now on, we will systematically suppress the helicity labels. The $T^a$ are fundamental representation $SU(N)$ colour matrices, normalised such that ${\rm Tr}(T^aT^b)= \delta^{ab}/2$.
${\cal A}^n_m(1,\cdots,n)$ denotes the $n$-loop colour ordered \textit{partial amplitude}.
It is gauge invariant, as well as being invariant under cyclic permutations of the gluons. For simplicity, we will frequently denote the momentum $p_j$ of gluon $j$ by $j$.

At leading colour,  the $n$-loop $(m+2)$-gluon contribution to the $M$-jet cross section is given by,
\begin{eqnarray}
\label{eq:Nmaster}
\dsigma&=&{\cal N}_{m+2}^n 
{\rm d}\Phi_{m}(p_3,\dots,p_{m+2};p_1,p_2)\frac{1}{m!}\nonumber \\
&& \times
\sum_{\sigma\in S_{m+2
}/Z_{m+2}}A_{m+2}^n(\sigma(1),\dots,\sigma(m+2))\JET_M^{(m)}(p_3,...,p_{m+2}) 
\end{eqnarray}
where  
\begin{equation}
\label{eq:Anm}
{A}^n_m(\sigma(1),\ldots,\sigma(m)) = \sum_{\rm helicities} \sum_{i=0,n}
{\cal A}_m^{i\dagger}(\sigma(1),\ldots,\sigma(m))
{\cal A}_m^{n-i}(\sigma(1),\ldots,\sigma(m)) 
+ {\cal O}\left(\frac{1}{N^2}\right)
\end{equation}
encodes the summed and squared matrix elements such that, for example, $A_m^2$ includes both the interference of two-loop graphs with tree-level as well as the square of the one-loop contribution.

The normalisation factor ${\cal N}_{m+2}^n$ includes the average over initial spins and colours and is given by,
\begin{eqnarray}
{\cal N}_{m+2}^n &=& {\cal N}_{LO} \times \left(\frac{\alpha_s N}{2\pi}\right)^{m-2} \frac{\bar{C}(\epsilon)^{m+n-2}}{C(\epsilon)^{m-2}},
\label{eq:NNLORR}
\end{eqnarray}
where for the $2\to 2$ Born process, ${\cal N}_{LO}$ is given in Eq.~\eqref{eq:NLOdef}.
As usual, we have converted $g^2$ into $\alpha_s$ using the useful factors $C(\epsilon)$ \eqref{eq:Cdef} and $\bar{C}(\epsilon)$ \eqref{eq:Cbar},
$$
g^2 N\, C(\epsilon) = \left(\frac{\alpha_s N}{2\pi}\right) \bar{C}(\epsilon).
$$

The dijet cross section is obtained by setting $M=2$ in~\eqref{eq:Nmaster}.  At NLO, $m+n = 3$;
the virtual contribution has $m=2$ and $n=1$, while the real radiation component has $m=3$ and $n=0$. At NNLO, $m+n = 4$ and the various contributions are obtained by setting $m=4$ and $n=0$ (double real), $m=3$ and $n=1$ (real-virtual) and $m=2$ and $n=2$ (double virtual). We will encounter both ${A}_4^2$ and ${A}_4^1$ when discussing the double virtual corrections to the NNLO dijet cross section in Sect.~\ref{sec:VVgluon}.

As discussed earlier, it is convenient to introduce two different topologies depending on the position of the initial state gluons in the colour ordered matrix elements. These are labelled by the colour ordering of initial and final state gluons. We denote the configurations where the two initial state gluons are colour-connected (i.e., adjacent) or not colour-connected as IIFF and IFIF respectively,
\begin{equation}
{\rm d}\hat\sigma_{LO} 
=
{\rm d}\hat\sigma_{LO}^{IIFF}+
{\rm d}\hat\sigma_{LO}^{IFIF},
\end{equation}
where,
\begin{eqnarray}
{\rm d}\hat\sigma_{LO}^{IIFF}&=& {\cal N}_{LO} \,
\dPStwo \, \JET_{2}^{(2)}(p_{3},p_{4})
 \, \nonumber\\
&\times &
 \sum_{P(i,j)\in(3,4)}
2\, A_{4}^{0} (\hat{1}_g,\hat{2}_g,i_g,j_g),
\label{eq:IIFF}\\
{\rm d}\hat\sigma_{LO}^{IFIF}&=& {\cal N}_{LO} \,
\dPStwo \, \JET_{2}^{(2)}(p_{3},p_{4})
 \, \nonumber\\
&\times &
\sum_{P(i,j)\in(3,4)}
A_{4}^0 (\hat{1}_g,i_g,\hat{2}_g,j_g),
\label{eq:IFIF}
\end{eqnarray}
where the sum runs over the 2! permutations of the final state gluons.

\subsection{Contributions to the gluonic final state at NLO}
\label{subsec:gluonNLO}

In this subsection, we collect up the terms that are relevant for the 2-particle final state at NLO.  There are three contributions, the renormalised interference of one-loop and tree graphs, the integrated subtraction term and the NLO mass factorisation contribution.   

\subsubsection{The four-gluon single virtual contribution ${\rm d}\hat\sigma_{NLO}^{V}$}
\label{subsec:four-V}
The leading colour four-gluon one-loop contribution to the NLO dijet cross section is obtained by setting $m=2$, $n=1$ and $M=2$ in (${\ref{eq:Nmaster}}$) such that
$\NNLOV={\cal N}_{4}^{1}$ and we have,
\begin{eqnarray}
{\rm d}\hat\sigma_{NLO}^{V,IIFF}&=& {\cal N}_{LO} \,
\left(\frac{\alpha_sN}{2\pi}\right)   \bar{C}(\epsilon)  
\dPStwozz \, \JET_{2}^{(2)}(p_{3},p_{4})
 \, \nonumber\\
&\times &
 \sum_{P(i,j)\in(3,4)}
2\, A_{4}^{1} (\hat{\bar{1}}_g,\hat{\bar{2}}_g,i_g,j_g)\,\delta(1-z_1)\,\delta(1-z_2),
\label{eq:IIFF-V}\\
{\rm d}\hat\sigma_{NLO}^{V,IFIF}&=& {\cal N}_{LO} \,
\left(\frac{\alpha_sN}{2\pi}\right)   \bar{C}(\epsilon)  
\dPStwozz \, \JET_{2}^{(2)}(p_{3},p_{4})
 \, \nonumber\\
&\times &
  \,
\sum_{P(i,j)\in(3,4)}
A_{4}^{1} (\hat{\bar{1}}_g,i_g,\hat{\bar{2}}_g,j_g)\,\delta(1-z_1)\,\delta(1-z_2).
\label{eq:IFIF-V}
\end{eqnarray}

We note that the singular part of the renormalised one-loop contribution involves the infrared singular operators,
\begin{eqnarray}
&&{A}^{1}_{4}(\hat{\bar{1}}_g,\hat{\bar{2}}_g,i_g,j_g)=   2\,{\bom I}^{(1)}(\epsilon; {\bar{1}}_g, {\bar{2}}_g,i_g,j_g) \,A_4^0(\hat{\bar{1}}_g,\hat{\bar{2}}_g,i_g,j_g)+{\cal O}(\eps^0),\\ 
&&{A}^{1}_{4}(\hat{\bar{1}}_g,i_g,\hat{\bar{2}}_g,j_g)=  2\,{\bom I}^{(1)}(\epsilon;\hat{\bar{1}}_g,i_g,\hat{\bar{2}}_g,j_g) \,A_4^0(\hat{\bar{1}}_g,i_g,\hat{\bar{2}}_g,j_g)+{\cal O}(\eps^0),
\label{eq:Vepoles}
\end{eqnarray}
where the infrared singular operators for the two permutations are given by,
\begin{eqnarray}
{\bom I}^{(1)}(\epsilon; {\bar{1}}_g, {\bar{2}}_g,i_g,j_g) 
&=& 
 {\bf I}_{gg}^{(1)}(\e,s_{\bar{1}\bar{2}})
+{\bf I}_{gg}^{(1)}(\e,s_{\bar{2}i})
+{\bf I}_{gg}^{(1)}(\e,s_{ij})
+{\bf I}_{gg}^{(1)}(\e,s_{j\bar{1}}), \\
{\bom I}^{(1)}(\epsilon;\hat{\bar{1}}_g,i_g,\hat{\bar{2}}_g,j_g) 
&=& 
 {\bf I}_{gg}^{(1)}(\e,s_{\bar{1}i})
+{\bf I}_{gg}^{(1)}(\e,s_{i\bar{2}})
+{\bf I}_{gg}^{(1)}(\e,s_{\bar{2}j})
+{\bf I}_{gg}^{(1)}(\e,s_{j\bar{1}}), 
\end{eqnarray}
with,
\begin{eqnarray}
{\bom I}_{gg}^{(1)}(\e,s_{gg})=-\frac{e^{\epsilon\gamma}}{2\Gamma(1-\epsilon)}
\left[\frac{1}{\epsilon^2}+
\frac{11}{6\epsilon}\right] \left(\frac{|s_{gg}|}{\mu^2}\right)^{-\epsilon}.
\label{eq:Ione}
\end{eqnarray}

\subsubsection{The integrated NLO subtraction term $\int_1 {\rm d}\hat\sigma_{NLO}^{S}$}
\label{subsec:four-S}

Within the antenna subtraction scheme, the NLO integrated subtraction terms for the IIFF and IFIF configurations are given by,
\begin{eqnarray}
\int_1 {\rm d}\hat\sigma_{NLO}^{S,IIFF}&=& {\cal N}_{LO} \left(\frac{\alpha_sN}{2\pi}\right)   \bar{C}(\epsilon)  
\dPStwozz 
\, \JET_{2}^{(2)}(p_{3},p_{4})
\nonumber\\
&\times &
\sum_{P(i,j)\in(3,4)}
\XX_3^0(  {\bar{1}}_g, {\bar{2}}_g,i_g,j_g;z_1,z_2)\,
2\, A_{4}^{0} (\hat{\bar{1}}_g,\hat{\bar{2}}_g,i_g,j_g),
\label{eq:IIFF-S}\\
\int_1 {\rm d}\hat\sigma_{NLO}^{S,IFIF}&=& {\cal N}_{LO} \left(\frac{\alpha_sN}{2\pi}\right)   \bar{C}(\epsilon)  
\dPStwozz 
\, \JET_{2}^{(2)}(p_{3},p_{4})\nonumber\\
&\times &
\sum_{P(i,j)\in(3,4)}
\XX_3^0(  {\bar{1}}_g,i_g, {\bar{2}}_g,j_g;z_1,z_2)\,
A_{4}^0 (\hat{\bar{1}}_g,i_g,\hat{\bar{2}}_g,j_g),
\label{eq:IFIF-S}
\end{eqnarray}
where the relevant $\XX_3^0$ are obtained by setting $n=0$ in Eqs.~\eqref{eq:f3sumIIFF} and \eqref{eq:f3sumIFIF} respectively.

\subsubsection{The NLO mass factorisation term ${\rm d}\hat\sigma_{NLO}^{MF}$}
\label{subsec:four-MF}

At NLO, the mass factorisation term is given by
\begin{eqnarray}
\int_1 {\rm d}\hat\sigma_{NLO}^{MF,IIFF}&=& -{\cal N}_{LO} \left(\frac{\alpha_sN}{2\pi}\right)   \bar{C}(\epsilon)  
\dPStwozz 
\, \JET_{2}^{(2)}(p_{3},p_{4})
\nonumber\\
&\times &
\sum_{P(i,j)\in(3,4)}
\GG1_{gg;gg}(z_1,z_2)\,
2\, A_{4}^{0} (\hat{\bar{1}}_g,\hat{\bar{2}}_g,i_g,j_g), \\
\label{eq:IIFF-MF} 
\int_1 {\rm d}\hat\sigma_{NLO}^{MF,IFIF}&=& -{\cal N}_{LO} \left(\frac{\alpha_sN}{2\pi}\right)   \bar{C}(\epsilon)  
\dPStwozz 
\, \JET_{2}^{(2)}(p_{3},p_{4})\nonumber\\
&\times &
\sum_{P(i,j)\in(3,4)}
\GG1_{gg;gg}(z_1,z_2)\, A_{4}^0 (\hat{\bar{1}}_g,i_g,\hat{\bar{2}}_g,j_g)  
\label{eq:IFIF-MF}
\end{eqnarray}
where
\begin{eqnarray}
\label{eq:gammagg1}
\GG1_{gg;gg}(z_1,z_2) &=& 
\delta(1-z_2) \,\GG1_{gg}(z_1)   
+\delta(1-z_1)\,\GG1_{gg}(z_2), 
\end{eqnarray}
with
\begin{equation}
\GG1_{gg}(z) = -\frac{1}{\e}p^0_{gg}(z).
\end{equation}

\subsubsection{Finiteness of the NLO two-particle final state contribution}
We see that the gluonic contribution to the  two-particle final state in the IIFF channel is given by combining Eqs.~\eqref{eq:IIFF-V}, \eqref{eq:IIFF-S} and \eqref{eq:IIFF-MF}, such that the combination
\begin{eqnarray}
2{\bom I}^{(1)}(\epsilon; \bar{1}_g,\bar{2}_g,i_g,j_g)
   \delta(1-z_1)\delta(1-z_2)
+ \XX_3^0(\bar{1}_g,\bar{2}_g,i_g,j_g;z_1,z_2)
- \GG1_{gg;gg}(z_1,z_2) 
\label{eq:NLOfin1} 
\end{eqnarray}
is finite, which is indeed the case.
Similarly, the combination of Eqs.~\eqref{eq:IFIF-V}, \eqref{eq:IFIF-S} and \eqref{eq:IFIF-MF}
\begin{eqnarray}
2{\bom I}^{(1)}(\epsilon; {\bar{1}}_g,i_g, {\bar{2}}_g,j_g)\delta(1-z_1)\delta(1-z_2)
+ \XX_3^0( {\bar{1}}_g,i_g, {\bar{2}}_g,j_g;z_1,z_2)
- \GG1_{gg;gg}(z_1,z_2)  
\end{eqnarray}
is also finite.

\section{Double Virtual corrections for gluon scattering at NNLO}
\label{sec:VVgluon}

As discussed in Sect.~\ref{sec:VVsub}, the four parton contribution consists of the genuine two-loop four parton scattering
matrix element together with doubly and singly integrated forms of the six-parton and five-parton subtraction terms. In this section, we give expressions for the contributions that enter in the implementation of the double virtual correction
and construct the subtraction term $\dsigma_{NNLO}^{U}$ for the IIFF and IFIF topologies. Here, and as in Refs.~\cite{Glover:2010im,GehrmannDeRidder:2011aa}, we focus on the leading colour contribution to the 
pure gluon channel.

\subsection{The four-gluon double virtual contribution ${\rm d}\hat\sigma_{NNLO}^{VV}$}
\label{subsec:four-VV}

The leading colour four-gluon double virtual contribution to the NNLO dijet cross section is obtained by setting $m=2$, $n=2$ and $M=2$ in (${\ref{eq:Nmaster}}$) such that
$\NNNLOVV={\cal N}_{4}^{2}$ and we have,
\begin{eqnarray}
{\rm d}\hat\sigma_{NNLO}^{VV,IIFF}&=& {\cal N}_{LO} \,
\left(\frac{\alpha_sN}{2\pi}\right)^2  \bar{C}(\epsilon)^2  
\dPStwozz \, \JET_{2}^{(2)}(p_{3},p_{4})
 \, \nonumber\\
&\times &
 \sum_{P(i,j)\in(3,4)}
2\, A_{4}^{2} (\hat{\bar{1}}_g,\hat{\bar{2}}_g,i_g,j_g)\,\delta(1-z_1)\,\delta(1-z_2),
\label{eq:IIFF-VV}\\
{\rm d}\hat\sigma_{NNLO}^{VV,IFIF}&=& {\cal N}_{LO} \,
\left(\frac{\alpha_sN}{2\pi}\right)^2  \bar{C}(\epsilon)^2  
\dPStwozz \, \JET_{2}^{(2)}(p_{3},p_{4})
 \, \nonumber\\
&\times &
  \,
\sum_{P(i,j)\in(3,4)}
A_{4}^{2} (\hat{\bar{1}}_g,i_g,\hat{\bar{2}}_g,j_g)\,\delta(1-z_1)\,\delta(1-z_2).
\label{eq:IFIF-VV}
\end{eqnarray}

Explicit expressions for the interference of two-loop amplitudes with tree-level and the self-interference of the one-loop amplitudes are given in Refs.~\cite{Glover:2001af} and \cite{Glover:2001rd}, respectively.

The renormalised singularity structure of the two-loop contribution in Eqs.~(\ref{eq:IIFF-VV}) and \eqref{eq:IFIF-VV} can be easily written in terms of the renormalised tree-level
and one-loop matrix elements~\cite{Catani:1998bh}.  For the 
IIFF ordering we have,
\begin{eqnarray}
\label{eq:catIIFF}
A_{4}^{2} (\hat{\bar{1}}_g,\hat{\bar{2}}_g,i_g,j_g) &=&
2 {\bom I}^{(1)}(\epsilon; \bar{1}_g, \bar{2}_g,i_g,j_g)\,
\left( A_{4}^{1} (\hat{\bar{1}}_g,\hat{\bar{2}}_g,i_g,j_g)
-\frac{b_0}{\epsilon} 
A_{4}^{0} (\hat{\bar{1}}_g,\hat{\bar{2}}_g,i_g,j_g)\right) \nonumber \\
&& - 2 {\bom I}^{(1)}(\epsilon; \bar{1}_g, \bar{2}_g,i_g,j_g)^2\,
A_{4}^{0} (\hat{\bar{1}}_g,\hat{\bar{2}}_g,i_g,j_g) \nonumber \\ 
&&+2 e^{-\epsilon\gamma}\frac{\Gamma(1-2\epsilon)}{\Gamma(1-\epsilon)}
\,\left(\frac{b_0}{\epsilon}+K\right) \,
{\bom I}^{(1)}(2\epsilon; \bar{1}_g, \bar{2}_g,i_g,j_g)\, A_{4}^{0} (\hat{\bar{1}}_g,\hat{\bar{2}}_g,i_g,j_g)\nonumber \\
&&+2 {\bom H}^{(2)}(\epsilon)\, A_{4}^{0} (\hat{\bar{1}}_g,\hat{\bar{2}}_g,i_g,j_g) +{\cal O}(\eps^0)
\end{eqnarray}
with an analogous equation for the other colour ordering.  At leading colour, the constants $K$ and ${\bom H}^{(2)}(\epsilon)$ are given by,
\begin{eqnarray}
K &=& \frac{67}{18}-\frac{\pi^2}{6},\nonumber \\
{\bom H}^{(2)}(\epsilon)
&=& \frac{e^{\epsilon\gamma}}{4\epsilon\Gamma(1-\epsilon)} \left(
2 \zeta_3 + \frac{5}{3} + \frac{11\pi^2}{36}\right). 
\end{eqnarray}

\subsection{The integrated NNLO subtraction term,
  $\int_2\dsigma_{NNLO}^{S}+\int_1\dsigma_{NNLO}^{VS,(b,c)}$} 
\label{subsec:four-intsub}

Combining terms as discussed in Sect.~2, the integrated four parton subtraction contribution for this topology can be written as,
\begin{eqnarray}
&& \left(\int_{2}\dsigma_{NNLO}^{S,IIFF}+\int_{1}\dsigma_{NNLO}^{VS,IIFF}\right)\nonumber\\
&&=\NNNLOVV\,\dPStwozz \,J_2^{(2)}(p_3,p_4) \, 2\,\sum_{P(i,j)\in(3,4)} 
\nonumber\\
&&\times\Bigg\{\Bigg[\phantom{+}
\FF_4^0( {\bar{1}}_g, {\bar{2}}_g,i_g,j_g;z_1,z_2)
+
\FF_3^1(  {\bar{1}}_g, {\bar{2}}_g,i_g,j_g;z_1,z_2)\nonumber \\
&&\phantom{\times\Bigg(}
+
\frac{b_0}{\e}\FF_{s,3}^0(  {\bar{1}}_g, {\bar{2}}_g,i_g,j_g;z_1,z_2)
-\overline{\left[\FF_3^0\otimes \FF_3^0\right]}(  {\bar{1}}_g, {\bar{2}}_g,i_g,j_g;z_1,z_2)
\nonumber \\
&&\phantom{\times\Bigg(}
-\frac{b_0}{\e}\FF_3^0( {\bar{1}}_g, {\bar{2}}_g,i_g,j_g;z_1,z_2)
+
\frac{1}{2}\left[\FF_3^0\otimes \FF_3^0\right](  {\bar{1}}_g, {\bar{2}}_g,i_g,j_g;z_1,z_2) \Bigg]
\,A_4^0(\hat{\bar{1}}_g,\hat{\bar{2}}_g,i_g,j_g)
\nonumber \\
&&\phantom{\times\Bigg(}+\FF_3^0(  {\bar{1}}_g, {\bar{2}}_g,i_g,j_g;z_1,z_2)
\,A_4^1(\hat{\bar{1}}_g,\hat{\bar{2}}_g,i_g,j_g)\Bigg\},
\end{eqnarray}
where the various combinations of integrated three parton antennae are given in Sect.~2.
A similar expression with the same structure is valid for the IFIF topology.

\subsection{The NNLO mass factorisation contribution  $\dsigma_{NNLO}^{MF,2}$} 
\label{subsec:four-MF2}

The four-particle NNLO mass factorisation term for the IIFF topology is
\begin{eqnarray}
\dsigma_{NNLO}^{MF,2,IIFF}&=&
\NNNLOVV\,\dPStwozz \,J_2^{(2)}(p_3,p_4) \, 2\,\sum_{P(i,j)\in(3,4)}
\nonumber \\
&\times&\Bigg[
-\GG2_{gg;gg}(z_1,z_2)\,A_4^0(\hat{\bar{1}}_g,\hat{\bar{2}}_g,i_g,j_g)
-\GG1_{gg;gg}(z_1,z_2)\,A_4^1(\hat{\bar{1}}_g,\hat{\bar{2}}_g,i_g,j_g)\nonumber \\
&& -\left[\GG1_{gg;gg}\otimes \FF_3^0\right]( {\bar{1}}_g, {\bar{2}}_g,i_g,j_g;z_1,z_2)\,A_4^0(\hat{\bar{1}}_g,\hat{\bar{2}}_g,i_g,j_g)
\nonumber \\
&&+
\left[\GG1_{gg;gg}\otimes \GG1_{gg;gg}\right](z_1,z_2)\,A_4^0(\hat{\bar{1}}_g,\hat{\bar{2}}_g,i_g,j_g)
\bigg],
\end{eqnarray}
where $\GG1_{gg;gg}(z_1,z_2)$ is given by Eq.~\eqref{eq:gammagg1} and $\GG2_{gg;gg}(z_1,z_2)$ is given by
$$
\GG2_{gg;gg}(z_1,z_2)=
\delta(1-z_{2})\,  \GG2_{gg}(z_{1})
+\delta(1-z_{1})\, \GG2_{gg}(z_{2}) +
\GG1_{gg}(z_{1})\GG1_{gg}(z_2),  
$$
with 
\begin{eqnarray}
\GG2_{gg}(z)&=&
\frac{1}{2\epsilon^2}
\left[\left(p_{gg}^{0}\otimes p_{gg}^{0}\right)(z)
+\beta_{0}\,p_{gg}^{0}(z)
\right]
-\frac{1}{2\epsilon}\,p_{gg}^{1}(z).
\end{eqnarray}

The convolutions are given by,
\begin{eqnarray}
 \left[\GG1_{gg;gg}\otimes \FF_3^0\right]( {\bar{1}}_g, {\bar{2}}_g,i_g,j_g;z_1,z_2)
&=&
\int  {\rm d}x_{1}{\rm d}x_{2}{\rm d}y_{1}{\rm d}y_{2}\,
\delta(z_{1}-x_{1}y_{1})\delta(z_{2}-x_{2}y_{2})\nonumber \\
&&\times \GG1_{gg;gg}(x_1,x_2)
\FF_3^0(  {\bar{1}}_g, {\bar{2}}_g,i_g,j_g;y_1,y_2),\nonumber \\
\end{eqnarray}
and
\begin{eqnarray}
\left[\GG1_{gg;gg}\otimes \GG1_{gg;gg}\right](z_1,z_2)
&=& \int  {\rm d}x_{1}{\rm d}x_{2}{\rm d}y_{1}{\rm d}y_{2}\,
\delta(z_{1}-x_{1}y_{1})\delta(z_{2}-x_{2}y_{2})\,\nonumber\\
&&\times 
\GG1_{gg;gg}(x_1,x_2)\,\GG1_{gg;gg}(y_1,y_2).\nonumber \\
\end{eqnarray}
It is straightforward to obtain a similar expression for the IFIF topology.

\subsection{The four-gluon subtraction term,  $\dsigma_{NNLO}^{U}$} 
\label{sec:dsigmaU}

By assembling the expressions given in the previous subsections, the two-particle subtraction term $\dsigma_{NNLO}^{U,IIFF}$ can now be written in the suggestive way,
\begin{eqnarray}
\label{eq:UIIFF}
\dsigma_{NNLO}^{U,IIFF}&=&
-\int_2 \dsigma_{NNLO}^{S,IIFF}
-\int_1 \dsigma_{NNLO}^{VS,IIFF}
-   \dsigma_{NNLO}^{MF,2,IIFF}\nonumber \\
&&=\NNNLOVV\,\dPStwozz \,J_2^{(2)}(p_3,p_4) \, 2\,\sum_{P(i,j)\in(3,4)} 
\nonumber\\
&&\times\Bigg\{ 
-\left[
\FF_3^0( {\bar{1}}_g, {\bar{2}}_g,i_g,j_g;z_1,z_2)
-\GG1_{gg;gg}(z_1,z_2)\right]\nonumber \\
&& \hspace{2cm}\times \left[
\,A_4^1(\hat{\bar{1}}_g,\hat{\bar{2}}_g,i_g,j_g)
-\frac{b_0}{\e}\,A_4^0(\hat{\bar{1}}_g,\hat{\bar{2}}_g,i_g,j_g)\right]\nonumber \\
&&\phantom{\times\Bigg(}
-
\frac{1}{2}\left[\left(\FF_3^0-\GG1_{gg;gg}\right)
\otimes \left(\FF_3^0-\GG1_{gg;gg}\right)\right](  {\bar{1}}_g, {\bar{2}}_g,i_g,j_g;z_1,z_2)  
\,A_4^0(\hat{\bar{1}}_g,\hat{\bar{2}}_g,i_g,j_g)
\nonumber \\
&&\phantom{\times\Bigg(}
-\bigg[\FF_4^0(  {\bar{1}}_g, {\bar{2}}_g,i_g,j_g;z_1,z_2)
+\FF_3^1(  {\bar{1}}_g,{\bar{2}}_g,i_g,j_g;z_1,z_2)\nonumber \\
&&\phantom{\times\Bigg(}
+
\frac{b_0}{\eps}\FF_{s,3}^0(  {\bar{1}}_g, {\bar{2}}_g,i_g,j_g;z_1,z_2)
-\overline{\left[ \FF_3^0 
\otimes  \FF_3^0\right]}(  {\bar{1}}_g, {\bar{2}}_g,i_g,j_g;z_1,z_2)\nonumber \\
&&\phantom{\times\Bigg(}
-\barGG2_{gg;gg}(z_1,z_2)\bigg]
\,A_4^0(\hat{\bar{1}}_g,\hat{\bar{2}}_g,i_g,j_g)
\Bigg\},
\end{eqnarray}
where we have eliminated $\GG2$ using,
\begin{eqnarray}
\GG2_{gg;gg}(z_1,z_2)  &=&
\frac{1}{2} \left[\GG1_{gg;gg}\otimes \GG1_{gg;gg}\right](z_1,z_2)
-\frac{b_0}{\e} \GG1_{gg;gg}(z_1,z_2)
+\barGG2_{gg;gg}(z_1,z_2),
\end{eqnarray}
with
\begin{eqnarray}
\barGG2_{gg;gg}(z_1,z_2) =
&=&
-\frac{1}{2\e}\left[\delta(1-z_2)p^1_{gg}(z_1)+
\delta(1-z_1)p^1_{gg}(z_2)\right]\nonumber \\
&&
-\frac{b_0}{2\e^2}\left[\delta(1-z_2)p^0_{gg}(z_1)+
\delta(1-z_1)p^0_{gg}(z_2)\right],
\end{eqnarray}
and the explicit form of the convolution $(\FF_3^0-\GG1_{gg;gg})\otimes(\FF_3^0-\GG1_{gg;gg})$ in this topology is given in~\eqref{eq:X3mGcIIFF}.

It is a key requirement of $\dsigma_{NNLO}^U$ that it explicitly cancels the infared poles present in the double virtual contribution $\dsigma_{NNLO}^{VV}$.
We previously observed that to cancel the infrared poles at NLO, the pole structure of the combination $\FF_3^0-\GG1_{gg;gg}$ must reproduce that appearing in the Catani ${\bom I}^{(1)}$ operator
as in~\eqref{eq:NLOfin1}.

With this in mind, we see that the first two lines of Eq.~\eqref{eq:UIIFF} correspond to the first two terms in the Catani pole structure 
of the two-loop contribution $A_4^2(\hat{\bar{1}}_g,\hat{\bar{2}}_g,i_g,j_g)$ given in Eq.~\eqref{eq:catIIFF}.
In fact these two lines give a contribution proportional to the finite part of $\FF_3^0$,
$$
\frac{b_0}{\eps} A_4^0 \times \Finite ( \FF_3^0 ).
$$
The remaining lines in Eq.~\eqref{eq:UIIFF} correspond to the remaining terms in the Catani pole structure where the fourth order poles cancel and, as expected, 
produce a deepest pole contribution of 
$$
-\frac{b_0}{\eps^3}A_4^0 + {\cal O}\left(\frac{1}{\eps^2}\right).
$$ 
We have checked that the  remaining terms in \eqref{eq:UIIFF} analytically cancel against the infrared poles (that start at ${\cal O}(1/\eps^{3})$)
coming from the
${\bom I}^{(1)}(2\epsilon)$ and ${\bom H}^{(2)}(\epsilon)$ terms in \eqref{eq:catIIFF}.

The subtraction term for the IFIF topology can be constructed in an analogous manner and we find an identical structure,
\begin{eqnarray}
\label{eq:UIFIF}
\dsigma_{NNLO}^{U,IFIF}&=&
-\int_2 \dsigma_{NNLO}^{S,IFIF}
-\int_1 \dsigma_{NNLO}^{VS,IFIF}
-   \dsigma_{NNLO}^{MF,2,IFIF}\nonumber \\
&&=\NNNLOVV\,\dPStwozz \,J_2^{(2)}(p_3,p_4) \,  \sum_{P(i,j)\in(3,4)} 
\nonumber\\
&&\times\Bigg\{ 
-\left[
\FF_3^0(  {\bar{1}}_g,i_g, {\bar{2}}_g,j_g;z_1,z_2)
-\GG1_{gg;gg}(z_1,z_2)\right]\nonumber \\
&& \hspace{2cm}\times \left[
\,A_4^1(\hat{\bar{1}}_g,i_g,\hat{\bar{2}}_g,j_g)
-\frac{b_0}{\e}\,A_4^0(\hat{\bar{1}}_g,i_g,\hat{\bar{2}}_g,j_g)\right]\nonumber \\
&&\phantom{\times\Bigg(}
-
\frac{1}{2}\left[\left(\FF_3^0-\GG1_{gg;gg}\right)
\otimes \left(\FF_3^0-\GG1_{gg;gg}\right)\right](  {\bar{1}}_g,i_g, {\bar{2}}_g,j_g;z_1,z_2)  
\,A_4^0(\hat{\bar{1}}_g,i_g,\hat{\bar{2}}_g,j_g)
\nonumber \\
&&\phantom{\times\Bigg(}
-\bigg[
\FF_4^0(  {\bar{1}}_g,i_g, {\bar{2}}_g,j_g;z_1,z_2)
+\FF_3^1(  {\bar{1}}_g,i_g, {\bar{2}}_g,j_g;z_1,z_2)\nonumber \\
&&\phantom{\times\Bigg(}
+
\frac{b_0}{\eps}
\FF_{s,3}^0( {\bar{1}}_g,i_g, {\bar{2}}_g,j_g;z_1,z_2)
-\overline{\left[ \FF_3^0 
\otimes  \FF_3^0\right]}(  {\bar{1}}_g,i_g, {\bar{2}}_g,j_g;z_1,z_2)\nonumber \\
&&\phantom{\times\Bigg(}
-\barGG2_{gg;gg}(z_1,z_2)\bigg]
\,A_4^0(\hat{\bar{1}}_g,i_g,\hat{\bar{2}}_g,j_g)
\Bigg\}.
\end{eqnarray}
Once again, we see that there is a correspondence with the Catani pole structure and we have checked explicitly that Eq.~\eqref{eq:UIFIF} precisely reproduces this pole structure.

\section{Conclusions} 
\label{sec:conc}

In this paper, we have generalised the antenna subtraction method for the calculation of higher order QCD
corrections to derive the double virtual subtraction term for exclusive collider observables for situations
with partons in the initial state to NNLO.  We focussed particular attention on the application of the
antenna subtraction formalism to construct the subtraction term relevant for the leading colour gluonic
double virtual contribution to dijet production. The gluon scattering channel is expected to be the dominant
contribution at NNLO. The subtraction term includes a mixture of integrated (tree-level three- and
four-parton and one-loop three-parton) antenna functions in final-final, initial-final and initial-initial
configurations. We note that the subtraction terms for processes involving quarks, as required for dijet or
vector boson plus jet processes, will make use of the same types of antenna building blocks as those
discussed here. By construction the counter-term removes the explicit infrared poles present on the two-loop
amplitude rendering the double virtual contribution locally finite over the whole of phase space.

The double virtual subtraction terms presented here provide a major step towards the NNLO evaluation of the
dijet observables at hadron colliders.  The expressions for $\dsigma_{NNLO}^U$ presented in Section~4,
completes the required set of subtraction terms $\dsigma_{NNLO}^{S}$~\cite{Glover:2010im}, $\dsigma_{NNLO}^{T}$~\cite{GehrmannDeRidder:2011aa}
and $\dsigma_{NNLO}^{U}$ required to render the  four-, five- and six-gluon channels explicitly finite and
well behaved in the single and double unresolved limits relevant for the gluon scattering at NNLO, thereby
enabling the construction of a full parton-level Monte Carlo implementation at leading order in the number of colours. The ultimate goal is the construction of a
numerical program to compute the  full NNLO QCD corrections to dijet production in hadron-hadron collisions.

\acknowledgments 

This research was supported in part by the UK Science and Technology Facilities
Council,  in part by the Swiss National Science Foundation (SNF) under contracts
PP00P2-139192 and 200020-138206 and in part by the European Commission through the ``LHCPhenoNet"
Initial Training Network PITN-GA-2010-264564. EWNG gratefully acknowledges the
support of the Wolfson Foundation, the Royal Society and the Pauli Center for Theoretical Studies. 

\appendix
\section{Momentum mappings}
\label{sec:appendixA}

The NNLO corrections to an $m$-jet final state receive contributions from processes with different numbers of final state particles.

In the antenna subtraction scheme, one is replacing antennae consisting of 
two hard radiators plus unresolved particles with two new hard radiators.
A key element of the antenna subtraction scheme is the factorisation of the matrix elements and phase space in the singular limits where one or more particles are unresolved.   This factorisation is guaranteed by the momentum mapping.    
 
If we denote the sets of momenta for the ${M}$-particle processs by $\{p\}_{M}$, then in order to subtract singular configurations from one final state, and add their integrated form back to final states with fewer particles, there needs to be a set of consistent momentum maps such that 
\begin{eqnarray}
\label{eq:map4to3}
\{p\}_{m+4} &\to &\{p\}_{m+3},\\
\label{eq:map4to2}
\{p\}_{m+4} &\to &\{p\}_{m+2},\\
\label{eq:map3to2}
\{p\}_{m+3} &\to &\{p\}_{m+2}.
\end{eqnarray} 

The single unresolved emissions corresponding to Eqs.~\eqref{eq:map4to3} and \eqref{eq:map3to2} have been the subject of previous publications and therefore we collect here
only the transformations that map singularities in the $(m+4)$-parton process 
to their integrated form in the $(m+2)$-parton process, i.e., ~Eq.~\eqref{eq:map4to2}.

If the antenna consists of two unresolved particles $j,k$ colour linked to two hard radiators
$i$ and $l$, then the mapping must produce two new hard radiators $I$ and $L$.  
Each mapping must conserve four-momentum and maintain the on-shellness of the particles involved.
There are three distinct cases,
\begin{align}
&{\rm final-final~configuration}   &  i j k l\to I L \nonumber\\
&{\rm initial-final~configuration} &  \hat i j k l\to \hat I L\nonumber \\ 
&{\rm initial-initial~configuration} &  \hat i j k\hat l \to \hat I \hat L\nonumber
\end{align}
where, as usual, initial state particles are denoted by a hat.
In principle, the momenta not involved in the antenna are also affected by the mapping.   For the final-final and initial-final maps, this is trivial.  Only in the initial-initial case are the spectator momenta actually modified.
The momentum transformations for these three mappings are described in
Refs.~\cite{Kosower:2002su,Daleo:2006xa} and are recalled below.

\subsection{Final-Final mapping}
\label{sec:appendixAFF}

The final-final mapping is given in 
\cite{Kosower:2002su} and reads,
\begin{eqnarray}
p_{I}^{\mu} \equiv p_{\widetilde{(ijk)}}^{\mu}&=&x\,p_{i}^{\mu}+r_1\,p_{j}^{\mu}+r_2\,p_{k}^{\mu}+z\,p_{l}^{\mu}\;,\nonumber\\
p_{L}^{\mu} \equiv p_{\widetilde{(jkl)}}^{\mu}&=&(1-x)\,p_{i}^{\mu}+(1-r_1)\,p_{j}^{\mu}+(1-r_2)\,p_{k}^{\mu}+(1-z)\,p_{l}^{\mu}\;,
\end{eqnarray}
with $p_{I}^2=p_{L}^2=0$.
Defining $s_{kl}=(p_{k}+p_{l})^2$, 
the coefficients $x,r_1,r_2,z$ are given by \cite{Kosower:2002su},
\begin{eqnarray}
r_1&=&\frac{s_{jk}+s_{jl}}{s_{ij}+s_{jk}+s_{jl}}\;,\nonumber\\
r_2&=&\frac{s_{kl}}{s_{ik}+s_{jk}+s_{kl}}\;,\nonumber\\
x&=&\frac{1}{2(s_{ij}+s_{ik}+s_{il})}\Big[(1+\rho)\,s_{ijkl}-r_1\,(s_{jk}+2\,s_{jl})   -r_2\,(s_{jk}+2\,s_{kl})  \nonumber\\
&&+(r_1-r_2)\frac{s_{ij}s_{kl}-s_{ik}s_{jl}}{s_{il}}     \Big]\;,\nonumber\\
z&=&\frac{1}{2(s_{il}+s_{jl}+s_{kl})}\Big[(1-\rho)\,s_{ijkl} -r_1\,(s_{jk}+2\,s_{ij})   -r_2\,(s_{jk}+2\,s_{ik})  \nonumber\\
&&-(r_1-r_2)\frac{s_{ij}s_{kl}-s_{ik}s_{jl}}{s_{il}}     \Big]\;,\nonumber\\
\rho&=&\Big[1+\frac{(r_1-r_2)^2}{s_{il}^2\,s_{ijkl}^2}\,\lambda(s_{ij}\,
s_{kl},s_{il}\,s_{jk},s_{ik}\,s_{jl})\nonumber\\
&&  +\frac{1}{s_{il}\,s_{ijkl}}\Big\{
2\,\big(r_1\,(1-r_2)+r_2(1-r_1)\big)\big( s_{ij}s_{kl}+s_{ik}s_{jl}-s_{jk}s_{il} \big)\nonumber\\
&&\qquad\qquad +\,4\,r_1\,(1-r_1)\,s_{ij} s_{jl}+4\,r_2\,(1-r_2)\,s_{ik}s_{kl}\Big\}\Big]^{\frac{1}{2}}\;,
\nonumber\\
\lambda(u,v,w)&=&u^2+v^2+w^2-2(uv+uw+vw)\;.\nonumber
\end{eqnarray}
This mapping smoothly interpolates all colour connected double unresolved singularities. It satisfies the following properties;
\begin{align} 
&p_{\widetilde{(ijk)}}\to p_{i}, 
&p_{\widetilde{(jkl)}}\to p_{l}\hspace{2.5cm} 
&\textrm{when $p_j,p_k\to0$,}\nonumber\\
&p_{\widetilde{(ijk)}}\to p_{i}+p_{j}+p_{k},
&p_{\widetilde{(jkl)}}\to p_{l}\hspace{2.5cm}  
&\textrm{when $p_i\parallel p_j\parallel p_k$,}\nonumber\\
&p_{\widetilde{(ijk)}}\to p_{i}, 
&p_{\widetilde{(jkl)}}\to p_{l}+p_{k}+p_{j}\hspace{0.9cm}   
&\textrm{when $p_j\parallel p_k\parallel p_l$,}\nonumber\\
&p_{\widetilde{(ijk)}}\to p_{i}, 
&p_{\widetilde{(jkl)}}\to p_{l}+p_{k}\hspace{1.7cm}  
&\textrm{when $p_j\to 0$ and $p_k\parallel p_l$,}\nonumber\\
&p_{\widetilde{(ijk)}}\to p_{i}+p_{j}, 
&p_{\widetilde{(jkl)}}\to p_{j}\hspace{2.5cm} 
&\textrm{when $p_k\to 0$ and $p_i\parallel p_j$,}\nonumber\\
&p_{\widetilde{(ijk)}}\to p_{i}+p_{j}, 
&p_{\widetilde{(jkl)}}\to p_{k}+p_{l} \hspace{1.7cm}
&\textrm{when $p_i\parallel p_j$ and $p_k\parallel p_l$.}\nonumber\\
\end{align}

\subsection{Initial-Final mapping}
\label{sec:appendixAIF}

The initial-final mapping is given in 
 \cite{Daleo:2006xa} and reads,
\begin{eqnarray}
p_{\hat I}^\mu \equiv \bar{p}_i^{\mu}&=&\hat{z}_i\,p_i^{\mu}\;,\nonumber\\
p_L^\mu \equiv p^\mu_{\widetilde{(jkl)}}&=&p_j^{\mu}+\,p_k^{\mu}+\,p_l^{\mu}-(1-\hat{z}_i)\,p_i^{\mu}\;,
\label{4to2IFmap}
\end{eqnarray}
with $p_{\hat I}^2=p_L^2=0$ and where the bar denotes a rescaling of the initial state parton and $\hat{z}_i$ is given by \cite{Daleo:2006xa},
\begin{eqnarray}
\hat{z}_i&=&\frac{s_{ijkl}}{s_{ij}+s_{ik}+s_{il}}\;.
\label{eq:xiIF}
\end{eqnarray}
The mapping satisfies the appropriate limits in all double singular configurations;
\begin{align} 
&{p_{\hat I}}\to p_{i}, 
&p_{\widetilde{(jkl)}}\to p_{l}\hspace{3cm} 
&\textrm{when $p_j,p_k\to0$,}\nonumber\\
&{p_{\hat I}}\to p_{i},
&p_{\widetilde{(jkl)}}\to p_{k}+p_{l}\hspace{2.2cm}  
&\textrm{when $p_j \to 0$ and $p_k\parallel p_l$,}\nonumber\\
&{p_{\hat I}}\to p_{i}, 
&p_{\widetilde{(jkl)}}\to p_{j}+p_{k}+p_{l}\hspace{1.4cm} 
&\textrm{when $p_j\parallel p_k\parallel p_l$,}\nonumber\\
&{p_{\hat I}}\to (1-w) p_{i}, 
&p_{\widetilde{(jkl)}}\to p_{l}\hspace{3.05cm}   
&\textrm{when $p_j\to w p_i$ and $p_k\to 0$,}\nonumber\\
&{p_{\hat I}}\to (1-w) p_{i}, 
&p_{\widetilde{(jkl)}}\to p_{l}+p_{k}\hspace{2.2cm}  
&\textrm{when $p_j\to w p_i$ and $p_k\parallel p_l$,}\nonumber\\
&{p_{\hat I}}\to (1-w) p_{i}, 
&p_{\widetilde{(jkl)}}\to p_{l} \hspace{3.05cm}
&\textrm{when $p_{j}+p_{k} = w p_{i}$,} 
\end{align}
and where the roles of partons $j$ and $k$ can be interchanged.

\subsection{Initial-Initial mapping}
\label{sec:appendixAII}
The initial-initial mapping for $il, jk \to IL$ is given in Ref.~\cite{Daleo:2006xa} and reads,
\begin{eqnarray}
p_{\hat I}^\mu \equiv \bar{p}_i^{\mu}&=&\hat{z}_i\,p_i^{\mu},\nonumber\\
p_{\hat L}^\mu \equiv \bar{p}_l^{\mu}&=&\hat{z}_l\,p_l^{\mu},\nonumber\\
\tilde{p}_m^{\mu}&=&p_m^{\mu}-\frac{2p_m\cdot (q+\tilde{q})}{(q+\tilde{q})^2}\;(q^{\mu}+\tilde{q}^{\mu})+\frac{2p_m\cdot q}{q^2}\;\tilde{q}^{\mu},\nonumber\\
q^{\mu}&=&p_i^{\mu}+p_l^{\mu}-p_j^{\mu}-p_k^{\mu}, \nonumber\\
\tilde{q}^{\mu}&=&\bar{p}_i^{\mu}+\bar{p}_l^{\mu}
\label{4to2IImap}
\end{eqnarray}
with $p_{\hat I}^2=p_{\hat L}^2=\tilde{p}_m^2=0$ and where the bar denotes rescaling of both of the initial state partons and $m$ runs over all the particles in the final state that are not actually part of the antenna but require boosting in order to restore momentum conservation.

The rescaling of the initial state momenta are given by the fractions $\hat{z}_i$ and $\hat{z}_l$ given by \cite{Daleo:2006xa},
\begin{eqnarray}
\hat{z}_i&=&
\sqrt{\frac{s_{il}+s_{jl}+s_{kl}}{s_{il}+s_{ij}+s_{ik}}}
\sqrt{\frac{s_{ijkl}}{s_{il}}}\;,\nonumber\\
\hat{z}_l&=&
\sqrt{\frac{s_{il}+s_{ij}+s_{ik}}{s_{il}+s_{jl}+s_{kl}}}
\sqrt{\frac{s_{ijkl}}{s_{il}}}\;. 
\label{eq:xiII}
\end{eqnarray}
With these definitions it is straightforward to check that the momenta mapping satisfies
the correct limits required for proper subtraction of infrared singularities.
Specifically, in the double unresolved limits;
\begin{align} 
&{p_{\hat I}}\to p_{i}, 
&{p_{\hat L}}\to p_{l}\hspace{3cm} 
&\textrm{when $p_j,p_k\to0$,}\nonumber\\
&{p_{\hat I}}\to (1-w_i) p_{i},
&{p_{\hat L}}\to p_{l}\hspace{3cm}  
&\textrm{when $p_j \to 0$ and $p_k=w_i p_i$,}\nonumber\\
&{p_{\hat I}}\to (1-w_i)p_{i}, 
&{p_{\hat L}}\to (1-w_l)p_{l}\hspace{1.65cm} 
&\textrm{when $p_j= w_ip_i$ and $p_k = w_l p_l$,}\nonumber\\
&{p_{\hat I}}\to (1-w_i) p_{i}, 
&{p_{\hat L}}\to p_{l} \hspace{3cm}
&\textrm{when $p_{j}+p_{k} = w_i p_{i}$,} 
\end{align}
together with the limits obtained from these by the exchange $p_i \leftrightarrow p_l$  and $p_j \leftrightarrow p_k$.

\section{Splitting functions}
\label{sec:appendixC}
In this section we collect the splitting functions that appear in the NLO and NNLO mass factorisation counter terms,  $\dsigma_{NLO}^{MF}$ and $\dsigma_{NNLO}^{MF,2}$, that are required for the construction 
of the double virtual subtraction term $\dsigma_{NNLO}^U$.

The leading colour contribution to the gluonic splitting functions read~\cite{Vogt:2004mw}
\begin{eqnarray}
\label{eq:pgg0}
p_{gg}^{0}(y)&=&  \frac{11}{6}\delta(1-y)+2{\cal D}_{0}(y)+\frac{2}{y}-4+2y-2y^2 ,\\
\label{eq:pgg0otimespgg0}
\l(p_{gg}^{0}\otimes p_{gg}^{0}\r)(y)&=&\l(\frac{121}{36}-\frac{2\pi^2}{3}\r)\delta(1-y)
	+\frac{11}{3}\l(2{\cal D}_{0}(y)+\frac{2}{y}-4+2y-2y^2\r)\nonumber\\&&
	+8{\cal D}_{1}(y)-\frac{4\h(0;y)}{1-y}+12-12y+\frac{44}{3}y^2-12y\,\h(0;y)+4y^2\,\h(0;y)\nonumber\\&&
	-\frac{44}{3y}-\frac{4\,\h(0;y)}{y}-4\h(1;y)\l(\frac{2}{y}-4+2y-2y^2\r)\\
\label{eq:pgg1}
p_{gg}^{1}(y) &=&  \l( \frac{8}{3} + 3 \* \z3 \r)\delta(1-y)
           + 27
          + (1+y) \* \bigg(
         \frac{11}{3} \* \h(0;y)
          + 8 \* \h(0,0;y)
       - \frac{27}{2}
          \bigg)\nonumber \\ &&
          + 2 \*\left(\frac{1}{1+y} -\frac{1}{y}-2-y-y^2\right)
\* \bigg(
            \h(0,0;y)
          - 2 \* \h(-1,0;y)
          - \z2
          \bigg)\nonumber \\ &&
          - \frac{67}{9} \* \bigg(\frac{1}{y}-y^2\bigg)
          - 12 \* \h(0;y)
       - \frac{44}{3} \*y^2\*\h(0;y)\nonumber \\ &&
          + \l(\frac{2}{y}-4+2y-2y^2\r) \bigg(
            \frac{67}{18}
          - \z2
          + \h(0,0;y)
          + 2 \* \h(1,0;y)
          + 2 \* \h(0,1;y)
          \bigg)\nonumber\\&&
          +\frac{2}{1-y}\l(\h(0,0;y)+2\h(1,0;y)+2\h(0,1;y)\r)+2\l(\frac{67}{18}-\z2\r){\cal D}_{0}(y)\nonumber\\&&          
\end{eqnarray}
where we have introduced the distributions
\beq
{\cal D}_{n}(y)=\left(\frac{\ln^{n}(1-y)}{1-y}\right)_{+}
\label{eq:Dn}
\eeq
and where the notation for the harmonic polylogarithms $\h(m_1,...,m_w;y)$, 
$m_j = 0,\pm 1$ follows Ref.~\cite{Vogt:2004mw,Remiddi:1999ew}. The lowest-weight ($w = 1$) functions $\h(m;y)$ are 
given by
\beq
\label{eq:hpol1}
  \h(0;y)       \: = \: \ln y \:\: , \quad\quad
  \h(\pm 1;y) \: = \: \mp \, \ln (1 \mp y) \:\: .
\eeq
The higher-weight ($w \geq 2$) functions are recursively defined as
\beq
\label{eq:hpol2}
  \h(m_1,...,m_w;y) \: = \:
    \left\{ \begin{array}{cl}
    \displaystyle{ \frac{1}{w!}\,\ln^w y \:\: ,}
       & \quad {\rm if} \:\:\: m^{}_1,...,m^{}_w = 0,\ldots ,0 \\[2ex]
    \displaystyle{ \int_0^y \! dz\: f_{m_1}(z) \, \h(m_2,...,m_w;z)
       \:\: , } & \quad {\rm otherwise}
    \end{array} \right.
\eeq
with
\beq
\label{eq:hpolf}
  f_0(y)       \: = \: \frac{1}{y} \:\: , \quad\quad
  f_{\pm 1}(y) \: = \: \frac{1}{1 \mp y} \:\: .
\eeq

\section{Convolution integrals}
\label{sec:appendixC}
In this section we collect the convolution integrals that capture the infrared structure of the double virtual contribution. In the most general case, we find
convolution integrals between two integrated three-parton tree-level antennae that come from the integrated double real and real-virtual subtraction terms, but also
convolution integrals of an Altarelli-Parisi splitting kernel with an integrated three-parton tree-level antenna and convolution integrals of
two Altarelli-Parisi splitting kernels that are produced by the mass factorisation contribution. 

The convolution integrals described above are defined respectively as,
\begin{eqnarray}
\left[{\cal X}_{3}^{0}\otimes{\cal X}_{3}^{0}\right](\souter,\sinner;z_1,z_2)&=&\int {\rm d}x_{1}{\rm d}x_{2}\,{\rm d}y_{1}{\rm d}y_{2}\,{\cal X}_{3}^0(\souter,x_1,x_2)
\,{\cal X}_{3}^0(\sinner,y_1,y_2)
\nonumber\\
&\times&\delta(z_{1}-x_{1}y_{1})\delta(z_{2}-x_{2}y_{2})\,,\\
\left[\GG1_{ki}\otimes{\cal X}_{3}^{0}\right]_1(\souter;z_1,z_2)&=&\int {\rm d}x_{1}\,{\rm d}y_{1}\,\GG1_{ki}(x_{1})
\,{\cal X}_{3}^0(\souter,y_1,z_2)\delta(z_{1}-x_{1}y_{1})\,, \\
\left[\GG1_{ki}\otimes{\cal X}_{3}^{0}\right]_2(\souter;z_1,z_2)&=&\int {\rm d}x_{2}\,{\rm d}y_{2}\,\GG1_{ki}(x_{2})
\,{\cal X}_{3}^0(\souter;z_1,y_2)\delta(z_{2}-x_{2}y_{2})\,,\\
\left[\GG1_{kl}\otimes\GG1_{li}\right]_1(z_1,z_2)&=&\int {\rm d}x_{1}\,{\rm d}y_{1}\,\GG1_{kl}(x_{1})\,\GG1_{li}(y_{1})\delta(z_{1}-x_{1}y_{1})\delta(1-z_{2})\,,\\
\left[\GG1_{kl}\otimes\GG1_{li}\right]_2(z_1,z_2)&=&\int {\rm d}x_{2}\,{\rm d}y_{2}\,\GG1_{kl}(x_{2})\,\GG1_{li}(y_{2})\delta(z_{2}-x_{2}y_{2})\delta(1-z_{1})\,\nonumber\\
&=&\left[\GG1_{kl}\otimes\GG1_{li}\right]_1(z_2,z_1)\;.
\end{eqnarray}
To explicitly show the cancellation of $\e$-poles we perform the above integrals analytically. 
For the purpose of this paper the relevant antennae are those containing the pure gluon final state that we recall below.
The integrated final-final~\cite{GehrmannDeRidder:2005cm}, initial-final~\cite{Daleo:2006xa} and initial-initial~\cite{Daleo:2006xa} to ${\cal O}(\e)$ read, respectively,
\begin{eqnarray}
{\cal F}_{ijk}^0(s_{ik},x_{1},x_{2})&=&\left(\frac{s_{ik}}{\mu^2}\right)^{-\e}\delta(1-x_{1})\delta(1-x_{2})\l[\frac{3}{\e^2}+\frac{11}{2\e}+\l(\frac{73}{4}-\frac{7\pi^2}{4}\r)+ {\cal O}(\e)\r]\;,\nonumber\\
{\cal F}^0_{1,jk}(s_{\bar{1}k},x_{1},x_{2})&=&\left(\frac{|s_{\bar{1}k}|}{\mu^2}\right)^{-\e}\delta(1-x_{2})\Bigg[\frac{2}{\e^2}\delta(1-x_{1})+\frac{1}{\e}\Bigg(\frac{11}{6}\delta(1-x_{1})+4-\frac{2}{x_{1}}
-2x_{1}\nonumber\\
&&+2x_{1}^2-2{\cal D}_{0}(x_{1})\Bigg)+\delta(1-x_{1})\l(\frac{67}{18}-\frac{\pi^2}{2}\r)-\frac{11}{6}{\cal D}_{0}(x_{1})+2{\cal D}_{1}(x_{1})\nonumber\\
&&-\frac{11}{6x_{1}}+\h(0,x_{1})\Big(4-\frac{2}{x_{1}}-\frac{2}{1-x_{1}}-2x_{1}+2x_{1}^2\Big)\nonumber\\
&&+\h(1,x_{1})\Bigg(4-\frac{2}{x_{1}}-2x_{1}+2x_{1}^2\Bigg) + {\cal O}(\e)\Bigg]\;,\nonumber\\
{\cal F}^{0}_{12,j}(s_{\bar{1}\bar{2}},x_1,x_2)&=&\l(\frac{s_{\bar{1}\bar{2}}}{\mu^2}\r)^{-\e}\Bigg[\frac{1}{2\e^2}\delta(1-x_{1})\delta(1-x_{2})+\frac{1}{\e}\Bigg(\delta(1-x_{1})\Big(2-\frac{1}{x_{2}}
-x_{2}+x_{2}^2\nonumber\\
&&-{\cal D}_{0}(x_{2})\Big)\Bigg)-\delta(1-x_{1})\delta(1-x_{2})\frac{\pi^2}{24}+\delta(1-x_{1})\Big({\cal D}_{1}(x_{2})\nonumber\\
&&+(\h(-1,x_{2})-\log(2))
\Big(2-\frac{1}{x_{2}}-\frac{1}{1-x_{2}}-x_{2}+x_{2}^2\Big)\nonumber\\
&&+\h(1,x_{2})\Big(2-\frac{1}{x_{2}}-x_{2}+x_{2}^2\Big)\Big)-{\cal D}_{0}(x_{1})\Big(2-\frac{1}{x_{2}}-x_{2}+x_{2}^2\Big)\nonumber\\
&&+\frac{1}{2}{\cal D}_{0}(x_{1}){\cal D}_{0}(x_{2})+\frac{1}{2x_{1}x_{2}(1+x_{1})(1+x_{2})}\Big(
2 x_{1}^4 x_{2}^4+2 x_{1}^4 x_{2}^3-4 x_{1}^3 x_{2}^5\nonumber\\&&-2 x_{1}^3 x_{2}^4-2 x_{1}^3 x_{2}^2+10 x_{1}^2 x_{2}^6+6
   x_{1}^2 x_{2}^5+2 x_{1}^2 x_{2}^4+4 x_{1}^2 x_{2}^3+2 x_{1}^2 x_{2}^2+10 x_{1} x_{2}^6\nonumber\\&&+10 x_{1} x_{2}^5+4
   x_{1} x_{2}^4+2 x_{1} x_{2}^3+2 x_{1} x_{2}^2+x_{1} x_{2}+2 x_{2}+2
\Big)\nonumber\\
&&+\frac{1}{(x_{1}+x_{2})^4}\Big(
-10 x_{1}^3 x_{2}^5+x_{1}^3 x_{2}^3-22 x_{1}^2 x_{2}^6-5 x_{1}^2 x_{2}^4+5 x_{1}^2 x_{2}^2-18 x_{1} x_{2}^7\nonumber\\&&-4
   x_{1} x_{2}^5+2 x_{1} x_{2}^3-3 x_{1} x_{2}-5 x_{2}^8-x_{2}^6-4 x_{2}
\Big)+(x_{1}\leftrightarrow x_{2}) + {\cal O}(\e)\Bigg]\;.\nonumber
\end{eqnarray}
To evaluate the finite part of the convolutions consistently, the integrated antennae are needed through to ${\cal O}(\epsilon^2)$ and computer-readable expressions can be found in files attached to \cite{GehrmannDeRidder:2005cm,Daleo:2006xa}.

Convolution integrals involving the antennae above appear explicitly in the four-gluon subtraction term $\dsigma_{NNLO}^{U}$ derived in section~\ref{sec:dsigmaU}. 
In particular the explicit form of the convolution $(\FF_3^0-\GG1_{gg;gg})\otimes(\FF_3^0-\GG1_{gg;gg})$ for both the IIFF and IFIF topologies reads, 
\begin{eqnarray}
&&\left(\FF_3^0-\GG1_{gg;gg}\right)
\otimes \left(\FF_3^0-\GG1_{gg;gg}\right)(\b{1}_g,\b{2}_g,i_g,j_g;z_1,z_2)=\nonumber\\
&&
\phantom{+2}\l[{\cal F}_3^0\otimes{\cal F}_3^0\r](s_{\b{1}\b{2}},s_{\b{1}\b{2}};z_1,z_2)
+\frac{1}{4}\l[{\cal F}_3^0\otimes{\cal F}_3^0\r](s_{\b{2}i},s_{\b{2}i};z_1,z_2)
+\frac{1}{9}\l[{\cal F}_3^0\otimes{\cal F}_3^0\r](s_{ij},s_{ij};z_1,z_2)\nonumber\\
&&+\frac{1}{4}\l[{\cal F}_3^0\otimes{\cal F}_3^0\r](s_{j\b{1}},s_{j\b{1}};z_1,z_2)
\,\,+\,\,\l[{\cal F}_3^0\otimes{\cal F}_3^0\r](s_{\b{1}\b{2}},s_{\b{2}i};z_1,z_2)
+\frac{2}{3}\l[{\cal F}_3^0\otimes{\cal F}_3^0\r](s_{\b{1}\b{2}},s_{ij};z_1,z_2)\nonumber\\
&&+\phantom{\frac{1}{4}}\l[{\cal F}_3^0\otimes{\cal F}_3^0\r](s_{\b{1}\b{2}},s_{j\b{1}};z_1,z_2)
+\frac{1}{3}\l[{\cal F}_3^0\otimes{\cal F}_3^0\r](s_{\b{2}i},s_{ij};z_1,z_2)
+\frac{1}{2}\l[{\cal F}_3^0\otimes{\cal F}_3^0\r](s_{\b{2}i},s_{j\b{1}};z_1,z_2)\nonumber\\
&&+\frac{1}{3}\l[{\cal F}_3^0\otimes{\cal F}_3^0\r](s_{ij},s_{j\b{1}};z_1,z_2)
+\l[\GG1_{gg}\otimes\GG1_{gg}\r]_1(z_1,z_2)
+\l[\GG1_{gg}\otimes\GG1_{gg}\r]_2(z_1,z_2)\nonumber\\
&&+2\,\GG1_{gg}(z_1)\GG1_{gg}(z_2)-2\l[\GG1_{gg}\otimes{\cal F}_3^0\r]_1(s_{\b{1}\b{2}};z_1,z_2)
-2\l[\GG1_{gg}\otimes{\cal F}_3^0\r]_2(s_{\b{1}\b{2}};z_1,z_2)\nonumber\\
&&-\phantom{2}\l[\GG1_{gg}\otimes{\cal F}_3^0\r]_1(s_{\b{2}i};z_1,z_2)
-\l[\GG1_{gg}\otimes{\cal F}_3^0\r]_2(s_{\b{2}i};z_1,z_2)
-\frac{2}{3}\l[\GG1_{gg}\otimes{\cal F}_3^0\r]_1(s_{ij};z_1,z_2)\nonumber\\
&&
-\frac{2}{3}\l[\GG1_{gg}\otimes{\cal F}_3^0\r]_2(s_{ij};z_1,z_2)
-\l[\GG1_{gg}\otimes{\cal F}_3^0\r]_1(s_{j\b{1}};z_1,z_2)
-\l[\GG1_{gg}\otimes{\cal F}_3^0\r]_2(s_{j\b{1}};z_1,z_2)\;,
\label{eq:X3mGcIIFF}
\end{eqnarray}

\begin{eqnarray}
&&\left(\FF_3^0-\GG1_{gg;gg}\right)
\otimes \left(\FF_3^0-\GG1_{gg;gg}\right)(\b{1}_g,i_g,\b{2}_g,j;z_1,z_2)=\nonumber\\
&&
\phantom{+}\frac{1}{4}\l[{\cal F}_3^0\otimes{\cal F}_3^0\r](s_{\b{1}i},s_{\b{1}i};z_1,z_2)
+\frac{1}{4}\l[{\cal F}_3^0\otimes{\cal F}_3^0\r](s_{i\b{2}},s_{i\b{2}};z_1,z_2)
+\frac{1}{4}\l[{\cal F}_3^0\otimes{\cal F}_3^0\r](s_{\b{2}j},s_{\b{2}j};z_1,z_2)\nonumber\\
&&+\frac{1}{4}\l[{\cal F}_3^0\otimes{\cal F}_3^0\r](s_{j\b{1}},s_{j\b{1}};z_1,z_2)
+\frac{1}{2}\l[{\cal F}_3^0\otimes{\cal F}_3^0\r](s_{\b{1}i},s_{i\b{2}};z_1,z_2)
+\frac{1}{2}\l[{\cal F}_3^0\otimes{\cal F}_3^0\r](s_{\b{1}i},s_{\b{2}j};z_1,z_2)\nonumber\\
&&+\frac{1}{2}\l[{\cal F}_3^0\otimes{\cal F}_3^0\r](s_{\b{1}i},s_{j\b{1}};z_1,z_2)
+\frac{1}{2}\l[{\cal F}_3^0\otimes{\cal F}_3^0\r](s_{i\b{2}},s_{\b{2}j};z_1,z_2)
+\frac{1}{2}\l[{\cal F}_3^0\otimes{\cal F}_3^0\r](s_{i\b{2}},s_{j\b{1}};z_1,z_2)\nonumber\\
&&+\frac{1}{2}\l[{\cal F}_3^0\otimes{\cal F}_3^0\r](s_{\b{2}j},s_{j\b{1}};z_1,z_2)
+\l[\GG1_{gg}\otimes\GG1_{gg}\r]_1(z_1,z_2)
+\l[\GG1_{gg}\otimes\GG1_{gg}\r]_2(z_1,z_2)\nonumber\\
&&+2\,\GG1_{gg}(z_1)\GG1_{gg}(z_2)
-\l[\GG1_{gg}\otimes{\cal F}_3^0\r]_1(s_{\b{1}i};z_1,z_2)
-\l[\GG1_{gg}\otimes{\cal F}_3^0\r]_2(s_{\b{1}i};z_1,z_2)\nonumber\\
&&
-\l[\GG1_{gg}\otimes{\cal F}_3^0\r]_1(s_{i\b{2}};z_1,z_2)
-\l[\GG1_{gg}\otimes{\cal F}_3^0\r]_2(s_{i\b{2}};z_1,z_2)
-\l[\GG1_{gg}\otimes{\cal F}_3^0\r]_1(s_{\b{2}j};z_1,z_2)\nonumber\\
&&
-\l[\GG1_{gg}\otimes{\cal F}_3^0\r]_2(s_{\b{2}j};z_1,z_2)
-\l[\GG1_{gg}\otimes{\cal F}_3^0\r]_1(s_{j\b{1}};z_1,z_2)
-\l[\GG1_{gg}\otimes{\cal F}_3^0\r]_2(s_{j\b{1}};z_1,z_2)\;.
\label{eq:X3mGcIFIF}
\end{eqnarray}

The full expressions for the convolution integrals through the finite part are quite lengthy and are attached in computer-readable form to the arXiv submission of this paper. In this section
we give only the leading singular contributions.
For the convolutions involving two three-parton tree-level antennae we obtain,
\begin{eqnarray}
&&\left[{\cal F}_{ijk}^3\otimes{\cal F}_{nop}^3\right](s_{ik},s_{np};z_{1},z_{2})=\left(\frac{s_{ik}}{\mu^2}\right)^{-\e}\left(\frac{s_{np}}{\mu^2}\right)^{-\e}\delta(1-z_{1})\delta(1-z_{2})
\nonumber\\
&&\Bigg[\frac{9}{\e^4}+\frac{33}{\e^3}+\frac{1}{\e^2}\Big(\frac{559}{4}-\frac{21}{2}\pi^2\Big)+\frac{1}{\e}\Big(539-\frac{77}{2}\pi^2-150\zeta_3\Big)
+\Big(\frac{31625}{16}-\frac{3913}{24}\pi^2+\frac{87}{40}\pi^4\nonumber\\
&&-550\zeta_3\Big)\Bigg] + {\cal O}(\e),\\
&&\left[{\cal F}_{ijk}^3\otimes{\cal F}_{1,np}^3\right](s_{ik},s_{\b{1}p};z_{1},z_{2})=\left(\frac{s_{ik}}{\mu^2}\right)^{-\e}\left(\frac{|s_{\b{1}p}|}{\mu^2}\right)^{-\e}\delta(1-z_{2})\Bigg[\nonumber\\
&&\phantom{+}\frac{1}{\e^4}\Big[6\delta(1-z_{1})\Big]\nonumber\\
&&+\frac{1}{\e^3}\Big[\frac{33}{2}\delta(1-z_{1})+\Big(12-\frac{6}{z_{1}}-6z_{1}+6z_{1}^2-6{\cal D}_{0}(z_{1})\Big)\Big]\nonumber\\
&&+\frac{1}{\e^2}\Big[\delta(1-z_{1})\Big(\frac{231}{4}-5\pi^2\Big)-\frac{33}{2}{\cal D}_{0}(z_{1})+6{\cal D}_{1}(z_{1})+22-\frac{33}{2z_{1}}-11z_{1}+11z_{1}^2\nonumber\\
&&\hspace{1.0cm}+\h(1,z_{1})\Big(12-\frac{6}{z_{1}}-6z_{1}+6z_{1}^2\Big)+\h(0,z_{1})\Big(12-\frac{6}{z_{1}}-\frac{6}{1-z_{1}}-6z_{1}+6z_{1}^2\Big)\Big]\nonumber\\
&&+{\cal O}(\e^{-1})\Bigg],\\
&&\left[{\cal F}_{ijk}^3\otimes{\cal F}_{12,p}^3\right](s_{ik},s_{\b{1}\b{2}};z_{1},z_{2})=\left(\frac{s_{ik}}{\mu^2}\right)^{-\e}\left(\frac{s_{\b{1}\b{2}}}{\mu^2}\right)^{-\e}\Bigg[\nonumber\\
&&\phantom{+}\frac{1}{\e^4}\Big[3\delta(1-z_{1})\delta(1-z_{2})\Big]\nonumber\\
&&+\frac{1}{\e^3}\Big[\frac{11}{2}\delta(1-z_{1})\delta(1-z_{2})+\delta(1-z_{1})\Big(6-\frac{3}{z_{2}}-3z_{2}+3z_{2}^2-3{\cal D}_{0}(z_{2})\Big)\nonumber\\
&&\hspace{1.0cm}+\delta(1-z_{2})\Big(6-\frac{3}{z_{1}}-3z_{1}+3z_{1}^2-3{\cal D}_{0}(z_{1})\Big)\Big]\nonumber\\
&&+{\cal O}(\e^{-2})\Bigg],\\
&&\left[{\cal F}_{1,jk}^3\otimes{\cal F}_{1,np}^3\right](s_{\b{1}k},s_{\b{1}p};z_{1},z_{2})=\left(\frac{|s_{\b{1}k}|}{\mu^2}\right)^{-\e}\left(\frac{|s_{\b{1}p}|}{\mu^2}\right)^{-\e}\Bigg[\nonumber\\
&&\phantom{+}\frac{1}{\e^4}\Big[4\delta(1-z_{1})\delta(1-z_{2})\Big]\nonumber\\
&&+\frac{1}{\e^3}\Big[\frac{22}{3}\delta(1-z_{1})\delta(1-z_{2})+\delta(1-z_{2})\Big(16-\frac{8}{z_{1}}-8z_{1}+8z_{1}^2-8{\cal D}_{0}(z_{1})\Big)\Big]\nonumber\\
&&+\frac{1}{\e^2}\Big[\delta(1-z_{1})\delta(1-z_{2})\Big(\frac{73}{4}-\frac{8}{3}\pi^2\Big)+\delta(1-z_{2})\Big(-\frac{44}{3}{\cal D}_{0}(z_{1})+16{\cal D}_{1}(z_{1})
+\frac{80}{3}-\frac{88}{3z_{1}}\nonumber\\
&&\hspace{1.0cm}-\frac{58}{3z_{1}}+22z_{1}^2+\h(1,z_{1})\Big(32-\frac{16}{z_{1}}-16z_{1}+16z_{1}^2\Big)+\h(0,z_{1})\Big(16-\frac{12}{z_{1}}
-\frac{12}{1-z_{1}}\nonumber\\
&&\hspace{1.0cm}-20z_{1}^2+12z_{1}^2\Big)\Big)\Big]\nonumber\\
&&+{\cal O}(\e^{-1})\Bigg],\\
&&\left[{\cal F}_{1,jk}^3\otimes{\cal F}_{2,np}^3\right](s_{\b{1}k},s_{\b{2}p};z_{1},z_{2})=\left(\frac{|s_{\b{1}k}|}{\mu^2}\right)^{-\e}\left(\frac{|s_{\b{2}p}|}{\mu^2}\right)^{-\e}\Bigg[\nonumber\\
&&\phantom{+}\frac{1}{\e^4}\Big[4\delta(1-z_{1})\delta(1-z_{2})\Big]\nonumber\\
&&+\frac{1}{\e^3}\Big[\frac{22}{3}\delta(1-z_{1})\delta(1-z_{2})+\delta(1-z_{1})\Big(8-\frac{4}{z_{2}}-4z_{2}+4z_{2}^2-4{\cal D}_{0}(z_{2})\Big)\nonumber\\
&&\hspace{1.0cm}+\delta(1-z_{2})\Big(8-\frac{4}{z_{1}}-4z_{1}+4z_{1}^2-4{\cal D}_{0}(z_{1})\Big)\Big]\nonumber\\
&&+\frac{1}{\e^2}\Big[\delta(1-z_{1})\delta(1-z_{2})\Big(\frac{73}{4}-2\pi^2\Big)+\delta(1-z_{1})\Big(-\frac{22}{3}{\cal D}_{0}(z_{2})+4{\cal D}_{1}(z_{2})+\frac{22}{3}-\frac{22}{3z_{2}}-\frac{11}{3}z_{2}\nonumber\\
&&\hspace{1.0cm}+\frac{11}{3}z_{2}^2+\h(1,z_{2})\Big(8-\frac{4}{z_{2}}-4z_{2}+4z_{2}^2\Big)+\h(0,z_{2})\Big(8-\frac{4}{z_{2}}-\frac{4}{1-z_{2}}-4z_{2}+4z_{2}^2\Big)\Big)\nonumber\\
&&\hspace{1.0cm}+\delta(1-z_{2})\Big(-\frac{22}{3}{\cal D}_{0}(z_{1})+4{\cal D}_{1}(z_{1})+\frac{22}{3}-\frac{22}{3z_{1}}-\frac{11}{3}z_{1}
+\frac{11}{3}z_{1}^2+\h(1,z_{1})\Big(8-\frac{4}{z_{1}}\nonumber\\
&&\hspace{1.0cm}-4z_{1}+4z_{1}^2\Big)+\h(0,z_{1})\Big(8-\frac{4}{z_{1}}-\frac{4}{1-z_{1}}-4z_{1}+4z_{1}^2\Big)\Big)-{\cal D}_{0}(z_{1})\Big(8-\frac{4}{z_{2}}-4z_{2}\nonumber\\
&&\hspace{1.0cm}+4z_{2}^2\Big)-{\cal D}_{0}(z_{2})\Big(8-\frac{4}{z_{1}}-4z_{1}+4z_{1}^2\Big)+4{\cal D}_{0}(z_{1}){\cal D}_{0}(z_{2})
+16+\frac{4}{z_{1}z_{2}}-\frac{8}{z_{1}}+\frac{4z_2}{z_1}\nonumber\\
&&\hspace{1.0cm}-\frac{4z_2^2}{z_1}-\frac{8}{z_2}-8z_2+8z_2^2+\frac{4z_1}{z_2}-8z_1+4z_1z_2-4z_1z_2^2-\frac{4z_1^2}{z_2}+8z_1^2-4z_1^2z_2
+4z_1^2z_2^2\Big]\nonumber\\
&&+{\cal O}(\e^{-1})\Bigg],\\
&&\left[{\cal F}_{\b{1},jk}^3\otimes{\cal F}_{\b{1}\b{2},p}^3\right](s_{\b{1}k},s_{\b{1}\b{2}};z_{1},z_{2})=\left(\frac{|s_{\b{1}k}|}{\mu^2}\right)^{-\e}\left(\frac{s_{\b{1}\b{2}}}{\mu^2}\right)^{-\e}\Bigg[\nonumber\\
&&\phantom{+}\frac{1}{\e^4}\Big[2\delta(1-z_{1})\delta(1-z_{2})\Big]\nonumber\\
&&+\frac{1}{\e^3}\Big[\frac{11}{6}\delta(1-z_{1})\delta(1-z_{2})+\delta(1-z_{1})\Big(4-\frac{2}{z_{2}}-2z_{2}+2z_{2}^2-2{\cal D}_{0}(z_{2})\Big)\nonumber\\
&&\hspace{1.0cm}+\delta(1-z_{2})\Big(8-\frac{4}{z_{1}}-4z_{1}+4z_{1}^2-4{\cal D}_{0}(z_{1})\Big)\Big]\nonumber\\
&&+{\cal O}(\e^{-2})\Bigg],\\
&&\left[{\cal F}_{\b{1}\b{2},j}^3\otimes{\cal F}_{\b{1}\b{2},p}^3\right](s_{\b{1}\b{2}},s_{\b{1}\b{2}};z_{1},z_{2})=\left(\frac{s_{\b{1}\b{2}}}{\mu^2}\right)^{-\e}\left(\frac{s_{\b{1}\b{2}}}{\mu^2}\right)^{-\e}\Bigg[\nonumber\\
&&+\frac{1}{\e^4}\Big[\delta(1-z_{1})\delta(1-z_{2})\Big]\nonumber\\
&&+\frac{1}{\e^3}\Big[\delta(1-z_{1})\Big(4-\frac{2}{z_{2}}-2z_{2}+2z_{2}^2-2{\cal D}_{0}(z_{2})\Big)+\delta(1-z_{2})\Big(4-\frac{2}{z_{1}}-2z_{1}+2z_{1}^2
-2{\cal D}_{0}(z_{1})\Big)\Big]\nonumber\\
&&\hspace{1.0cm}\nonumber\\
&&+{\cal O}(\e^{-2})\Bigg].
\end{eqnarray}

For the convolution integrals involving an integrated three-parton tree-level antenna and an Altarelli-Parisi splitting kernel we obtain,
\begin{eqnarray}
&&\left[\GG1_{gg}\otimes{\cal F}_{nop}^3\right]_1(s_{np};z_{1},z_{2})=\left(\frac{s_{np}}{\mu^2}\right)^{-\e}\delta(1-z_{2})\Bigg[
\nonumber\\
&&\phantom{+}\frac{1}{\e^3}\Big[-\frac{11}{2}\delta(1-z_{1})+12-\frac{6}{z_1}-6z_1+6z_1^2-6{\cal D}_{0}(z_1)\Big]\nonumber\\
&&+\frac{1}{\e^2}\Big[-\frac{121}{12}\delta(1-z_{1})+22-\frac{11}{z_{1}}-11z_{1}+11z_{1}^2-11{\cal D}_{0}(z_{1})\Big]\nonumber\\
&&+\frac{1}{\e}\Big[\delta(1-z_{1})\Big(-\frac{803}{24}+\frac{77}{24}\pi^2\Big)+73
-\frac{73}{2z_1}-\frac{73}{2}z_1+\frac{73}{2}z_1^2-\frac{73}{2}{\cal D}_{0}(z_1)\nonumber\\
&&\hspace{1.0cm}-\pi^2\Big(7-\frac{7}{2z_1}-\frac{7}{2}z_1+\frac{7}{2}z_1^2-\frac{7}{2}{\cal D}_{0}(z_1)\Big)\Big]\nonumber\\
&&+{\cal O}(\e^{0})\Bigg],\\
&&\left[\GG1_{gg}\otimes{\cal F}_{1,jk}^3\right]_1(s_{\b{1}k};z_{1},z_{2})=\left(\frac{|s_{\b{1}k}|}{\mu^2}\right)^{-\e}\delta(1-z_{2})\Bigg[
\nonumber\\
&&\phantom{+}\frac{1}{\e^3}\Big[-\frac{11}{3}\delta(1-z_{1})+8-\frac{4}{z_1}-4z_1+4z_1^2-4{\cal D}_{0}(z_1)\Big]\nonumber\\
&&+\frac{1}{\e^2}\Big[\delta(1-z_{1})\Big(-\frac{121}{36}-\frac{2}{3}\pi^2\Big)+\h(1,z_{1})\Big(16-\frac{8}{z_1}-8z_1+8z_1^2\Big)
-\h(0,z_1)\Big(\frac{4}{z_1}+\frac{4}{1-z_1}\nonumber\\
&&\hspace{1.0cm}+12z_1-4z_1^2\Big)+12-\frac{44}{3z_1}-12z_1+\frac{44}{3}z_1^2+8{\cal D}_{1}(z_1)\Big]\nonumber\\
&&+\frac{1}{\e}\Big[\delta(1-z_{1})\Big(-\frac{737}{108}+\frac{11}{36}\pi^2-4\zeta_3\Big)+\h(1,1,z_1)\Big(24-\frac{12}{z_1}-12z_1+12z_1^2\Big)\nonumber\\
&&\hspace{1.0cm}+\h(1,0,z_1)\Big(16-\frac{8}{z_1}-\frac{8}{1-z_1}-8z_1+8z_1^2\Big)+\h(0,1,z_1)\Big(8-\frac{8}{z_1}-\frac{8}{1-z_1}-16z_1\nonumber\\
&&\hspace{1.0cm}+8z_1^2\Big)+\h(0,0,z_1)\Big(-\frac{4}{z_1}-\frac{4}{1-z_1}-12z_1+4z_1^2\Big)
+\h(1,z_1)\Big(12-\frac{55}{3z_1}-12z_1\nonumber\\
&&\hspace{1.0cm}+\frac{44}{3}z_1^2\Big)+\h(0,z_1)\Big(14-\frac{22}{3z_1}-4z_1+\frac{44}{3}z_1^2\Big)-6{\cal D}_{2}(z_1)+\frac{11}{3}{\cal D}_{1}(z_1)
-\frac{49}{12}{\cal D}_{0}(z_1)\nonumber\\
&&\hspace{1.0cm}+\frac{571}{18}-\frac{73}{4z_1}-\frac{437}{18}z_1+\frac{389}{18}z_1^2
+\pi^2\Big(\frac{5}{3z_1}-2+3z_1-\frac{5}{3}z_1^2+{\cal D}_{0}(z_1)\Big)\Big]\nonumber\\
&&+{\cal O}(\e^{0})\Bigg],\\
&&\left[\GG1_{gg}\otimes{\cal F}_{1,jk}^3\right]_2(s_{\b{1}k};z_{1},z_{2})=\left(\frac{|s_{\b{1}k}|}{\mu^2}\right)^{-\e}\Bigg[
\nonumber\\
&&\phantom{+}\frac{1}{\e^3}\Big[-\frac{11}{3}\delta(1-z_{1})\delta(1-z_{2})+\delta(1-z_{1})\Big(8-\frac{4}{z_2}-4z_2+4z_2^2-4{\cal D}_{0}(z_2)\Big)\Big]\nonumber\\
&&+\frac{1}{\e^2}\Big[\-\frac{121}{36}\delta(1-z_{1})\delta(1-z_{2})+\delta(1-z_{1})\Big(\frac{22}{3}-\frac{11}{3z_2}-\frac{11}{3}z_2+\frac{11}{3}z_2^2
-\frac{11}{3}{\cal D}_{0}(z_2)\Big)\Big]\nonumber\\
&&\hspace{1.0cm}+\delta(1-z_{2})\Big(-\frac{22}{3}+\frac{11}{3z_1}+\frac{11}{3}z_1-\frac{11}{3}z_1^2
+\frac{11}{3}{\cal D}_{0}(z_1)\Big)-{\cal D}_{0}(z_{1})\Big(8-\frac{4}{z_2}-4z_2+4z_2^2\Big)\nonumber\\
&&\hspace{1.0cm}-{\cal D}_{0}(z_{2})\Big(8-\frac{4}{z_1}-4z_1+4z_1^2\Big)+4{\cal D}_{0}(z_{1}){\cal D}_{0}(z_{2})+16+\frac{4}{z_1z_2}-\frac{8}{z_1}
+\frac{4z_2}{z_1}-\frac{4z_2^2}{z_1}-\frac{8}{z_2}\nonumber\\
&&\hspace{1.0cm}-8z_2+8z_2^2+\frac{4z_1}{z_2}-8z_1+4z_1z_2-4z_1z_2^2-\frac{4z_1^2}{z_2}+8z_1^2-4z_1^2z_2+4z_1^2z_2^2\Big]\nonumber\\
&&+{\cal O}(\e^{-1})\Bigg],\\
&&\left[\GG1_{gg}\otimes{\cal F}_{12,j}^3\right]_1(s_{\b{1}\b{2}};z_{1},z_{2})=\left(\frac{s_{\b{1}\b{2}}}{\mu^2}\right)^{-\e}\Bigg[
\nonumber\\
&&\phantom{+}\frac{1}{\e^3}\Big[-\frac{11}{6}\delta(1-z_{1})\delta(1-z_{2})+\delta(1-z_{2})\Big(4-\frac{2}{z_1}-2z_1+2z_1^2-2{\cal D}_0(z_1)\Big)\Big]\nonumber\\
&&+\frac{1}{\e^2}\Big[-\frac{1}{3}\pi^2\delta(1-z_1)\delta(1-z_2)+\delta(1-z_1)\Big(-\frac{11}{3}+\frac{11}{6z_2}+\frac{11}{6}z_2-\frac{11}{6}z_2^2+\frac{11}{6}{\cal D}_{0}(z_2)\Big)
\nonumber\\
&&\hspace{1.0cm}+\delta(1-z_{2})\Big(\h(1,z_1)\Big(8-\frac{4}{z_1}-4z_1+4z_1^2\Big)+\h(0,z_1)\Big(-\frac{2}{z_1}-\frac{2}{1-z_1}-6z_1+2z_1^2\Big)
\nonumber\\
&&\hspace{1.0cm}+4{\cal D}_{1}(z_1)+\frac{11}{6}{\cal D}_{0}(z_1)+\frac{7}{3}-\frac{11}{2z_1}-\frac{25}{6}z_1+\frac{11}{2}z_1^2\Big)+{\cal D}_{0}(z_1)\Big(-4+\frac{2}{z_2}+2z_2-2z_2^2\Big)
\nonumber\\
&&\hspace{1.0cm}+{\cal D}_{0}(z_2)\Big(-4+\frac{2}{z_1}+2z_1-2z_1^2\Big)+2{\cal D}_{0}(z_1){\cal D}_{0}(z_2)+8+\frac{2}{z_1z_2}-\frac{4}{z_1}+\frac{2z_2}{z_1}-\frac{2z_2^2}{z_1}-\frac{4}{z_2}\nonumber\\
&&\hspace{1.0cm}-4z_2+4z_2^2
+\frac{2z_1}{z_2}-4z_1+2z_1z_2-2z_1z_2^2-\frac{2z_1^2}{z_2}+4z_1^2-2z_1^2z_2+2z_1^2z_2^2\Big]\nonumber\\
&&+{\cal O}(\e^{-1})\Bigg].
\end{eqnarray}

In all the cases where an initial-final antenna involves the opposite initial state leg, the resulting
expression can be obtained with $(z_1\leftrightarrow z_2)$ interchange.
\begin{eqnarray}
&&\left[{\cal F}_{ijk}^3\otimes{\cal F}_{2,np}^3\right](s_{ik},s_{\b{2}p};z_{1},z_{2})=\left[{\cal F}_{ijk}^3\otimes{\cal F}_{1,np}^3\right](s_{ik},s_{\b{2}p};z_{2},z_{1})\;,\\
&&\left[{\cal F}_{2,jk}^3\otimes{\cal F}_{2,np}^3\right](s_{\b{2}k},s_{\b{2}p};z_{1},z_{2})=\left[{\cal F}_{1,jk}^3\otimes{\cal F}_{1,np}^3\right](s_{\b{2}k},s_{\b{2}p},z_{2},z_{1})\;,\\
&&\left[{\cal F}_{\b{2},jk}^3\otimes{\cal F}_{\b{1}\b{2},p}^3\right](s_{\b{1}k},s_{\b{1}\b{2}};z_{1},z_{2})=\left[{\cal F}_{\b{1},jk}^3\otimes{\cal F}_{\b{1}\b{2},p}^3\right](s_{\b{2}k},s_{\b{1}\b{2}},z_{2},z_{1})
\;,\\
&&\left[\GG1_{gg}\otimes{\cal F}_{nop}^3\right]_2(s_{np};z_{1},z_{2})=\left[\GG1_{gg}\otimes{\cal F}_{nop}^3\right]_1(s_{np},z_{2},z_{1})\;,\\
&&\left[\GG1_{gg}\otimes{\cal F}_{2,jk}^3\right]_1(s_{\b{2}k};z_{1},z_{2})=\left[\GG1_{gg}\otimes{\cal F}_{1,jk}^3\right]_2(s_{\b{2}k},z_{2},z_{1})\;,\\
&&\left[\GG1_{gg}\otimes{\cal F}_{2,jk}^3\right]_2(s_{\b{2}k};z_{1},z_{2})=\left[\GG1_{gg}\otimes{\cal F}_{1,jk}^3\right]_1(s_{\b{2}k},z_{2},z_{1})\;,\\
&&\left[\GG1_{gg}\otimes{\cal F}_{12,j}^3\right]_2(s_{\b{1}\b{2}};z_{1},z_{2})=\left[\GG1_{gg}\otimes{\cal F}_{12,j}^3\right]_1(s_{\b{1}\b{2}},z_{2},z_{1})\;.
\end{eqnarray}

For completeness we give also the convolution of two Altarelli-Parisi splitting kernels,
\begin{eqnarray}
&&\left[\GG1_{gg}\otimes\GG1_{gg}\right]_1(z_{1},z_{2})=\nonumber\\
&&\phantom{+}\frac{1}{\e^2}\Big[\delta(1-z_1)\Big(\frac{121}{36}-\frac{2}{3}\pi^2\Big)+\h(1,z_1)\Big(16-\frac{8}{z_1}-8z_1+8z_1^2\Big)
+\h(0,z_1)\Big(-\frac{4}{z_1}-\frac{4}{1-z_1}\nonumber\\
&&\hspace{1.0cm}-12z_1+4z_1^2\Big)+8{\cal D}_{1}(z_1)+\frac{22}{3}{\cal D}_{0}(z_1)-\frac{8}{3}-\frac{22}{3z_1}-\frac{14}{3}z_1+\frac{22}{3}z_1^2\Big]\,\delta(1-z_{2})\;.
\end{eqnarray}

\subsection{Convolutions of plus distributions}
\label{sec:DnDm}
In this section we collect the results for the convolution of two plus distributions which appear in the convolution integrals of the previous section. They are defined as,
\begin{equation}
\left[{\cal D}_{n}\otimes{\cal D}_{m}\right](z)=\int {\rm d}x\,{\rm d}y \left(\frac{\ln^{n}(1-x)}{1-x}\right)_+\left(\frac{\ln^{n}(1-y)}{1-y}\right)_+\delta(z-xy)
\end{equation}
and read,
\begin{eqnarray}
\left[{\cal D}_{0}\otimes{\cal D}_{0}\right](z)&=&-\zeta_2\delta(1-z)+2{\cal D}_{1}(z)-\frac{\h(0,z)}{1-z}\;,\\
\left[{\cal D}_{1}\otimes{\cal D}_{0}\right](z)&=&\phantom{-}\zeta_3\delta(1-z)-\zeta_2{\cal D}_{0}(z)+\frac{3}{2}{\cal D}_{2}(z)+\frac{\h(0,1,z)}{1-z}+\frac{\h(1,0,z)}{1-z}\;,\\
\left[{\cal D}_{2}\otimes{\cal D}_{0}\right](z)&=&-2\zeta_4\delta(1-z)+2\zeta_3{\cal D}_{0}(z)-2\zeta_2{\cal D}_{1}(z)+\frac{4}{3}{\cal D}_{3}(z)-\frac{2\h(0,1,1,z)}{1-z}\;\nonumber\\
&&-\frac{2\h(1,0,1,z)}{1-z}-\frac{2\h(1,1,0,z)}{1-z}\;\\
\left[{\cal D}_{1}\otimes{\cal D}_{1}\right](z)&=&-\frac{\zeta_4}{4}\delta(1-z)+2\zeta_3{\cal D}_{0}(z)-2\zeta_2{\cal D}_{1}(z)+{\cal D}_{3}(z)-\frac{2\zeta_3}{1-z}-\frac{\h(0,1,0,z)}{1-z}\nonumber\\
&&-\frac{2\h(0,1,1,z)}{1-z}-\frac{2\h(1,0,1,z)}{1-z}-\frac{2\h(1,1,0,z)}{1-z}\;.
\end{eqnarray}

\bibliographystyle{JHEP-2}
\bibliography{ref}

\providecommand{\href}[2]{#2}\begingroup\raggedright\begin{thebibliography}{10}

\bibitem{cdfjet}
{\bf CDF} Collaboration, T.~Aaltonen {\em et.~al.}, {\it {Measurement of the
  Inclusive Jet Cross Section at the Fermilab Tevatron $p\bar p$ Collider Using
  a Cone-Based Jet Algorithm}},  {\em Phys. Rev.} {\bf D78} (2008) 052006
  [\href{http://arXiv.org/abs/0807.2204}{{\tt 0807.2204}}].

\bibitem{d0jet}
{\bf D0} Collaboration, V.~M. Abazov {\em et.~al.}, {\it {Measurement of the
  inclusive jet cross-section in $p\bar p$ collisions at $\sqrt{s}
  =1.96$~TeV}},  {\em Phys. Rev. Lett.} {\bf 101} (2008) 062001
  [\href{http://arXiv.org/abs/0802.2400}{{\tt 0802.2400}}].

\bibitem{atlasjet}
{\bf ATLAS} Collaboration, G.~Aad {\em et.~al.}, {\it {Measurement of inclusive
  jet and dijet cross sections in proton-proton collisions at 7 TeV
  centre-of-mass energy with the ATLAS detector}},  {\em Eur. Phys. J.} {\bf
  C71} (2011) 1512 [\href{http://arXiv.org/abs/1009.5908}{{\tt 1009.5908}}].

\bibitem{atlasjet2}
{\bf ATLAS} Collaboration, G.~Aad {\em et.~al.}, {\it {Measurement of inclusive
  jet and dijet production in $pp$ collisions at $\sqrt{s}=7$ TeV using the
  ATLAS detector}},  {\em Phys. Rev.} {\bf D86} (2012) 014022
  [\href{http://arXiv.org/abs/1112.6297}{{\tt 1112.6297}}].

\bibitem{cmsjet}
{\bf CMS} Collaboration, S.~Chatrchyan {\em et.~al.}, {\it {Measurement of the
  differential dijet production cross section in proton-proton collisions at
  $\sqrt{s}=7$ TeV}},  {\em Phys.Lett.} {\bf B700} (2011) 187--206
  [\href{http://arXiv.org/abs/1104.1693}{{\tt 1104.1693}}].

\bibitem{cmsjet2}
{\bf CMS} Collaboration, S.~Chatrchyan {\em et.~al.}, {\it {Measurement of the
  Inclusive Jet Cross Section in pp Collisions at sqrt(s) = 7 TeV}},  {\em
  Phys.Rev.Lett.} {\bf 107} (2011) 132001
  [\href{http://arXiv.org/abs/1106.0208}{{\tt 1106.0208}}].

\bibitem{cmsjet3}
{\bf CMS} Collaboration, S.~Chatrchyan {\em et.~al.}, {\it {Measurement of the
  inclusive production cross sections for forward jets and for dijet events
  with one forward and one central jet in pp collisions at sqrt(s) = 7 TeV}},
  {\em JHEP} {\bf 1206} (2012) 036 [\href{http://arXiv.org/abs/1202.0704}{{\tt
  1202.0704}}].

\bibitem{asjet}
{\bf D0} Collaboration, V.~M. Abazov {\em et.~al.}, {\it {Determination of the
  strong coupling constant from the inclusive jet cross section in $p\bar p$
  collisions at $\sqrt{s}=1.96$~TeV}},  {\em Phys. Rev.} {\bf D80} (2009)
  111107 [\href{http://arXiv.org/abs/0911.2710}{{\tt 0911.2710}}].

\bibitem{Giele:1995kb}
W.~T. Giele, E.~W.~N. Glover and J.~Yu, {\it {The Determination of $alpha_s$ at
  hadron colliders}},  {\em Phys. Rev.} {\bf D53} (1996) 120--130
  [\href{http://arXiv.org/abs/hep-ph/9506442}{{\tt hep-ph/9506442}}].

\bibitem{eks}
S.~D. Ellis, Z.~Kunszt and D.~E. Soper, {\it {Two jet production in hadron
  collisions at order $\alpha_s^3$ in QCD}},  {\em Phys. Rev. Lett.} {\bf 69}
  (1992) 1496--1499.

\bibitem{jetrad}
W.~T. Giele, E.~W.~N. Glover and D.~A. Kosower, {\it {The Two-Jet Differential
  Cross Section at ${\cal O}(\alpha_s^3)$ in Hadron Collisions}},  {\em Phys.
  Rev. Lett.} {\bf 73} (1994) 2019--2022
  [\href{http://arXiv.org/abs/hep-ph/9403347}{{\tt hep-ph/9403347}}].

\bibitem{nlojet1}
Z.~Nagy, {\it {Three-jet cross sections in hadron hadron collisions at
  next-to-leading order}},  {\em Phys. Rev. Lett.} {\bf 88} (2002) 122003
  [\href{http://arXiv.org/abs/hep-ph/0110315}{{\tt hep-ph/0110315}}].

\bibitem{nlojet2}
Z.~Nagy, {\it {Next-to-leading order calculation of three jet observables in
  hadron hadron collision}},  {\em Phys.Rev.} {\bf D68} (2003) 094002
  [\href{http://arXiv.org/abs/hep-ph/0307268}{{\tt hep-ph/0307268}}].

\bibitem{powheg2j}
S.~Alioli, K.~Hamilton, P.~Nason, C.~Oleari and E.~Re, {\it {Jet pair
  production in POWHEG}},  {\em JHEP} {\bf 1104} (2011) 081
  [\href{http://arXiv.org/abs/1012.3380}{{\tt 1012.3380}}].

\bibitem{meks}
J.~Gao {\em et.~al.}, {\it {MEKS: a program for computation of inclusive jet
  cross sections at hadron colliders}},
  \href{http://arXiv.org/abs/1207.0513}{{\tt 1207.0513}}.

\bibitem{Dittmaier:2012kx}
S.~Dittmaier, A.~Huss and C.~Speckner, {\it {Weak radiative corrections to
  dijet production at hadron colliders}},
  \href{http://arXiv.org/abs/1210.0438}{{\tt 1210.0438}}.

\bibitem{Catani:1996vz}
S.~Catani and M.~H. Seymour, {\it {A general algorithm for calculating jet
  cross sections in NLO QCD}},  {\em Nucl. Phys.} {\bf B485} (1997) 291--419
  [\href{http://arXiv.org/abs/hep-ph/9605323}{{\tt hep-ph/9605323}}].

\bibitem{Frixione:1995ms}
S.~Frixione, Z.~Kunszt and A.~Signer, {\it {Three jet cross-sections to
  next-to-leading order}},  {\em Nucl. Phys.} {\bf B467} (1996) 399--442
  [\href{http://arXiv.org/abs/hep-ph/9512328}{{\tt hep-ph/9512328}}].

\bibitem{Binoth:2000ps}
T.~Binoth and G.~Heinrich, {\it {An automatized algorithm to compute infrared
  divergent multi-loop integrals}},  {\em Nucl. Phys.} {\bf B585} (2000)
  741--759 [\href{http://arXiv.org/abs/hep-ph/0004013}{{\tt hep-ph/0004013}}].

\bibitem{Heinrich:2002rc}
G.~Heinrich, {\it {A numerical method for NNLO calculations}},  {\em Nucl.
  Phys. Proc. Suppl.} {\bf 116} (2003) 368--372
  [\href{http://arXiv.org/abs/hep-ph/0211144}{{\tt hep-ph/0211144}}].

\bibitem{Anastasiou:2003gr}
C.~Anastasiou, K.~Melnikov and F.~Petriello, {\it {A new method for real
  radiation at NNLO}},  {\em Phys. Rev.} {\bf D69} (2004) 076010
  [\href{http://arXiv.org/abs/hep-ph/0311311}{{\tt hep-ph/0311311}}].

\bibitem{Binoth:2004jv}
T.~Binoth and G.~Heinrich, {\it {Numerical evaluation of phase space integrals
  by sector decomposition}},  {\em Nucl. Phys.} {\bf B693} (2004) 134--148
  [\href{http://arXiv.org/abs/hep-ph/0402265}{{\tt hep-ph/0402265}}].

\bibitem{Anastasiou:2010pw}
C.~Anastasiou, F.~Herzog and A.~Lazopoulos, {\it {On the factorization of
  overlapping singularities at NNLO}},  {\em JHEP} {\bf 03} (2011) 038
  [\href{http://arXiv.org/abs/1011.4867}{{\tt 1011.4867}}].

\bibitem{qtsub}
S.~Catani and M.~Grazzini, {\it {An NNLO subtraction formalism in hadron
  collisions and its application to Higgs boson production at the LHC}},  {\em
  Phys. Rev. Lett.} {\bf 98} (2007) 222002
  [\href{http://arXiv.org/abs/hep-ph/0703012}{{\tt hep-ph/0703012}}].

\bibitem{GehrmannDeRidder:2005cm}
A.~Gehrmann-De~Ridder, T.~Gehrmann and E.~W.~N. Glover, {\it {Antenna
  Subtraction at NNLO}},  {\em JHEP} {\bf 09} (2005) 056
  [\href{http://arXiv.org/abs/hep-ph/0505111}{{\tt hep-ph/0505111}}].

\bibitem{stripper1}
M.~Czakon, {\it {A novel subtraction scheme for double-real radiation at
  NNLO}},  {\em Phys. Lett.} {\bf B693} (2010) 259--268
  [\href{http://arXiv.org/abs/1005.0274}{{\tt 1005.0274}}].

\bibitem{stripper2}
M.~Czakon, {\it {Double-real radiation in hadronic top quark pair production as
  a proof of a certain concept}},  {\em Nucl. Phys.} {\bf B849} (2011) 250--295
  [\href{http://arXiv.org/abs/1101.0642}{{\tt 1101.0642}}].

\bibitem{babishiggs1}
C.~Anastasiou, K.~Melnikov and F.~Petriello, {\it {Higgs boson production at
  hadron colliders: Differential cross sections through next-to-next-to-leading
  order}},  {\em Phys. Rev. Lett.} {\bf 93} (2004) 262002
  [\href{http://arXiv.org/abs/hep-ph/0409088}{{\tt hep-ph/0409088}}].

\bibitem{babishiggs2}
C.~Anastasiou, G.~Dissertori and F.~Stockli, {\it {NNLO QCD predictions for the
  $H \to WW \to l l \nu \nu$ signal at the LHC}},  {\em JHEP} {\bf 09} (2007)
  018 [\href{http://arXiv.org/abs/0707.2373}{{\tt 0707.2373}}].

\bibitem{babishiggs3}
C.~Anastasiou, F.~Herzog and A.~Lazopoulos, {\it {The fully differential decay
  rate of a Higgs boson to bottom-quarks at NNLO in QCD}},
  \href{http://arXiv.org/abs/1110.2368}{{\tt 1110.2368}}.

\bibitem{kirilldy}
K.~Melnikov and F.~Petriello, {\it {Electroweak gauge boson production at
  hadron colliders through O($\alpha_s^2$)}},  {\em Phys. Rev.} {\bf D74}
  (2006) 114017 [\href{http://arXiv.org/abs/hep-ph/0609070}{{\tt
  hep-ph/0609070}}].

\bibitem{grazzinihiggs}
M.~Grazzini, {\it {NNLO predictions for the Higgs boson signal in the $H\to WW
  \to \ell\nu \ell\nu$ and $H\to ZZ\to 4\ell$ decay channels}},  {\em JHEP}
  {\bf 02} (2008) 043 [\href{http://arXiv.org/abs/0801.3232}{{\tt 0801.3232}}].

\bibitem{grazzinidy1}
S.~Catani, L.~Cieri, G.~Ferrera, D.~de~Florian and M.~Grazzini, {\it {Vector
  boson production at hadron colliders: a fully exclusive QCD calculation at
  NNLO}},  {\em Phys. Rev. Lett.} {\bf 103} (2009) 082001
  [\href{http://arXiv.org/abs/0903.2120}{{\tt 0903.2120}}].

\bibitem{grazzinidy2}
S.~Catani, G.~Ferrera and M.~Grazzini, {\it {W boson production at hadron
  colliders: the lepton charge asymmetry in NNLO QCD}},  {\em JHEP} {\bf 05}
  (2010) 006 [\href{http://arXiv.org/abs/1002.3115}{{\tt 1002.3115}}].

\bibitem{grazziniwh}
G.~Ferrera, M.~Grazzini and F.~Tramontano, {\it {Associated WH production at
  hadron colliders: a fully exclusive QCD calculation at NNLO}},
  \href{http://arXiv.org/abs/1107.1164}{{\tt 1107.1164}}.

\bibitem{grazzinigg}
S.~Catani, L.~Cieri, D.~de~Florian, G.~Ferrera and M.~Grazzini, {\it {Diphoton
  production at hadron colliders: a fully- differential QCD calculation at
  NNLO}},  \href{http://arXiv.org/abs/1110.2375}{{\tt 1110.2375}}.

\bibitem{scettop}
J.~Gao, C.~S. Li and H.~X. Zhu, {\it {Top Quark Decay at Next-to-Next-to
  Leading Order in QCD}},  \href{http://arXiv.org/abs/1210.2808}{{\tt
  1210.2808}}.

\bibitem{our3j1}
A.~Gehrmann-De~Ridder, T.~Gehrmann, E.~W.~N. Glover and G.~Heinrich, {\it
  {Infrared structure of $e^+e^- \to 3$~jets at NNLO}},  {\em JHEP} {\bf 11}
  (2007) 058 [\href{http://arXiv.org/abs/0710.0346}{{\tt 0710.0346}}].

\bibitem{our3j2}
A.~Gehrmann-De~Ridder, T.~Gehrmann, E.~W.~N. Glover and G.~Heinrich, {\it {Jet
  rates in electron-positron annihilation at O($\alpha_s^3$) in QCD}},  {\em
  Phys. Rev. Lett.} {\bf 100} (2008) 172001
  [\href{http://arXiv.org/abs/0802.0813}{{\tt 0802.0813}}].

\bibitem{weinzierl3j1}
S.~Weinzierl, {\it {NNLO corrections to 3-jet observables in electron-positron
  annihilation}},  {\em Phys. Rev. Lett.} {\bf 101} (2008) 162001
  [\href{http://arXiv.org/abs/0807.3241}{{\tt 0807.3241}}].

\bibitem{weinzierl3j2}
S.~Weinzierl, {\it {The infrared structure of $e^+ e^- \to 3$~jets at NNLO
  reloaded}},  {\em JHEP} {\bf 07} (2009) 009
  [\href{http://arXiv.org/abs/0904.1145}{{\tt 0904.1145}}].

\bibitem{weinzierl3j3}
S.~Weinzierl, {\it {Jet algorithms in electron-positron annihilation:
  Perturbative higher order predictions}},  {\em Eur. Phys. J.} {\bf C71}
  (2011) 1565 [\href{http://arXiv.org/abs/1011.6247}{{\tt 1011.6247}}].

\bibitem{ourevent1}
A.~Gehrmann-De~Ridder, T.~Gehrmann, E.~W.~N. Glover and G.~Heinrich, {\it
  {Second-order QCD corrections to the thrust distribution}},  {\em Phys. Rev.
  Lett.} {\bf 99} (2007) 132002 [\href{http://arXiv.org/abs/0707.1285}{{\tt
  0707.1285}}].

\bibitem{ourevent2}
A.~Gehrmann-De~Ridder, T.~Gehrmann, E.~W.~N. Glover and G.~Heinrich, {\it {NNLO
  corrections to event shapes in $e^+e^-$ annihilation}},  {\em JHEP} {\bf 12}
  (2007) 094 [\href{http://arXiv.org/abs/0711.4711}{{\tt 0711.4711}}].

\bibitem{weinzierlevent1}
S.~Weinzierl, {\it {Event shapes and jet rates in electron-positron
  annihilation at NNLO}},  {\em JHEP} {\bf 06} (2009) 041
  [\href{http://arXiv.org/abs/0904.1077}{{\tt 0904.1077}}].

\bibitem{ourevent3}
A.~Gehrmann-De~Ridder, T.~Gehrmann, E.~W.~N. Glover and G.~Heinrich, {\it {NNLO
  moments of event shapes in $e^+e^-$ annihilation}},  {\em JHEP} {\bf 05}
  (2009) 106 [\href{http://arXiv.org/abs/0903.4658}{{\tt 0903.4658}}].

\bibitem{weinzierlevent2}
S.~Weinzierl, {\it {Moments of event shapes in electron-positron annihilation
  at NNLO}},  {\em Phys. Rev.} {\bf D80} (2009) 094018
  [\href{http://arXiv.org/abs/0909.5056}{{\tt 0909.5056}}].

\bibitem{czakontop1}
P.~Baernreuther, M.~Czakon and A.~Mitov, {\it {Percent Level Precision Physics
  at the Tevatron: First Genuine NNLO QCD Corrections to $q \bar{q} \to t
  \bar{t} + X$}},  {\em Phys.Rev.Lett.} {\bf 109} (2012) 132001
  [\href{http://arXiv.org/abs/1204.5201}{{\tt 1204.5201}}].

\bibitem{czakontop2}
M.~Czakon and A.~Mitov, {\it {NNLO corrections to top-pair production at hadron
  colliders: the all-fermionic scattering channels}},
  \href{http://arXiv.org/abs/1207.0236}{{\tt 1207.0236}}.

\bibitem{Czakon:2012pz}
M.~Czakon and A.~Mitov, {\it {NNLO corrections to top pair production at hadron
  colliders: the quark-gluon reaction}},
  \href{http://arXiv.org/abs/1210.6832}{{\tt 1210.6832}}.

\bibitem{Daleo:2006xa}
A.~Daleo, T.~Gehrmann and D.~Maitre, {\it {Antenna subtraction with hadronic
  initial states}},  {\em JHEP} {\bf 04} (2007) 016
  [\href{http://arXiv.org/abs/hep-ph/0612257}{{\tt hep-ph/0612257}}].

\bibitem{Daleo:2009yj}
A.~Daleo, A.~Gehrmann-De~Ridder, T.~Gehrmann and G.~Luisoni, {\it {Antenna
  subtraction at NNLO with hadronic initial states: initial-final
  configurations}},  {\em JHEP} {\bf 01} (2010) 118
  [\href{http://arXiv.org/abs/0912.0374}{{\tt 0912.0374}}].

\bibitem{Boughezal:2010mc}
R.~Boughezal, A.~Gehrmann-De~Ridder and M.~Ritzmann, {\it {Antenna subtraction
  at NNLO with hadronic initial states: double real radiation for
  initial-initial configurations with two quark flavours}},  {\em JHEP} {\bf
  02} (2011) 098 [\href{http://arXiv.org/abs/1011.6631}{{\tt 1011.6631}}].

\bibitem{Gehrmann:2011wi}
T.~Gehrmann and P.~F. Monni, {\it {Antenna subtraction at NNLO with hadronic
  initial states: real-virtual initial-initial configurations}},  {\em JHEP}
  {\bf 1112} (2011) 049 [\href{http://arXiv.org/abs/1107.4037}{{\tt
  1107.4037}}].

\bibitem{GehrmannDeRidder:2012ja}
A.~Gehrmann-De~Ridder, T.~Gehrmann and M.~Ritzmann, {\it {Antenna subtraction
  at NNLO with hadronic initial states: double real initial-initial
  configurations}},  {\em JHEP} {\bf 1210} (2012) 047
  [\href{http://arXiv.org/abs/1207.5779}{{\tt 1207.5779}}].

\bibitem{Abelof:2011jv}
G.~Abelof and A.~Gehrmann-De~Ridder, {\it {Antenna subtraction for the
  production of heavy particles at hadron colliders}},  {\em JHEP} {\bf 04}
  (2011) 063 [\href{http://arXiv.org/abs/1102.2443}{{\tt 1102.2443}}].

\bibitem{Abelof:2011ap}
G.~Abelof and A.~Gehrmann-De~Ridder, {\it {Double real radiation corrections to
  $t\bar{t}$ production at the LHC: the all-fermion processes}},  {\em JHEP}
  {\bf 1204} (2012) 076 [\href{http://arXiv.org/abs/1112.4736}{{\tt
  1112.4736}}].

\bibitem{Abelof:2012rv}
G.~Abelof and A.~Gehrmann-De~Ridder, {\it {Double real radiation corrections to
  $t\bar{t}$ production at the LHC: the $gg\rightarrow t\bar{t}q\bar{q}$
  channel}},  \href{http://arXiv.org/abs/1207.6546}{{\tt 1207.6546}}.

\bibitem{Abelof:2012he}
G.~Abelof, A.~Gehrmann-De~Ridder and O.~Dekkers, {\it {Antenna subtraction with
  massive fermions at NNLO: Double real initial-final configurations}},
  \href{http://arXiv.org/abs/1210.5059}{{\tt 1210.5059}}.

\bibitem{Glover:2010im}
E.~W.~N. Glover and J.~Pires, {\it {Antenna subtraction for gluon scattering at
  NNLO}},  {\em JHEP} {\bf 06} (2010) 096
  [\href{http://arXiv.org/abs/1003.2824}{{\tt 1003.2824}}].

\bibitem{GehrmannDeRidder:2011aa}
A.~Gehrmann-De~Ridder, E.~W.~N. Glover and J.~Pires, {\it {Real-Virtual
  corrections for gluon scattering at NNLO}},  {\em JHEP} {\bf 1202} (2012) 141
  [\href{http://arXiv.org/abs/1112.3613}{{\tt 1112.3613}}].

\bibitem{Catani:1998bh}
S.~Catani, {\it {The singular behaviour of {QCD} amplitudes at two-loop
  order}},  {\em Phys. Lett.} {\bf B427} (1998) 161--171
  [\href{http://arXiv.org/abs/hep-ph/9802439}{{\tt hep-ph/9802439}}].

\bibitem{Sterman:2002qn}
G.~Sterman and M.~E. Tejeda-Yeomans, {\it {Multi-loop amplitudes and
  resummation}},  {\em Phys. Lett.} {\bf B552} (2003) 48--56
  [\href{http://arXiv.org/abs/hep-ph/0210130}{{\tt hep-ph/0210130}}].

\bibitem{Berends:1987cv}
F.~A. Berends and W.~Giele, {\it {The Six Gluon Process as an Example of
  Weyl-Van Der Waerden Spinor Calculus}},  {\em Nucl. Phys.} {\bf B294} (1987)
  700.

\bibitem{Kosower:1987ic}
D.~A. Kosower, B.-H. Lee and V.~P. Nair, {\it {Multi gluon scattering: a string
  based calculation}},  {\em Phys. Lett.} {\bf B201} (1988) 85.

\bibitem{Mangano:1987xk}
M.~L. Mangano, S.~J. Parke and Z.~Xu, {\it {Duality and Multi - Gluon
  Scattering}},  {\em Nucl. Phys.} {\bf B298} (1988) 653.

\bibitem{Mangano:1990by}
M.~L. Mangano and S.~J. Parke, {\it {Multi-Parton Amplitudes in Gauge
  Theories}},  {\em Phys. Rept.} {\bf 200} (1991) 301--367
  [\href{http://arXiv.org/abs/hep-th/0509223}{{\tt hep-th/0509223}}].

\bibitem{Bern:1994zx}
Z.~Bern, L.~J. Dixon, D.~C. Dunbar and D.~A. Kosower, {\it {One-Loop n-Point
  Gauge Theory Amplitudes, Unitarity and Collinear Limits}},  {\em Nucl. Phys.}
  {\bf B425} (1994) 217--260 [\href{http://arXiv.org/abs/hep-ph/9403226}{{\tt
  hep-ph/9403226}}].

\bibitem{Glover:2001af}
E.~W.~N. Glover, C.~Oleari and M.~E. Tejeda-Yeomans, {\it {Two loop QCD
  corrections to gluon-gluon scattering}},  {\em Nucl.Phys.} {\bf B605} (2001)
  467--485 [\href{http://arXiv.org/abs/hep-ph/0102201}{{\tt hep-ph/0102201}}].

\bibitem{Glover:2001rd}
E.~W.~N. Glover and M.~Tejeda-Yeomans, {\it {One loop QCD corrections to
  gluon-gluon scattering at NNLO}},  {\em JHEP} {\bf 0105} (2001) 010
  [\href{http://arXiv.org/abs/hep-ph/0104178}{{\tt hep-ph/0104178}}].

\bibitem{Kosower:2002su}
D.~A. Kosower, {\it {Multiple singular emission in gauge theories}},  {\em
  Phys. Rev.} {\bf D67} (2003) 116003
  [\href{http://arXiv.org/abs/hep-ph/0212097}{{\tt hep-ph/0212097}}].

\bibitem{Vogt:2004mw}
A.~Vogt, S.~Moch and J.~A.~M. Vermaseren, {\it {The three-loop splitting
  functions in QCD: The singlet case}},  {\em Nucl. Phys.} {\bf B691} (2004)
  129--181 [\href{http://arXiv.org/abs/hep-ph/0404111}{{\tt hep-ph/0404111}}].

\bibitem{Remiddi:1999ew}
E.~Remiddi and J.~A.~M. Vermaseren, {\it {Harmonic polylogarithms}},  {\em Int.
  J. Mod. Phys.} {\bf A15} (2000) 725--754
  [\href{http://arXiv.org/abs/hep-ph/9905237}{{\tt hep-ph/9905237}}].

\end{thebibliography}\endgroup
\end{document}